\documentclass[aps,prl,noshowpacs,twocolumn,superscriptaddress,nobibnotes]{revtex4-1}
\usepackage{graphicx}
\usepackage{amsmath,amsfonts}
\usepackage{hyperref}
\usepackage[utf8] {inputenc}
\usepackage{color}  
\usepackage[normalem]{ulem}
\usepackage[T1]{fontenc} 
\usepackage{multirow}
\usepackage{booktabs}
\usepackage{threeparttable}
\usepackage{tabularx}
\renewcommand\appendix{\setcounter{secnumdepth}{0}}

\begin{document}
	\title{Geometric renormalization unravels \\self-similarity of the multiscale human connectome}
	
	\author{Muhua Zheng}
	\affiliation{Departament de F\'isica de la Mat\`eria Condensada, Universitat de Barcelona, Mart\'i i Franqu\`es 1, 08028 Barcelona, Spain}
	\affiliation{Universitat de Barcelona Institute of Complex Systems (UBICS), Universitat de Barcelona, Barcelona, Spain}
	
	\author{Antoine Allard}
	\affiliation{D\'epartement de physique, de g\'enie physique et d'optique, Universit\'e Laval, Qu\'ebec , Canada G1V 0A6}
	\affiliation{ Centre interdisciplinaire de mod\'elisation math\'ematique, Universit\'e Laval,
		Qu\'ebec, Canada G1V 0A6}
	
	\author{Patric Hagmann}
	\affiliation{Connectomics Lab, Department of Radiology, Lausanne University Hospital and University of Lausanne (CHUV-UNIL), 1011 Lausanne, Switzerland}
	
	\author{Yasser Alem\'an-G\'omez}
	\affiliation{Connectomics Lab, Department of Radiology, Lausanne University Hospital and University of Lausanne (CHUV-UNIL), 1011 Lausanne, Switzerland}
	\affiliation{ Center for Psychiatric Neurosciences, Department of Psychiatry, Lausanne University Hospital and University of Lausanne (CHUV-UNIL), 1008 Prilly, Switzerland}
	\affiliation{ Medical Image Analysis Laboratory, Lausanne University Hospital and University of Lausanne (CHUV-UNIL), 1011 Lausanne, Switzerland}
	
	\author{M. {\'A}ngeles Serrano}
	\email[]{marian.serrano@ub.edu}
	\affiliation{Departament de F\'isica de la Mat\`eria Condensada, Universitat de Barcelona, Mart\'i i Franqu\`es 1, 08028 Barcelona, Spain}
	\affiliation{Universitat de Barcelona Institute of Complex Systems (UBICS), Universitat de Barcelona, Barcelona, Spain}
	\affiliation{Catalan Institution for Research and Advanced Studies (ICREA), 08010 Barcelona, Spain}
	\date{\today}
\begin{abstract}
  Structural connectivity in the brain is typically studied by reducing its observation to a single spatial resolution. However, the brain possesses a rich architecture organized over multiple scales linked to one another. We explored the multiscale organization of human connectomes using datasets of healthy subjects reconstructed at five different resolutions. We found that the structure of the human brain remains self-similar when the resolution of observation is progressively decreased by hierarchical coarse-graining of the anatomical regions. Strikingly, a geometric network model, where distances are not Euclidean, predicts the multiscale properties of connectomes, including self-similarity. The model relies on the application of a geometric renormalization protocol which decreases the resolution by coarse-graining and averaging over short similarity distances. Our results suggest that simple organizing principles underlie the multiscale architecture of human structural brain networks, where the same connectivity law dictates short- and long-range connections between different brain regions over many resolutions. The implications are varied and can be substantial for fundamental debates, such as whether the brain is working near a critical point, as well as for applications including advanced tools to simplify the digital reconstruction and simulation of the brain.
\end{abstract}
\maketitle
\let\oldaddcontentsline\addcontentsline
\renewcommand{\addcontentsline}[3]{}

\textbf{
The architecture of the human brain underlies human behavior and is extremely complex with multiple scales interacting with one another. However, research efforts are typically focused on a single spatial scale. We explored the spatial multiscale organization of the human brain by using two high-quality datasets with connectomes at five anatomical resolutions for 84 healthy subjects. We found that the zoomed out layers remain self-similar, and that a geometric network model, where distances are not Euclidean, predicts the observations by application of a renormalization protocol. Our results prove that the same principles organize brain connectivity at different scales and lead to efficient decentralized communication. Implications extend to debates, like criticality in the brain, and applications, including tools for brain simulation.}

\section*{Introduction}
Extensive study of the topology of the human connectome~\cite{Bullmore2009,Rubinov2010,Fornito2016} has revealed characteristic features of complex networks, including the small-world phenomenon~\cite{Watts1998,Sporns2004,He2007,Hagmann2007}, high levels of clustering~\cite{Hagmann2007}, heterogeneous degree distributions (even though not scale-free)~\cite{Gong2008,Crossley2014}, rich club effect~\cite{VandenHeuvel2011}, and community structure~\cite{Sporns2016,Meunier2010}. These structural features have been typically observed at specific scales fixed by the resolution of the experimental imaging technique. Only very recently did brain networks began to be considered simultaneously at multiple resolutions. Notable contributions investigated network scaling effects in human resting-state fMRI data~\cite{fornito2010network}, introduced techniques for the reconstruction of multiscale connectomes~\cite{Cammoun2012}, proposed network algorithms for multiscale community detection~\cite{Betzel2013,Betzel2017}, and analyzed network metrics across parcellation scales~\cite{lacy2019effects}. However, other features remain unexplored, and modeling approaches are still missing. Novel methodological advances are needed to understand the multiscale organization of the human brain and, in particular, how the different scales are interrelated.

In the context of complex networks, the study of the multiscale problem~\cite{Song2005,Garcia2018}---and related concepts like scale invariance and self-similarity~\cite{Mandelbrot1982}---is built upon the renormalization technique of statistical physics~\cite{Wilson1983,Kadanoff2000}, which successfully explained the universality of critical behavior in phase transitions~\cite{Stanley1971}. The method allows a systematic investigation of the changes of a physical system as viewed at different length scales. More specifically, block spin renormalization~\cite{efrati2014real} gives a clear procedure to link the different scales by a transformation that aggregates components at short distances to define components at larger distances, eliminating from the system degrees of freedom whose scale is smaller than the new component size. These theories inspired a geometric renormalization (GR) transformation for real networks that allows to explore them at different resolutions~\cite{Garcia2018}.

The GR transformation is based on a geometric network model that positions nodes in a hidden metric space, thereby defining a map, such that the closer two nodes are in the space, the more likely is that they are connected~\cite{Serrano2008}. The model explains universal features of real networks ---including the small-world property, heterogeneous degree distributions, and clustering--- as well as fundamental mechanisms ---including preferential attachment in growing networks~\cite{Papadopoulos2012} and the emergence of communities~\cite{Zuev:2015aa,GarciaPerez:2018aa}--- by assuming the hyperbolic plane as the natural geometry to embed hierarchical topologies, and hence complex networks~\cite{Krioukov2009}. Hyperbolic maps of real networks can be obtained using statistical inference techniques~\cite{Boguna2010,Guille2019}, and have been observed to sustain efficient navigation~\cite{Boguna2010}. These result are also valid for connectomes of different species~\cite{Allard:2018}, implying that distances between brain regions in the hyperbolic plane offer a more accurate interpretation of the structure of connectomes as compared to results based on physical distances in Euclidean space. This is in line with recent findings that the brain's Euclidean embedding has a major but not definitive role in shaping neuronal network architecture~\cite{vertes2012simple,henderson2013using,henderson2014relations,Roberts:2016,betzel2016generative,stiso2018spatial}. Finally, maps of real networks provide effective distances to apply GR. The transformation unfolds real scale-free networks ---the Internet, word adjacencies in Darwin's {\it The Origin of Species}, the human metabolic network, and more--- into a shell of layers that distinguishes the coexisting scales and their interactions, revealing self-similarity over multiple scales as a ubiquitous symmetry~\cite{Garcia2018}.

In this work, we reconstructed multiscale human (MH) connectomes at five anatomical resolutions for a total of 84 healthy subjects in two different datasets. First, we measured network properties of the connectomes at each scale and found that their structure remained self-similar as the resolution of observation was progressively reduced. Second, we obtained the hyperbolic map of the highest resolution layer of each connectome and applied GR to obtain a multiscale unfolding. At each scale, we found a striking congruency between the empirical observations and the predictions given by the model. Third, we explored the impact of impairing the geometric properties of connectomes on self-similarity and navigation. Altogether, our results indicate that the same rules explain the formation of short-range and long-range connections in the brain---within the range of length scales covered by the datasets---, and support GR as a valid archetypical model for the multiscale structure of the human brain.
%
%
\section*{Empirical evidence for the self-similarity of the multiscale human connectome}
%
We used two different datasets with a total of 84 healthy human subjects. The first dataset (UL, University of Lausanne) contains diffusion spectrum MRI data of 40 subjects scanned at the University of Lausanne. Neural fibers connecting pairs of regions were tracked by following directions of maximum diffusion. The second dataset (HCP, Human Connectome Project)~\cite{Van2012}, used to cross-validate the results, contains the multiscale connectomes of 44 healthy subjects of the Test-Retest subsample. The fiber bundles were estimated by employing the intravoxel fiber Orientation Distribution Functions (fODFs) computed by a constrained spherical deconvolution technique~\cite{Tournier2012}. All connectomes in the two datasets were reconstructed using deterministic streamline tractography, and the multiscale parcellation of the cortex was obtained with the approach proposed in Ref.~\cite{Cammoun2012}. Details on the acquisition and processing of the datasets, and justification for the convenience of using deterministic algorithms for our purposes, are described in Materials and Methods. Even if the UL dataset is significantly sparser than HCP, see SI Appendix, Tables S1 and S2, similar results were found for both cohorts.

\begin{figure}[t]
	\centering
	\includegraphics[width=1\linewidth]{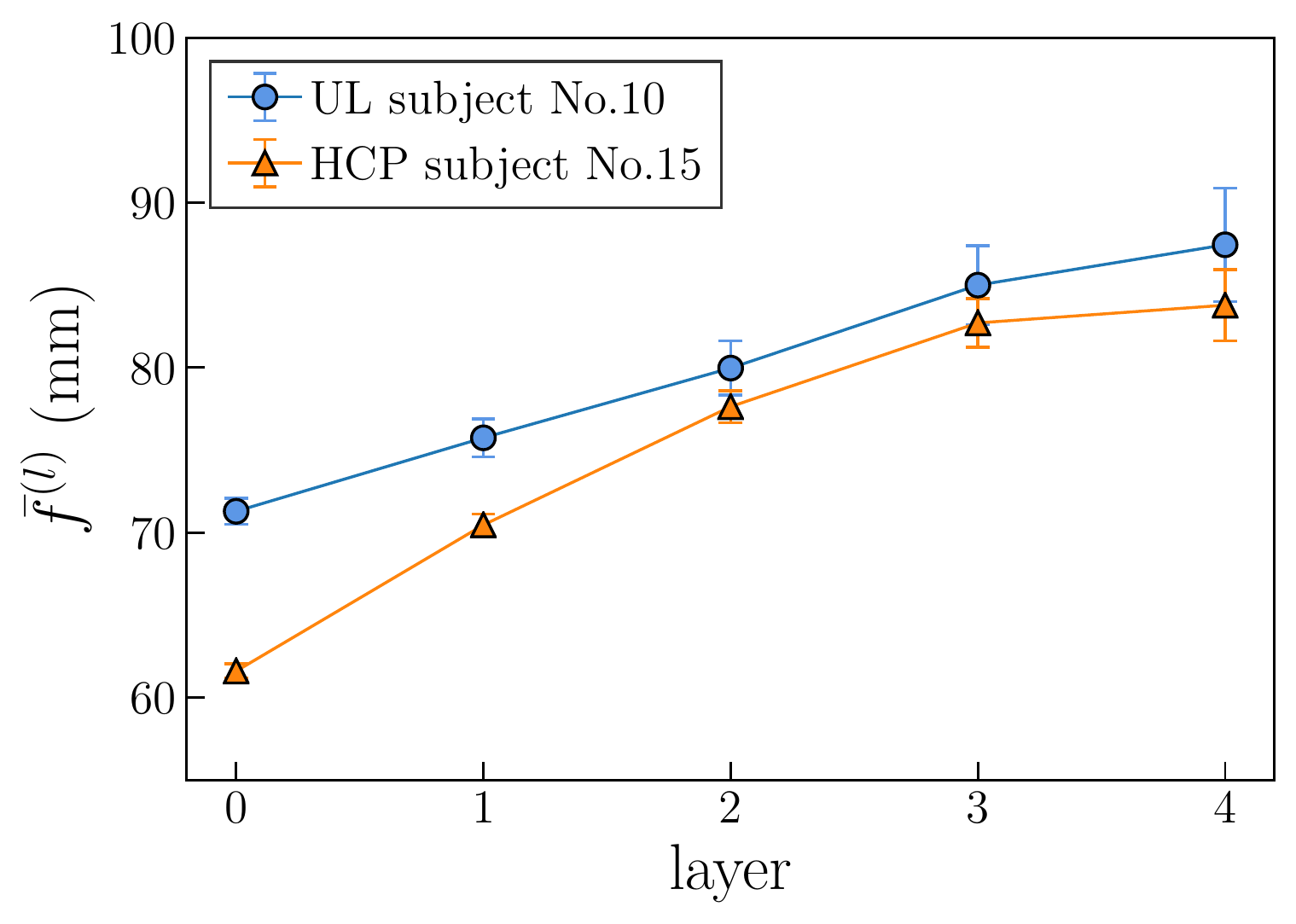}
	\caption{\label{fig:Fig0}
		\textbf{Average fiber length of connections in MH connectomes at different resolutions} for UL subject No.~10 and HCP subject No.~15 (see SI Appendix, Fig.~S3 and S20 for all subjects). Error bars indicate the $2$ standard error interval around the mean.}
\end{figure}

For each subject, we reconstructed a multiscale connectome organized in five layers with different anatomical resolutions following Ref.~\cite{Cammoun2012}; details can be found in Materials and Methods. Nodes in each layer correspond to parcels in the cortical and subcortical regions (the brainstem is excluded), and connections denote the presence of fibers between them. The multiscale parcelation is anatomically hierarchical, and was obtained by iterating a coarse-graining operation starting at layer $l=0$ to produce a subsequent layer with a reduced resolution. The technique consists in grouping sets of $2$ or $3$ neighboring brain regions to build a new brain partition and recomputing connection densities between each pair of the resulting parcels. The layers contain roughly 1014, 462, 233, 128, and 82 nodes (these numbers slightly fluctuate across subjects, see SI Appendix, Tables~S1 and S2), and are labeled $l=0, 1, 2, 3, 4$, respectively. The hierarchical anatomical coarse-graining determines the sequence of length scales characterizing the multiscale connectomes. As the resolution decreases, each node corresponds to a larger parcel of the brain, and the average fiber length of connections, computed from streamline tractography, also increases since short-distance connections are absorbed inside coarse-grained parcels, see Fig.~\ref{fig:Fig0}.

\begin{figure*}[t]
	\centering
	\includegraphics[width=1\linewidth]{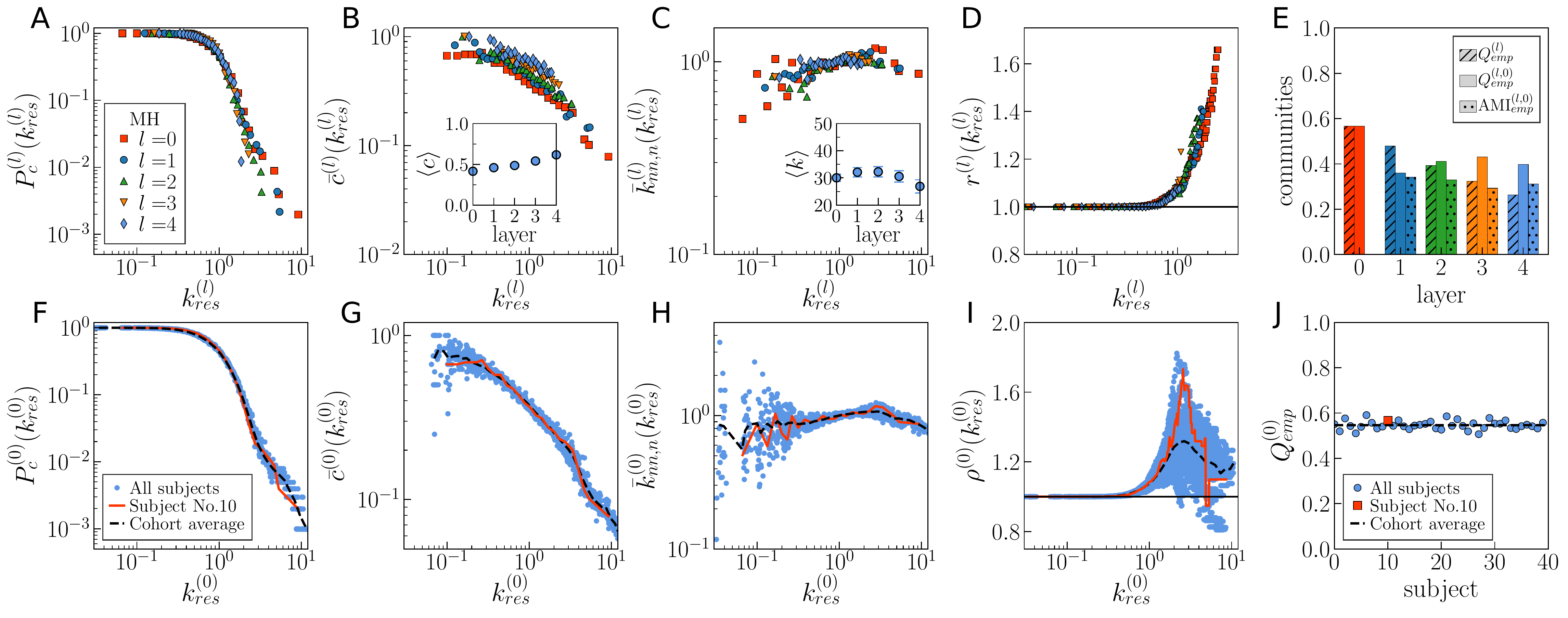}
	\caption{\label{fig:Fig1}\textbf{Self-similarity of the MH connectome across different resolutions}. (\textit{A-E}) Results for UL subject No.~10.
		(\textit{A}) Complementary cumulative degree distribution $P_c^{(l)}(k_{res}^{(l)})$.
		(\textit{B}) Degree dependent clustering coefficient $\bar{c}^{(l)}(k_{res}^{(l)})$. Inset: average clustering coefficient $\langle c \rangle$ across layers.
		(\textit{C}) Normalized average degree of nearest neighbors $\bar{k}_{nn,n}^{(l)} (k_{res}^{(l)})$. Inset: average degree $\langle k\rangle$ across layers.
		In the two insets in (\textit{B})-(\textit{C}), error bars indicate the $2$ standard error interval around the mean; note that some of the bars are smaller than symbols.
		(\textit{D}) Rich club coefficient $r^{(l)}(k_{res}^{(l)})$ for low and intermediate values of the rescaled threshold degrees.
		(\textit{E}) Community structure. $Q^{(l)}$ is the modularity in layer $l$, $Q^{(l,0)}$ is the modularity that the community structure of layer $l$ induces in layer $0$, and $\textrm{AMI}^{(l,0)}$ is the adjusted mutual information between the latter and the community partition directly detected in layer $0$, see Materials and Methods. The subscript $emp$ indicates the empirical MH connectome.
		(\textit{F--J}) Variability of topological properties in the UL dataset. Blue symbols correspond to the properties of layer $0$ for each of the 40 subjects. The red lines correspond to UL subject No.~10. The black dashed line represents the average value across the 40 subjects in the cohort. Degrees have been rescaled by the average degree of the corresponding layer $k^{(l)}_{res} = k^{(l)}/ \langle k^{(l)}\rangle$.}
\end{figure*}

For each layer $l$ of each subject, we measured the following properties: complementary cumulative degree distribution $P_c^{(l)}(k_{res}^{(l)})$, degree-degree correlations using the normalized average degree of nearest neighbors $\bar{k}_{nn,n}^{(l)} (k_{res}^{(l)}) = \bar{k}_{nn}^{(l)} (k_{res}^{(l)}) \langle  k^{(l)}\rangle/\langle(k^{(l)})^2\rangle$, degree-dependent clustering coefficient $\bar{c}^{(l)}(k_{res}^{(l)})$, rich club coefficient $r^{(l)}(k_{res}^{(l)})$~\cite{Colizza2006}, average degree and average clustering coefficient.  These quantities were calculated as a function of the rescaled degree $k^{(l)}_{res} = k^{(l)}/ \langle k^{(l)}\rangle$ to account for the variation of the average degree across layers. Figure~\ref{fig:Fig1} shows the results for a typical subject in the UL dataset (see SI Appendix, Figs.~S4-S11 and Figs.~S21-S30 for all connectomes in the UL and HCP datasets, respectively). The overlap of the curves in Figs.~\ref{fig:Fig1}A--\ref{fig:Fig1}D denotes self-similarity across layers for the degree distribution, degree-degree correlations, clustering, and rich club effect (note that Fig.~\ref{fig:Fig1}D omits the values corresponding to high degree thresholds to avoid cluttering the plot since the corresponding subgraphs are typically small and thus very noisy, see SI Appendix, Figs.~S7 for the complete curves).

The insets on Figs.~\ref{fig:Fig1}B~and~\ref{fig:Fig1}C display the average clustering coefficient, $\langle c \rangle$, and the average degree, $\langle k \rangle$, across the 5 layers of the MH connectome.  We see that $\langle c \rangle$ increases, first midly then more pronouncedly, as the resolution decreases (i.e. as $l$ goes from 0 to 4), which explains the shift observed in the corresponding $\bar{c}^{(l)}(k_{res}^{(l)})$ curves in Fig.~\ref{fig:Fig1}B. On the other hand, $\langle k \rangle$ first increases slightly---compatible with an almost constant average degree--- and then decreases more markedly in layers $3$ and $4$.  Values for the standard error of these average values are given in SI Appendix, Tables S1 and S2 for the UL and the HCP connectomes, respectively. The last two layers in the MH connectomes are more prone to finite-size effects due to their smaller number of nodes, and are also affected by a higher variability in the surface area of the anatomical regions, which may cause biases in streamline determination, see SI Appendix, Fig.~S1 and S2.

Finally, we also inferred the community partition using the Louvain method~\cite{Blondel2008}. The modularity $Q^{(l)}_{emp}$ of the detected partitions are shown in Fg.~\ref{fig:Fig1}E, along with the adjusted mutual information $\textrm{AMI}_{emp}^{(l)}$ between the community partition detected in layer $0$ and the community partition induced in layer $0$ by that in layer $l$ ---with modularity $Q^{(l,0)}_{emp}$--- (see Methods). The overlap between communities at different resolutions remains important even if the modularity is slightly weakened, especially in the last two layers where the finite-size effects are stronger due to their reduced size.

Altogether, our results strongly support the self-similarity of each MH connectome in the datasets. Moreover, we found the connectomes of different subjects in each dataset to be similar to one another. Figures~\ref{fig:Fig1}F--\ref{fig:Fig1}J show the properties measured in layer 0 of every subject in the UL dataset with the subject used in Figs.~\ref{fig:Fig1}A--\ref{fig:Fig1}E highlighted. Information about the other layers in the UL dataset can be found in SI Appendix, Figs.~S10 and S11, and SI Appendix, Figs.~S29-S30 provide results for all layers of the HCP dataset. The homogeneity across subjects within a dataset is further supported by the results of statistical tests that evaluate the congruency of each connectome at $l=0$ with the cohort average. For each connectome, we compared the degree of the nodes, the sum of the degree of their neighbors, and the number of triangles to which each node participates against the corresponding cohort averages. This cohort average was obtained by computing, for each brain region, the mean and standard deviation $\sigma$ of these three properties over all subjects. SI Appendix, Tables~S3~and~S4 provide the values for the Pearson correlation coefficient $\rho$, the $\chi^2$ test ($\chi^2=\sum_{i}^{N}(\frac{value_{real}-value_{group}}{\sigma_{group}})^2$) normalized by the number of nodes, and the score $\zeta$ quantifying the fraction of nodes with values outside the $2\sigma$ confidence interval around the mean.

\section*{Geometric renormalizaton of the human connectome}
%

We now show that the observed scale invariance of the real MH connectomes can be explained by a geometric network model, where distances are not Euclidean, that includes a renormalization protocol~\cite{Garcia2018}.

\subsection*{Geometric description of connectomes}
\begin{figure*}[ht]
	\centering
	\includegraphics[width=0.9\linewidth]{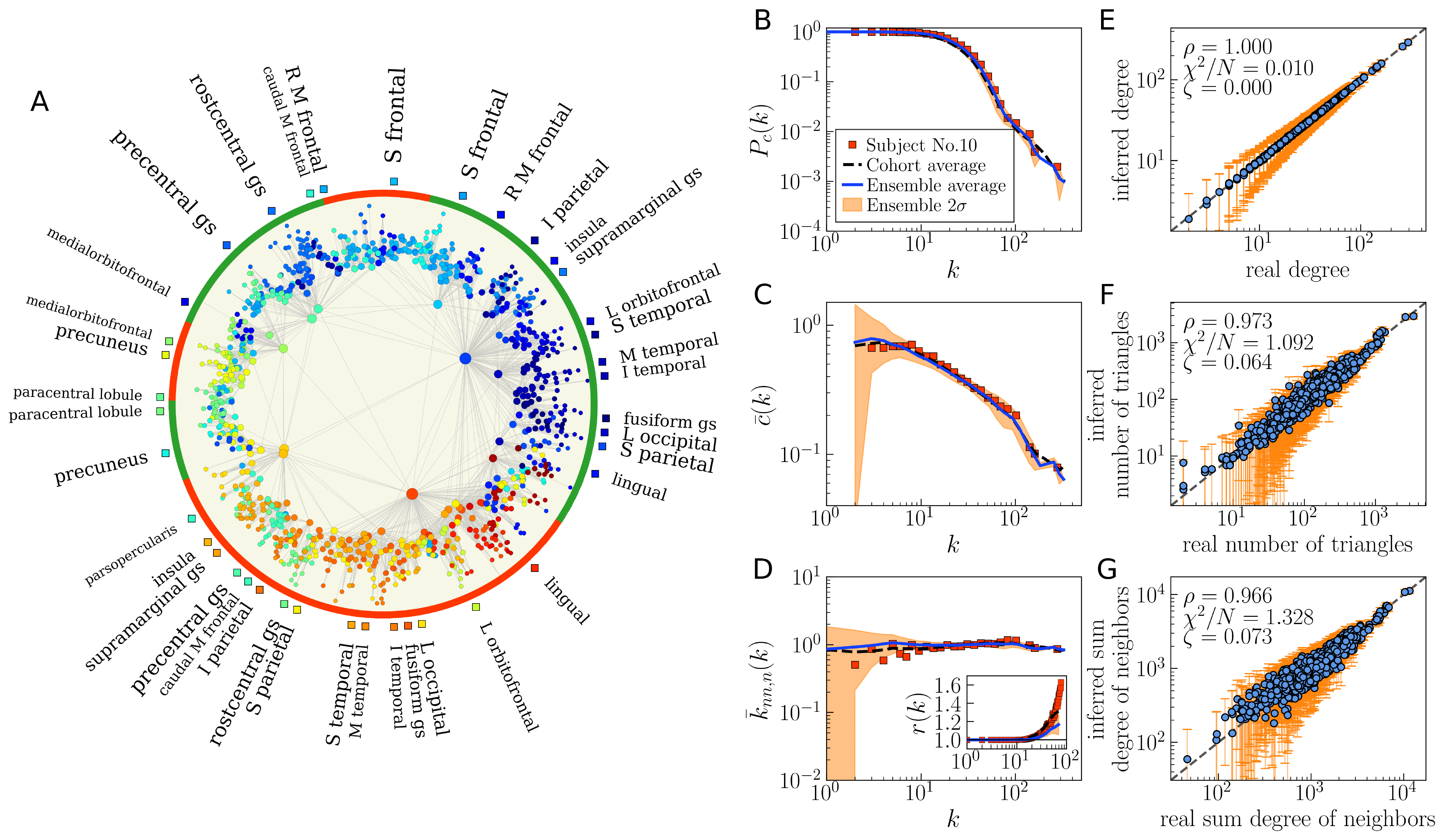}
	\caption{\label{fig:Fig2}  \textbf{Hyperbolic connectome map of UL subject No.~10}.
		(\textit{A}) Embedding of $l=0$ in the hyperbolic disk. Nodes are colored according to the 82 coarse-grained regions in layer $l=4$. Only links with connection probability greater than $0.5$ are shown. The size of each node is proportional to the logarithm of its degree, and the font size of the names of brain regions is proportional to the logarithm of the number of nodes in the regions (only regions with more than 10 nodes are shown). Red and green frames indicate left and right hemispheres, respectively. A=anterior, I=inferior, L=lateral, M=middle, R=rostral, S=superior, gs=gyrus.
		(\textit{B})--(\textit{D}) Network properties of $l=0$ compared to the model predictions, (\textit{B}) complementary cumulative degree distribution, (\textit{C}) degree-dependent clustering coefficient, (\textit{D}) average degree of nearest neighbors, and (\textit{D})-inset rich club coefficient. Red symbols correspond to subject No.~10, and the black dashed lines correspond to the group average across the 40 subjects in the UL dataset. The blue lines correspond to the average value obtained from $100$ synthetic networks generated with the $\mathbb{S}^1$ model using the coordinates and parameters inferred by Mercator~\cite{Guille2019}, and the orange regions show the $2\sigma$ confidence interval around the expected value.
		(\textit{E})--(\textit{G}) Comparison of the predictions of the $\mathbb{S}^1$ model (average over the ensemble of $100$ synthetic networks) with the actual values for (\textit{E}) degrees, (\textit{F}) number of triangle attached to each node, and (\textit{G}) sum of degrees of neighbors. Error bars show the $2\sigma$ confidence interval around the average values. Statistical tests for the goodness of fit --- Pearson correlation coefficient $\rho$, $\chi^2$ test normalized by the number of nodes $N$, and $\zeta$ score--- are reported in each subfigure.}
\end{figure*}

Connectome maps~\cite{Allard:2018} are based on the $\mathbb{S}^1$ network model \cite{Serrano2008}. Each brain region $i$  is characterized by two random variables: a hidden degree $\kappa_i$, that quantifies its \textit{popularity} and sets its scale of connectivity, and an angular position $\theta_i$ in a one-dimensional sphere (circle), or \textit{similarity} space, aggregating all other attributes that modulate the likelihood of connections including, but not limited to, the Euclidean physiological three-dimensional embedding of the brain.

Connections are pairwise in the $\mathbb{S}^1$ model, and their probability takes the form of the gravity law
\begin{align} \label{eq:con_pro}
p_{ij} = \frac{1}{1+\lambda_{ij}^\beta}=\frac{1}{1+\left(\frac{d_{ij}}{\mu \kappa_i \kappa_j}\right)^\beta}, \
\end{align}
Hence, the likelihood of a link between two nodes increases with the product of their hidden degrees and decreases with their angular distance (therefore increasing with their similarity). Parameter $\mu$ controls the average degree of synthetic connectomes produced by the model while $\beta$ controls the level of clustering, and so the coupling strength between the topology of the network and its underlying geometry (clustering is the topological signature of the triangle inequality in the underlying metric space). The angular distance $\Delta\theta_{ij}=\pi-|\pi-|\theta_i-\theta_j||$, combined with the radius $R$ of the similarity subspace (we set $R=N/2\pi$ to fix the density of nodes on the circle to $1$), gives the similarity distance $d_{ij}=R\Delta \theta{ij}$. By assigning hidden variables to the nodes---the hidden degrees are typically drawn from some heterogeneous distribution---, the model produces networks which are simultaneously small-world, highly clustered, with heterogenous degree distributions and rich clubs. One of the important features of Eq.~(\ref{eq:con_pro}) is that it encodes simultaneously the likelihood of long- and short-range connections at all distances, which therefore need not be described by different mechanisms. Another relevant property of the model is that the expected degree of a node $i$ is proportional to its hidden degree, $\kappa_i$.

Similarity captures affinity between brain regions so that when two brain regions are close in similarity space, they are more similar and are more likely to form connections. One of the consequences is that groups of nodes that are close in similarity space tend to be more strongly interconnected as compared with the rest of the network. Hence, the inferred angular positions of brain regions in the similarity subspace offer information about the community structure of the analyzed connectomes and correlate with neuroanatomical information, with nodes belonging to the same neuroanatomical region strongly concentrated in a narrow angular section of the similarity space~\cite{Allard:2018}. Notice that Euclidean distance is certainly an important factor but not the only one determining similarity distance, that is also inversely related but different from homophilic attraction measures used in generative models of the brain~\cite{betzel2016generative}. See SI Appendix, Fig.~S38 for the comparison with different similarity index.

The $\mathbb{S}^1$ model has an isomorphic purely geometric formulation---the $\mathbb{H}^2$ model \cite{Krioukov2009}---in which the popularity and similarity dimensions are combined into a single distance in the hyperbolic plane by transforming the hidden degrees into radial coordinates. The popularity and similarity coordinates of the nodes in a real connectome, i.e. its hyperbolic map, can be obtained using statistical inference to find the coordinates that maximize the likelihood that our geometric model reproduces the structure of the real connectome~\cite{Boguna2010,Papadopoulos2015a}. We used the tool Mercator~\cite{Guille2019} to infer hyperbolic connectomes maps; see Materials and Methods for more details.

The embedding of UL subject No.~10 is shown on Fig.~\ref{fig:Fig2} (results for HCP subject No.~15 are in SI Appendix, Fig.~S31). Figure~\ref{fig:Fig2}A displays the map at $l=0$ with nodes colored according to the 82 coarse-grained regions in layer $l=4$. The left and right hemispheres, indicated by the red and green frames on the edge of the disk, are naturally separated, and nodes belonging to the same brain region appear clustered in nearby angular positions. This is consistent with previous results~\cite{Cacciola2017,Allard:2018}. To test the accuracy of the embedding, we used the set of inferred coordinates $\{\kappa_i, \theta_i \}$ and parameters $\beta$ and $\mu$ to generate an ensemble of $100$ synthetic networks using \eqref{eq:con_pro}. We compared topological properties of this ensemble with those measured on the real connectome. Specifically, Fig.~\ref{fig:Fig2}B--\ref{fig:Fig2}D show the results for the complementary cumulative degree distribution $P_c(k)$, the degree-dependent clustering coefficient $\bar{c}(k)$, the average degree of nearest neighbors $\bar{k}_{nn,n}(k)$, and the rich club coefficient $r(k)$. In Fig.~\ref{fig:Fig2}E--\ref{fig:Fig2}G, we show the good agreement between local properties in the real connectomes---degree, sum of degrees of neighbors, and number of triangles to which a node participates--- and in the synthetic ensemble. We also report the results of the statistical tests described in the previous section (i.e. $\rho$, $\chi^2$, and $\zeta$); see SI Appendix, Tables S5 and S6 for all subjects. The results confirm that the generated networks reproduce the topological properties with remarkable precision.

\begin{figure*}[t]
	\centering
	\includegraphics[width=0.85\linewidth]{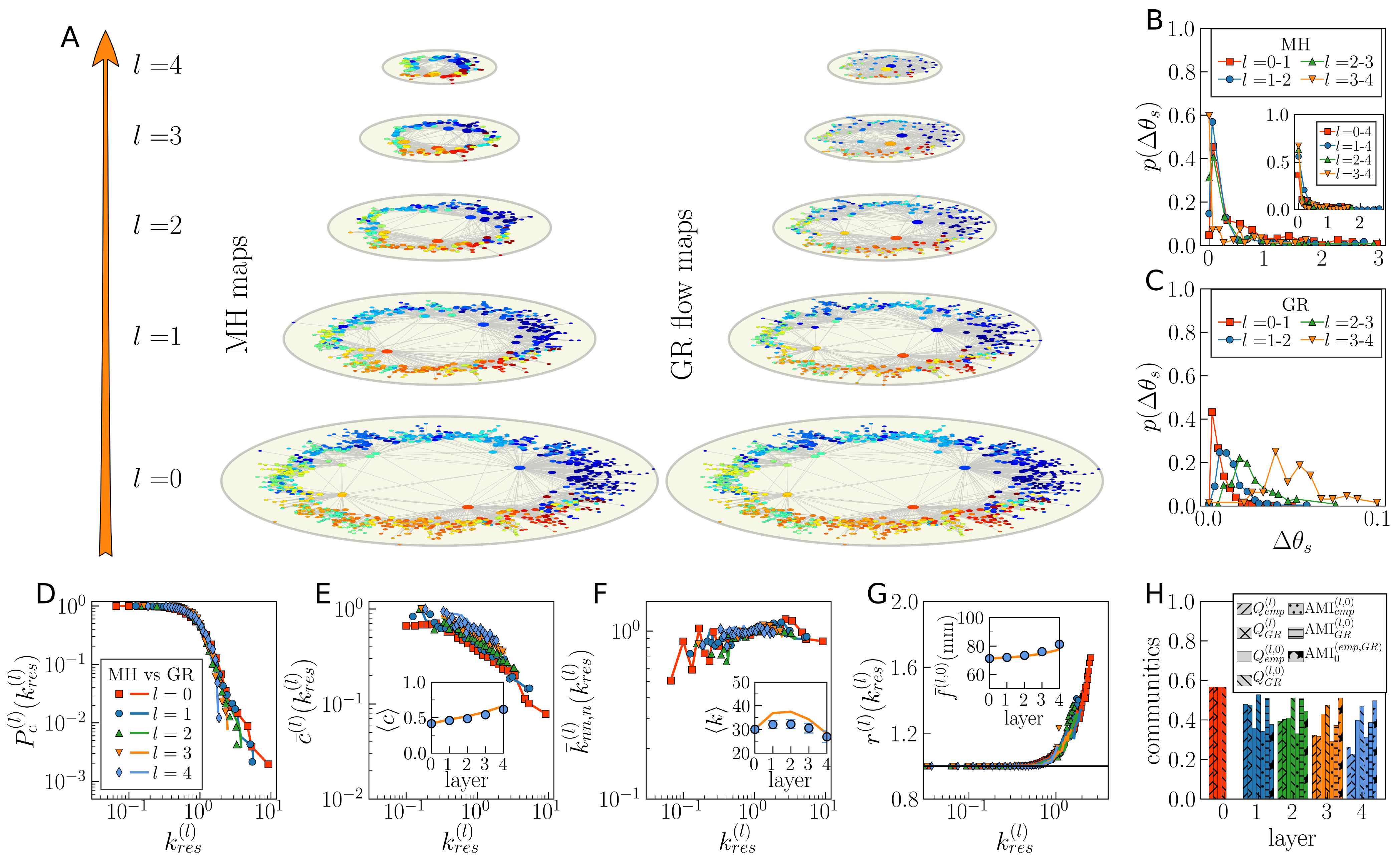}
	\caption{\label{fig:Fig2_part2} \textbf{Hyperbolic maps of the MH connectome and of the GR flow}. (\textit{A}) The bottom layer $l=0$ corresponds to the hyperbolic map of the highest resolution connectome for subject No.~10. The upper maps on the left are obtained by embedding independently each layer in the MH connectome. The different colors are given to the nodes according to the 82 coarse-grained regions defined in layer $l=4$. In the GR flow on the right, the maps are obtained by renormalizing layer $l=0$. A supernode in layer $l>0$ inherits the color of its subnode in layer $l-1$ positioned at its left in the similarity space (similar results are obtained if the color of the node on the right was chosen, or if the color was chosen at random between the two subnodes). For visualization purposes, we only represent links if their probability of connection according to \eqref{eq:con_pro} is larger than $0.5$. (\textit{B}) and (\textit{C}). Distribution $p(\Delta \theta_s)$ of the average angular separation between subnodes of coarse-grained supernodes from one layer to the next in MH and GR, respectively. The inset in (\textit{B}) shows the distribution $p(\Delta \theta_s)$ of subnodes in layer $l$ that correspond to a same supernode in layer-$4$ of MH. (\textit{D-H}) Comparison of topological properties in the empirical MH connectome of UL subject No.~10 (symbols) and GR predictions (lines).
		(\textit{D}) Complementary cumulative degree distribution $P_c^{(l)}(k_{res}^{(l)})$.
		(\textit{E}) Degree dependent clustering coefficient $\bar{c}^{(l)}(k_{res}^{(l)})$. Inset: average clustering coefficient $\langle c \rangle$ across layers.
		(\textit{F}) Degree-degree correlations $\bar{k}_{nn,n}^{(l)} (k_{res}^{(l)})$. Inset: average degree $\langle k\rangle$ across layers.
		(\textit{G}) Rich club coefficient $r^{(l)}(k_{res}^{(l)})$ for low and intermediate values of the rescaled threshold degrees. Inset: average fiber length $\bar{f}^{(l,0)}$ in layer $0$ of links outside supernodes in layer $l$, where supernodes are defined by the anatomical coarse-graining in the MH connectome or by the coarse-graining in the similarity dimension in the GR case. In the three insets in (\textit{E})--(\textit{G}), error bars show the $2\sigma$ confidence interval around the mean; the bars may be smaller than the symbol.
		(\textit{H}) Community structure of the multiscale connectomes. The subcripts $\{emp,GR\}$ indicate the empirical MH connectome and the GR shell, respectively. $\textrm{AMI}_{0}^{(emp,GR)}$ is the adjusted mutual information between topological communities in the empirical MH connectomes at each layer and the GR flow measured in their projection over layer 0. }
\end{figure*}

\subsection*{GR transformation}
Given that connectomes are well described by the $\mathbb{S}^1$ model, similarity distances in connectome maps at $l=0$ allow the application of renormalization techniques for a systematic investigation of their properties at different resolutions. Given a connectome map, the GR transformation introduced in Ref.~\cite{Garcia2018} produces a self-similar scaled-down replica with a reduced resolution by capturing longer range connections between coarse-grained groups of nodes, such that the average length of connections in similarity space, which determines the length scale, grows~\cite{Garcia2018}.

After obtaining the embedding of layer $l=0$, the GR transformation works by defining non-overlapping blocks of consecutive nodes of size $r=2$ along the similarity circle, that are coarse-grained to form supernodes. The supernodes are assigned an angular coordinate within the regions in the similarity subspace defined by the nodes in the block, so that the original angular ordering is preserved. Second, two supernodes $i$ and $j$ are connected in the new layer if and only if at least one node in block $i$ is connected to at least one node in block $j$ in the original layer. The resulting layer has a geometric description that is maximally congruent with the $\mathbb{S}^1$ model if the hidden variables of supernodes in the new map, $\kappa^\prime$ and $\theta^\prime$, are
\begin{eqnarray} \label{eq:kappa}
\kappa'=\left[  \sum_{j=1}^{r}(\kappa_j)^\beta \right] ^{1/\beta} \text{and  \hspace{0.1cm} }
\theta'=\left[  \frac{\sum_{j=1}^{r}(\theta_j\kappa_j)^\beta }{\sum_{j=1}^{r}(\kappa_j)^\beta}\right] ^{1/\beta} \ ,
\end{eqnarray}
where the sums are over the nodes that are merged into each supernode. Global parameters are rescaled as $\mu'= \mu/r$, $\beta'= \beta$, and $R'= R/r$. The transformation is stable and keeps the original distribution of degrees, the ordering of nodes in the similarity space, and the probability of connection between two supernodes $i$ and $j$ is still given by \eqref{eq:con_pro}. As a consequence, the $\mathbb{S}^1$ model is a renormalizable model and is self-similar under GR. The procedure can be iterated to produce a multiscale unfolding of a connectome in a sequence of self-similar, scaled-down replicas.

\subsection*{Multiscale GR shell of the human connectome}
We applied the GR transformation to the connectome map at layer $0$ of each subject. The transformation was applied iteratively 4 times to generate a 5-layer multiscale shell of each connectome. Since layer $0$ contains roughly 1014 nodes, each subsequent layer generated by GR has 507, 254, 127, and 64 nodes, to be compared with 462, 233, 128, and 82 nodes in the layers of the MH connectomes.

We also embedded each layer of the MH connectomes separately. The collection of maps from the individual embedding of each MH layer and the GR shell derived from layer $l=0$ of one subject is displayed in Figure~\ref{fig:Fig2_part2}A. The nodes are colored according to the 82 neuroanatomical regions represented by the nodes in layer $l=4$ of the MH maps. Remarkably, nodes corresponding to the same region remain angularly close in all MH maps. To support this claim, for each supernode in layer $l+1$ we measured the average angular separation of its subnodes in layer $l$, defined as $\Delta \theta_s^{(l+1)}=\frac{2} {N_s(N_s-1)} \sum_{i,j\in s} \Delta \theta_{ij}^{(l)}$, where $N_s$ is the number of nodes coarse-grained into supernode $s$. Values close to $0$ indicate that coarse-grained regions have similar angular positions. As shown in Fig.~\ref{fig:Fig2_part2}B~and~\ref{fig:Fig2_part2}C, the distribution $p(\Delta \theta_s)$ are similar in MH and GR maps (see SI Appendix, Fig.~S32 for HCP subject No.~15). All distributions are peaked around low average angular separation in both cases, even if MH distributions can reach large values.  Angular separation of subnodes within supernodes also remain small, on average, when we compare the angular distribution of subnodes in any layer with that of the corresponding supernodes in layer $l=4$ (inset Fig.~\ref{fig:Fig2_part2}B). The preservation of low average angular separation within coarse-grained anatomical regions, that is, the preservation of similarity in MH maps, indicates that the inferred coordinates are consistent across scales and encode significant information on the hierarchical anatomical structure of the connectomes, even if each layer was embedded independently. This feature is well reproduced by the GR flow, as expected, given that supernodes are produced by coarse-graining neighboring nodes in the similarity space and therefore preserve the original ordering.

We compared topological properties of the MH connectome shown in Fig.~\ref{fig:Fig1} with those computed for each layer in the corresponding GR shell, see Fig.~\ref{fig:Fig2_part2}D--\ref{fig:Fig2_part2}H and SI Appendix, Figs.~S4-S9 and S22-S28 for the remaining subjects. Strikingly, we observe a high congruency between the empirical curves of the different topological properties ---degree distributions, degree-degree correlations, clustering spectrum, rich club, and average degree and clustering--- and those predicted by GR at every scale (note that the perfect overlap for $l=0$ is trivial). The GR model is a renormalizable model~\cite{Garcia2018} and, therefore, unfolds a network into a shell of self-similar scaled-down versions of the original network. Hence, the good agreement between empirical and model curves gives a further proof of the self-similarity of the empirical data. In the case of the rich club effect, the GR prediction is even able to reproduce the whole curves despite finite-size effects, see SI Appendix, Figs.~S7 and S22-S27. To see if the model is also able to reproduce the progressive increase in fiber length of empirical connections as the resolution of observation is decreased, Fig.~\ref{fig:Fig0}, we calculated the average fiber length of links in layer $0$ that remain outside supernodes at each upper layer, see inset in Fig.~\ref{fig:Fig2_part2}G (see SI Appendix, Figs.~S12 and S33 for the remaining subjects). The two curves are in excellent agreement, meaning that the range of length scales covered in the real multiscale connectomes and in the model are consistent. Additionally, Fig.~\ref{fig:Fig2_part2}H shows the modularities $Q^{(l)}_{GR}$, $Q^{(l,0)}_{GR}$ and the adjusted mutual information $\textrm{AMI}^{(l,0)}_{GR}$ in the GR shell (see Methods). The community structure is preserved to a great extent in the flow with values for the adjusted mutual information similar to those measured in the MH connectome. We also report the overlap between topological communities in the MH connectomes at each layer and the GR flow measured in their projection over layer 0 ($\textrm{AMI}_{0}^{(emp,GR)}$; comparison via layer 0 avoids the problem of comparing networks with slightly different number of nodes). Here again, the community structure of real connectomes is also well approximated by the GR shell.

\begin{figure}[t]
	\centering
	\includegraphics[width=1\linewidth]{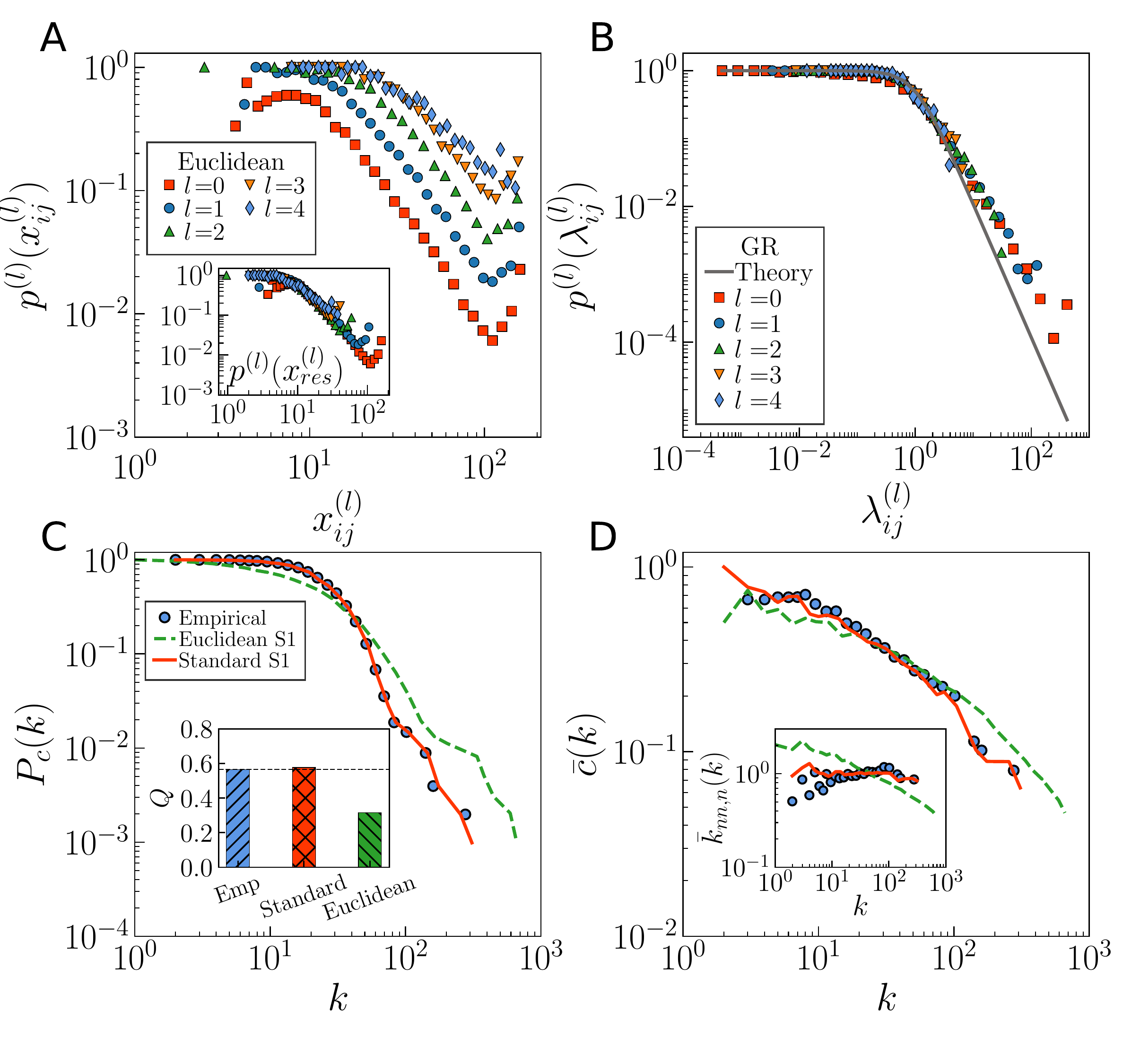}%
	\caption{\label{fig:Fig_x_px}
		\textbf{Empirical vs theoretical probability of connection.} Results for UL subject No. 10.
		(\textit{A}) Empirical connection probabilities $p^{(l)}(x_{ij}^{(l)})$ as a function of the Euclidean distances $x_{ij}^{(l)}$ in the MH connectome. Inset shows the empirical connection probabilities $p^{(l)}(x_{res}^{(l)})$ as a function of rescaled Euclidean distances $x_{res}^{(l)}=x^{(l)}/a^{(l)}$ with $a^{(l)}=[1.0, 1.5, 2.6, 3.8, 4.0]$.
		(\textit{B}) Empirical versus theoretical connection probability  $p^{(l)}(\lambda_{ij}^{(l)})$, Eq.~(\ref{eq:con_pro}),  in the GR shell.
		(\textit{C}) Complementary cumulative degree distribution $P_c(k)$. Modularity $Q$, as measured by the Louvain method, is shown in the inset.
		(\textit{D}) Degree dependent clustering coefficient  $\bar{c}(k)$. Inset: average degree of neighbors $\bar{k}_{nn,n}(k)$. The filled symbols correspond to the empirical connectome of Subject No.~10. Green dashed lines are generated using the $\mathbb{S}^1$ model with Euclidean distances and parameters $\beta=2.75$ and $\mu=0.0117$. Red lines correspond to the standard $\mathbb{S}^1$ model ($\beta=1.96, \mu=0.0104$).}
\end{figure}

Finally, Fig.~\ref{fig:Fig_x_px} shows the empirical connection probabilities as a function of Euclidean distance (3D separation between region centers) in the MH connectome, and as a function of the effective hyperbolic distance in the GR shell (see SI Appendix, Fig.~S13 and S14 for all UL subjects and Fig.~S34-S36 for HCP dataset). Finite-size effects aside, the curves show scale invariance in Euclidean and in hyperbolic spaces, as expected given the self-similarity of the topological properties shown in Figs.~\ref{fig:Fig1}A--\ref{fig:Fig1}D. In Euclidean space, the curves overlap only when distances are rescaled by specific values obtained ad hoc (see caption of Fig.~\ref{fig:Fig_x_px}). Interestingly, the curves in the GR shell overlap naturally due to the renormalizability of the geometric network model on which the GR technique is based.

Despite the scaling of the probability of connection in Fig.~\ref{fig:Fig_x_px}A, Euclidean distances alone do not contain enough information to explain the connectivity properties of the MH connectome~\cite{Kaiser2006}. In fact, networks generated by a purely geometric model based on Euclidean distances would have the small-world property if and only if $p_{ij} \sim x_{ij}^{-\beta}$, with $\beta\in(d,2d)$ and $d$ the dimension of the underlying space~\cite{Boguna:2019a}. However, these networks would be homogeneous in their node degree distribution, in contrast to the observed heterogeneity in human connectomes. To take it into consideration, we used the connection probability \eqref{eq:con_pro} as in the $\mathbb{S}^1$ model but using Euclidean distances $x_{ij}$ instead of similarity distances $d_{ij}$. The hidden degrees $\kappa$ can be approximated by the actual degrees $k$ (the two are very similar, see SI Appendix, Fig. S15), and the values of $\beta$ and $\mu$ were adjusted to match the clustering and average degree of the empirical connectome. As shown in Fig.~\ref{fig:Fig_x_px}C--\ref{fig:Fig_x_px}D, the model based on Euclidean distances cannot reproduce the empirical observations. Euclidean distance is certainly an important factor but not the only one determining similarity distance needed to reliably reproduce the topological properties of the MH connectome. In contrast, the fit of the $\mathbb{S}^1$ model based on similarity distance, underlying the geometric renormalization technique, is very good.

Altogether, these results indicate that GR naturally and accurately predicts the scale invariance and the self-similarity of MH connectomes. Indeed, GR provides rescaled layers that statistically mimic the structure of the brain at larger scales, using only structural information measured at one single resolution. Let us stress once more that no new information about the anatomical coarse-graining of brain regions in the MH connectome was used when going from one resolution to another in the GR renormalization process, we just inferred a geometric map from the highest resolution empirical data, and used consecutive nodes in this space to produce the structure of each renormalized layer.

\section*{Self-similarity and navigability}

Hyperbolic network maps sustain efficient navigation~\cite{Boguna2010}, a remarkable finding that is also valid for the brain~\cite{Allard:2018}. To check the navigability properties of connectomes at different resolutions, we implemented greedy routing, a decentralized communication protocol in which a source node transmits a message along to its neighbor that is the closest to a target node in the metric space~\cite{Kleinberg2006}. The performance of greedy routing is measured by the success rate, $p_s$, and the average stretch of successful greedy paths, $\bar{s}$. The success rate counts the fraction of successful greedy paths when considering $10000$ random node pairs source-target. Note that greedy routing does not guarantee that a message will reach its target node; the message may reach an already visited node and therefore may get trapped in a loop. The average stretch of successful paths consists in the ratio of the number of links in the successful greedy path and the number of links in the topological shortest path, averaged over all successful greedy paths. Navigation is considered maximally efficient if the success rate and the stretch are both equal to one, meaning that all messages have reached destination and have done so by following the shortest paths.

\begin{figure}[t!]
	\centering
	\includegraphics[width=1\linewidth]{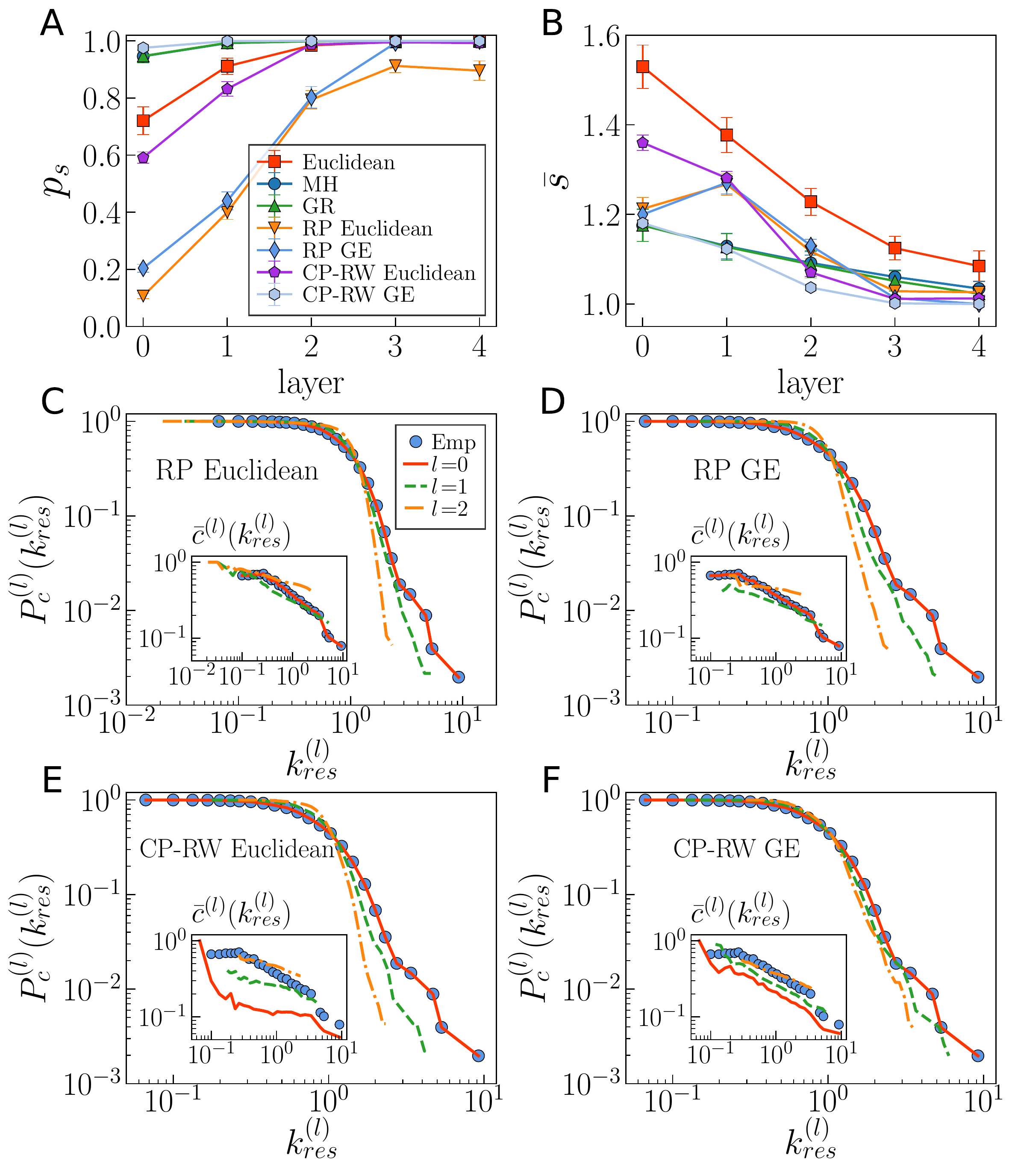}%
	\caption{\label{fig:Navigation}
		\textbf{Navigability of the MH connectomes and the GR shells at different resolutions.} (\textit{A}) average success rate and (\textit{B}) average stretch for all UL subjects. Navigation performance was benchmarked against $4$ different random null models. The error bars show the $2\sigma$ confidence interval around the expected values. (\textit{C})--(\textit{F}) The loss of self-similarity in the four ensembles of random null
		models can be seen through the loss of self-similarity of their complementary
		cumulative degree distribution and degree-dependent clustering coefficient
		(Insets) (see SI Appendix, Figs. S18 and S19 for more results). For each null
		model, we generated 100 multiscale surrogates for UL subject no. 10.}
\end{figure}

We studied navigation in the anatomical Euclidean embeddings, the collection of individual MH embeddings in the hyperbolic plane, and the GR shell for all UL subjects, results are summarized in Fig.~\ref{fig:Navigation} (see SI Appendix, Fig.~S37 for HCP dataset). Remarkably, the variability between subjects is very low (as shown by the error bars). In both geometries, there is a systematic trend towards a more efficient navigation, both in terms of success rate and stretch, as the resolution is decreased and longer range connections progressively dominate, and as the density of connections increases. However, the navigability of the hyperbolic maps is higher, as reported previously in~\cite{Allard:2018}, with larger success rates and lower stretch values. Note that the last two layers are very dense, which ensures a very high navigability in both Euclidean and hyperbolic geometries. Interestingly, the efficiency of the navigation protocol in each layer of the GR shell is close to that in the embedding of the corresponding MH layer (see SI Appendix, Fig.~S16). Similar results were obtained with the multiscale navigation protocol introduced in~\cite{Garcia2018} where we see that the success rate increases even more as more layers are used to guide navigation. In fact, it becomes very close to $100\%$ with the inclusion of just two renormalized layers, and this improvement comes at the expense of only a mild increase of the stretch of successful paths (see SI Appendix, Fig. S17).

To understand the interplay between self-similarity and navigability, we compared these results with results obtained with ensembles of multiscale null models. We used four null models~\cite{Seguin2018} to obtain randomized versions of layer $l=0$ (see \textit{Materials and Methods} for a description of the models). Repositioning Euclidean (RP Euclidean) and Repositioning Geometric (RP GE) preserve the topology but randomize the geometry by shuffling the position of the nodes in Euclidean space and in the similarity dimension, respectively. Coordinate-Preserving Edge-Rewiring Euclidean (CP-ER Euclidean) and Coordinate-Preserving Edge-Rewiring Geometric (CP-ER GE) preserve the geometry in terms of the total distance between connected nodes (measured respectively in the Euclidean space and in the hyperbolic space) while the topology is rewired. Further details are provided in Materials and Methods. In RP Euclidean and CP-RW Euclidean, the renormalized layers were constructed following the anatomical coarse-graining of the MH connectomes, while the ones in RP GE and CP-RW GE were obtained by GR. Because the topology-preserving null models, RP Euclidean and RP GE, destroy the ordering of nodes in their respective geometry, self-similarity disappears and the efficiency of greedy routing decreases dramatically. This means that geometry influences self-similarity, and therefore that both geometry and self-similarity play a crucial role in navigability. Turning our attention to geometry-preserving null models, we see that the navigability performances observed in human connectomes cannot be attributed solely to the underlying geometry, and therefore that the explicit wiring of these connectomes plays also a determinant role. Self-similarity is also lost in CP-RW Euclidean surrogates, even if 99\% of network cost was preserved, which resulted in significantly less navigability than in the original connectomes. In contrast, CP-RW GE surrogates still display a high level of self-similarity, meaning that surrogates do not depart significantly from the original network (disregarding layers 3 and 4 strongly affected by finite-size effects), and high navigation performance. See SI Appendix, Fig.~S18 and S19 for more results. These results suggests an interplay between the self-similar multiscale organization of brain connectomes and their navigability.

\section*{Discussion}
%
The structure of the human brain spans over a range of length scales, which magnifies its complexity, otherwise constrained by overarching patterns. We uncovered self-similarity as one of such patterns in the multiscale structure of the human connectome which, paradoxically, introduces simplicity as an organizing principle. In our work, simplicity has a very precise meaning. It states that the structure of connectomes and the underlying connectivity rules that explain this structure are independent of the scale of observation (at least within the scales covered in this work). In other words, a specific set of rules for each scale is not required.

On the one hand, we showed that the MH connectome is self-similar when the resolution of observation is progressively decreased by hierarchical coarse-graining of the anatomical regions. We reconstructed high-quality MH connectomes at five spatial resolutions for a total of 84 healthy subjects that display a remarkable level of homogeneity in the results. The fact that the smallest layers, with $128$ and $82$ nodes, are affected by finite-size effects suggests an upper bound on the scale for which the self-similarity of the MH connectome can be observed. Higher resolution datasets with more refined brain parcellations will be required to investigate possible lower bounds.

Using a geometric model based on a simple connectivity law, Eq.~(1), encoding simultaneously long- and short-range connections, our results show that the GR approach is able to explain the self-similar structure of human connectomes at different scales. This implies that the same principles govern the connectivity between brain regions at different length scales and that a different connectivity law at each resolution is not needed. Notice that the self-similarity of the MH connectomes and the goodness of the GR model to explain this self-similarity happens at the individual level of each single subject. The high fidelity of the GR shell in replicating the empirical data, as supported by the result of several statistical tests, suggests that connectivity at lower resolutions can be inferred from observations at higher resolutions.

As all questions related to symmetries, the ultimate reason for self-similarity in the brain is a profound matter that will require full understanding of how structure is related to function in the brain. Multiscale self-similarity implies a higher compression of the information needed to encode the architecture of the brain, and it may be related to the need of efficient communication between brain regions. While communication in the brain has been modeled using different protocols, from shortest paths to random diffusion~\cite{Graham2014,Goni2014,Avena2018,Seguin2018}, its control and regulation is a subject of ongoing debates~\cite{Avena2018}. Shortest path navigation relies on the unrealistic assumption that neural elements possess global knowledge of the network topology while random diffusion needs bias to travel via efficient routes. At the same time, evidence indicates that targeted information processing may play an important role in brain communication dynamics and greedy routing protocols could, nonetheless, offer a simplified yet fundamental illustration~\cite{Gulyas2015,Seguin2018,Allard:2018}. For instance, hippocampal neurons can transmit distinct behavior-contingent information selectively to different target areas~\cite{Ciocchi2015}.

Hyperbolic space and geometric renormalization might also be relevant for brain network development or evolution. In this respect, we have now supporting evidence that the GR transformation based on our geometric network models in hyperbolic space is directly related with a branching growth mechanism that is able to explain the self-similar evolution of real complex network over long time spans~\cite{Muhua:2020a}. Those results and the obtained in the present work pose the intriguing questions of whether the observed self-similarity of the human brain connectomes could be related to mechanisms driving their growth and, therefore, whether their evolution could be conceptualized within the framework of the geometric renormalization group as observed for other real networks.

The implications of our findings are, thus, varied and can affect fundamental debates, such as whether the connectome is a system near the critical point of a phase transition~\cite{Gollo:2018}. According to results in~\cite{Garcia2018}, successive GR transformations of networks with heterogeneous degree distributions and high levels of clustering typically flow towards a fully connected graph as a fixed point, while networks with more homogeneous distributions and very high levels of clustering tend towards a one-dimensional ring. In the former case, the average degree increases in the flow while it decreases in the later case. Right at the continuous transition between the two phases, networks would have a scale-invariant average degree with a constant value preserved under GR transformations. Our results show that the average degree of otherwise scale-invariant human brain connectomes increases slightly in the GR process, and according to statistical tests in a way that is compatible with an almost constant average degree. This implies that human brain connectomes are near this critical structural transition between small-world and non-small-world regimes.

At the level of applications, both the scale invariance of the brain structure and the existence of a model that unravels its self-similarity may have an important impact in the development of advanced tools that simplify its digital reconstruction and simulation. At the same time, our results suggest that the number of regions in a brain atlas is an important question. Specific details at the smallest of the considered scales could be redundant when informing about the large-scale structural organization of the brain while an insufficient number of regions could bias the observations. Another potential advantage of the self-similarity of brain connectomes is that it can be used to detect possible biases, depending on length scales, associated with the different data preprocessing methods and brain mapping techniques. Finally, immediate follow-ups of our work include studies to assert the renormalizability of functional brain networks and alterations in renormalizability produced by normal aging or possible brain disorders.


\section*{Methods}
	\subsection*{UL dataset} Informed written consent in accordance with the Institutional guidelines (protocol approved by the Ethics Committee of Canton de
	Vaud (CER-VD) was obtained for all subjects. Forty healthy subjects (16 females; $25.3\pm4.9$ years old) underwent an MRI session on a 3T Siemens Trio scanner with a 32-channel head coil. Magnetization prepared rapid acquisition with gradient echo (MPRAGE) sequence was 1 mm in-plane resolution and 1.2 mm slice thickness. The diffusion spectrum imaging
	(DSI) sequence included 128 diffusion weighted volumes $+1$ reference $b_0$ volume, maximum $b$ value=8000 s/mm$^2$, and $2.2 \times 2.2 \times 3.0$ mm as voxel size. The echo-planar imaging sequence was 2.2 mm in-plane resolution and 3.3-mm slice thickness with TR=$1920$ ms. DSI and MPRAGE data were processed using the Connectome Mapper Toolkit~\cite{Daducci:2012}. Each participant's gray and white matter compartments were segmented from the MPRAGE volume. The grey matter volume was subdivided into 68 cortical and 15 subcortical anatomical regions, according to the Desikan-Killiany atlas~\cite{Desikan:2006}, defining 83 anatomical regions. Each cortical region was subdivided into smaller region of interest (ROIs) of approximately identical surface such that the total number of regions was 1015 including both hemispheres. The ROIs were regrouped iteratively into bigger ROIs to create 5 different parcellations with 1015, 463, 234, 129, and 83  ROIs respectively, corresponding to five different resolution scales~\cite{Cammoun2012}. In layers $0$, $1$ and $2$ the surface areas of ROIs remain approximately equal, while for layers $3$ and $4$ the sizes are more disperse, see SI Appendix, Fig.~S1 and S2. The parcelations at different resolutions are spatially hierarchical, with a correspondence between the nodes at different length scales as defined by the coarse-grained regions. The hierarchical decomposition was obtained by grouping sets of $2$ or $3$ neighboring brain regions to build a partition with decreased resolution, and the operation was repeated several times until the 83 parcels at the lower resolution scale were recovered.
	
	At each scale and for each individual subject, connection weights between pairs of regions in the corresponding parcellation were quantified as fiber density~\cite{Hagmann:2008}. To track wires between brain regions, whole brain deterministic streamline tractography was performed on reconstructed DSI data, initiating 32 streamline propagations (seeds) per diffusion direction, per white matter voxel~\cite{Wedeen:2008}. Within each voxel, seeds were randomly placed and for each seed, a fiber streamline was grown in two opposite directions with a 1 mm fixed step. Fibers were stopped if a change in direction was greater than 60 degrees/mm. The process was complete when both ends of the fiber left the white matter mask. The connection weight between the pair of brain regions $\{u,v\}$ captures the average number of streamlines per unit surface between $u$ and $v$, corrected by the average length of the streamlines connecting such brain regions. The aim of these corrections is to control for the variability in cortical region size and the linear bias toward longer streamlines introduced by the tractography algorithm.
	
	Fiber densities were used to construct individual subject structural connectivity matrices at the five different resolutions. Each matrix is modeled as a weighted adjacency matrix $W=w_{ij}$ of a network G = {V,G} with nodes $V = {v_1,...,v_n}$ representing regions at the corresponding scale, and weighted, undirected edges $E = {e_1,...,e_m}$ representing anatomical connections with their fiber densities. The present study considers the unweighted version of the connectivity adjacency matrices at each scale and discards the brainstem (one node) for all subjects. We have also removed nodes that were isolated in the original dataset due to fluctuation in the data acquisition experiment, nodes that became isolated after the removal of the brainstem region, and nodes that were only connected to themselves by a self-loop. These adjustments cause negligible variations in the number of nodes of the highest resolution layer from subject to subject. The highest resolution layer comprises typically 1014 equal sized regions of interest (ROI), which are then coarse-grained into 462, 233, 128, and 82 regions at lower resolutions.
	
	\subsection*{Human Connectome Project dataset} For cross-validation, we used T1-weighted and corrected diffusion-weighted magnetic resonance images (DWIs) of 44 subjects from the Human Connectome Project (HCP)~\cite{Van2012}. The corrected diffusion weighted image for each subject was employed to fit a second order tensor for each voxel and its different voxelwise scalar maps (fractional anisotropy (FA) and mean diffusivity (MD)) by using Dipy~\cite{Garyfallidis2014}. The DWIs were also used to estimate the intravoxel fiber distribution function (fODF) by using the Constrained Spherical Deconvolution (CSD)~\cite{Tournier2007} approach implemented in MRtrix3 (https://www.mrtrix.org/). This technique, based on high-angular resolution diffusion imaging, estimates the orientation of multiple intravoxel fiber populations within regions of complex white matter architecture. This fODFs were used by the SDSTREAM (Streamlines by using Spherical Deconvolution)~\cite{Tournier2012} deterministic fiber tracking algorithm to obtain the streamlines distribution for each subject. The structural connectivity matrices were then computed, defining the connection strength between each pair of regions as the number of streamlines connecting them. Finally, multiscale structural connectivity matrices were obtained using the same hierarchical anatomical coarse-graining method described above for the UL dataset.
	
	Notice that, in this work, we always used deterministic streamline tractography algorithms yielding sparse connectomes, that give higher specificity and lower sensitivity as compared with probabilistic algorithms. Connectomes with high sensitivity and high specificity are unattainable with current axonal fiber reconstruction methods. Sparse connectomes contain only a subset of the possible projections in the fiber orientation distribution, whereas the probabilistic algorithms yield denser connectomes at the price of low specificity due to false positives (FPs). The network science methods that we use in our study, and in particular the embedding technique, require that connectomes are sparse and reliable. Hence, deterministic streamline tractography is more appropriate for our purposes. In addition, as argued in~\cite{Zalesky:2016}, connectome specificity is paramount since false positives are at least twice as detrimental as false negatives when estimating key topological properties of brain networks including clustering and modularity.
	
	\subsection*{$\mathbb{H}^2$ model and hyperbolic maps}
	In the $\mathbb{H}^2$ representation, the angular coordinates remain the same as in the $\mathbb{S}^1$ model, but the hidden degrees are transformed into radial coordinates using
	\begin{eqnarray} \label{eq:r_i}
	r_i=R_{\mathbb{H}^2}-2\ln\frac{\kappa_i}{\kappa_0}\ ,
	\end{eqnarray}
	where the radius of the hyperbolic disk is $R_{\mathbb{H}^2}=2\ln N/(\pi\mu\kappa_0^2)$ with $\kappa_0=\mathrm{min}(\{\kappa_i\})$. Higher degree nodes are therefore located closer to the center of the $\mathbb{H}^2$ disk.
	
	The hyperbolic MH maps, used as the starting point of the GR process, were obtained using the algorithm introduced in Ref.~\cite{Guille2019}. More precisely, these maps were inferred by finding the hidden degree and angular position of each node, $\{\kappa_i\}$ and $\{\theta_i\}$, that maximize the likelihood
	\begin{align}
	\mathcal{L} = \prod_{i<j} \left[ p_{ij} \right]^{a_{ij}} \left[ 1 - p_{ij} \right]^{1 - a_{ij}}
	\end{align}
	that the structure of the network was generated by the $\mathbb{S}^1$ model, where $\{a_{ij}\}$ are the entries of the adjacency matrix of the network. In Figs.~\ref{fig:Fig2}B--\ref{fig:Fig2}G, we show the topological validation of the embedding of the highest resolution network for UL subject No.~10.

	\subsection*{Renormalization flow of community structure}
	To asses how the community structure of the empirical MH connectomes and of the GR unfolding changes with the resolution scale, we obtained the community partitions $P_{emp}^{(l)}$ and $P_{GR}^{(l)}$ and the corresponding modularities $Q^{(l)}_{emp}$ and $Q^{(l)}_{GR}$ for every layer $l$ using the Louvain method \cite{Blondel2008}. We also defined the partition induced by $P_{emp/GR}^{(l)}$ on layer $0$, $P_{emp/GR}^{(l,0)}$, obtained by considering that if two nodes $i$ and $j$
	in layer $l$ belong to the same community in $P_{emp/GR}^{(l)}$, then all the nodes in layer $0$ belonging hierarchically to coase-grained regions $i$ and $j$ are in the same community in $P_{emp/GR}^{(l,0)}$. We can calculate the modularities $Q^{(l,0)}_{emp/GR}$ of $P_{emp/GR}^{(l,0)}$, and the adjusted mutual information $\textrm{AMI}_{emp/GR}^{(l,0)}$ between the induced community partition $P_{emp/GR}^{(l,0)}$, and the original community partition directly detected in layer $0$. We also report the overlap with the adjusted mutual information $\textrm{AMI}_{0}^{(emp,GR)}$ between topological communities in the MH connectomes at each layer and the GR flow measured in their projection over layer 0, i.e., $P_{emp/GR}^{(l,0)}$.
	
	\subsection*{Null models}
	We used the following null models~\cite{Seguin2018} to obtain randomized versions of layer $l=0$:
	\begin{itemize}
		\item RP Euclidean, repositioning nodes in Euclidean space by swapping coordinates of pairs of nodes selected at random;
		\item RP GE, repositioning nodes in the geometric embedding by swapping angular coordinates of pairs of nodes selected at random;
		\item CP-RW Euclidean, rewiring edges while preserving coordinates, degrees and the total cost in the network defined as the sum of the Euclidean distances between connected nodes. More specifically, two selected edges $A-B$ and $C-D$ with Euclidean distances $d_{AB}$ and $d_{CD}$ are swapped to $A-D$ and $B-C$ with Euclidean distances $d_{AD}$ and $d_{BC}$ if $|(d_{AB}+d_{CD})-(d_{AD}+d_{BC})|<\epsilon$, so that connection swaps that do not alter the resulting connectome cost by more than the tolerance $\epsilon$ (set to 1mm). Self-connections and multiple links are forbidden in the rewiring process. The departure from the cost of the original connectome grows with the number of swaps. We keep $\Delta cost =1-|1-cost_{null}/cost_{emp}|>99\%$.
		\item CP-RW GE, rewiring edges while preserving coordinates, degrees and the the total cost in the network defined as the sum of hyperbolic distances in the geometric embedding. We implemented with a routine similar to the Euclidean case but replacing Euclidean distance with the hyperbolic one $h_{ij}$. We set the tolerance $\epsilon$ to $0.478$ so that the ratio $\epsilon/\text{min}(h_{ij})$ is equal to the one in CP-RW Euclidean.
	\end{itemize}
	
	\subsection*{Data availability}
	The UL and HCP multiscale human connectomes obtained for this study, and their maps in hyperbolic space and GR shells will be available via the Zenodo platform at https://doi.org/10.5281/zenodo.3766139 upon publication.

\section*{Acknowledgements}
We thank Filip Miscevic, Olaf Sporns and Mari\'an Bogu\~{n}\'a for helpful discussions. We acknowledge the Department of Psychiatry of Lausanne University Hospital and particularly Professor Philippe Conus, Professor Kim Do Cuenod, Mrs Martine Cleusix and Mr Raoul Jenni for having helped with the recruitment process of the study volunteers of the Lausanne Dataset and Centre d'Imagerie BioMédicale (CIBM) of the UNIL, UNIGE, HUG, CHUV, EPFL and the Leenaards and Jeantet Foundations for supporting image acquisition. M.\'A.S. acknowledges support from a James S. McDonnell Foundation Scholar Award in Complex Systems; Ministerio de Ciencia, Innovaci\'on y Universidades of Spain project no. FIS2016-76830-C2-2-P (AEI/FEDER, UE); the project {\it Mapping Big Data Systems: embedding large complex networks in low-dimensional hidden metric spaces} -- Ayudas Fundaci\'on BBVA a Equipos de Investigaci\'on Cient\'{\i}fica 2017; and Generalitat de Catalunya grant No. 2017SGR1064. A.A. acknowledges financial support from the Sentinelle Nord initiative of the Canada First Research Excellence Fund, from the Natural Sciences and Engineering Research Council of Canada (project 2019-05183) and from the Spanish ``Juan de la Cierva-incorporaci\'on'' program (IJCI-2016-30193). Y.A-G. and P.H. were financially supported by Swiss National Science Foundation Grant 1115 51NF40–185897.

%
\bibliographystyle{apsrev4-1}
\bibliography{reference}

\onecolumngrid
\appendix
\newpage
\clearpage

\setcounter{page}{0}
\pagenumbering{arabic}
\newcommand*{\thd}[1]{\multicolumn{1}{c}{#1}}

\clearpage
\begin{minipage}[h]{\textwidth}
	\begin{center}
		\large{\textbf{Supplementary Materials for\\ Geometric renormalization unravels \\self-similarity of the multiscale human connectome}}\\
		\vspace{0.5cm}
		Muhua Zheng$^{1,2}$, Antoine Allard$^{3,4}$, Patric Hagmann$5$, Yasser Alem\'an-G\'omez$5,6,7$ \& M. \'Angeles Serrano$^{1,2,8*}$\\ 
		\vspace{1cm}
		
		\small{1. Departament de F{\'\i}sica de la Mat\`eria Condensada,\\ Universitat de Barcelona, Mart\'{\i} i Franqu\`es 1, 08028 Barcelona, Spain}\\
		\small{2. Universitat de Barcelona Institute of Complex Systems (UBICS), \\Universitat de Barcelona, Barcelona, Spain}\\
		\small{3. D\'epartement de physique, de g\'enie physique et d'optique, Universit\'e Laval, Qu\'ebec , Canada G1V 0A6}\\
		\small{4. Centre interdisciplinaire de mod\'elisation math\'ematique, Universit\'e Laval,
			Qu\'ebec, Canada G1V 0A6}		\\
		\small{5. Connectomics Lab, Department of Radiology, Lausanne University Hospital and University of Lausanne (CHUV-UNIL), 1011 Lausanne, Switzerland}\\
			
		\small{6. Center for Psychiatric Neurosciences, Department of Psychiatry, Lausanne University Hospital and University of Lausanne (CHUV-UNIL), 1008 Prilly, Switzerland}\\
		\small{7. Medical Image Analysis Laboratory, Lausanne University Hospital and University of Lausanne (CHUV-UNIL), 1011 Lausanne, Switzerland}\\
		
		\small{8. Catalan Institution for Research and Advanced Studies (ICREA), 08010 Barcelona, Spain}\\

		\small{*Correspondence and requests for materials should be addressed to M.A.S. (marian.serrano@ub.edu)}
		\\

	\end{center}
	
\end{minipage}

\vspace{2cm}



\tableofcontents
\let\addcontentsline\oldaddcontentsline

\newpage
\renewcommand\thefigure{S\arabic{figure}}
\renewcommand\thetable{S\arabic{table}}
\setcounter{figure}{0}    
\setcounter{table}{0}

\section{Results for UL dataset}

\subsection{Parcellation and distribution of areas}
\begin{figure*}[!h]
	\centering
	\includegraphics[width=0.9\linewidth]{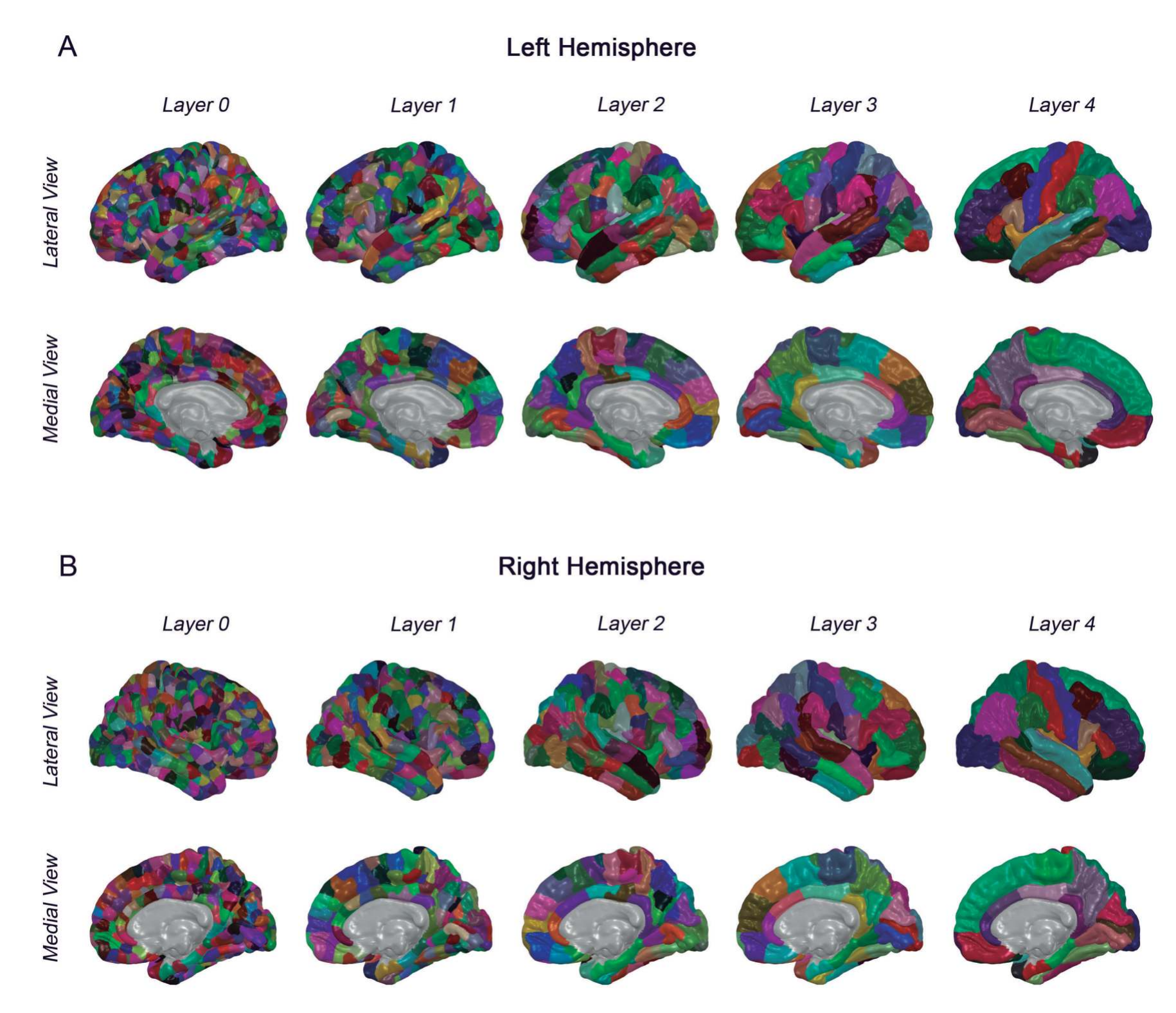}
	\caption{\textbf{Lateral and medial views of the multi-scale cortical parcellation.} }
\end{figure*}

\begin{figure*}[h]
	\centering
	\includegraphics[width=1\linewidth]{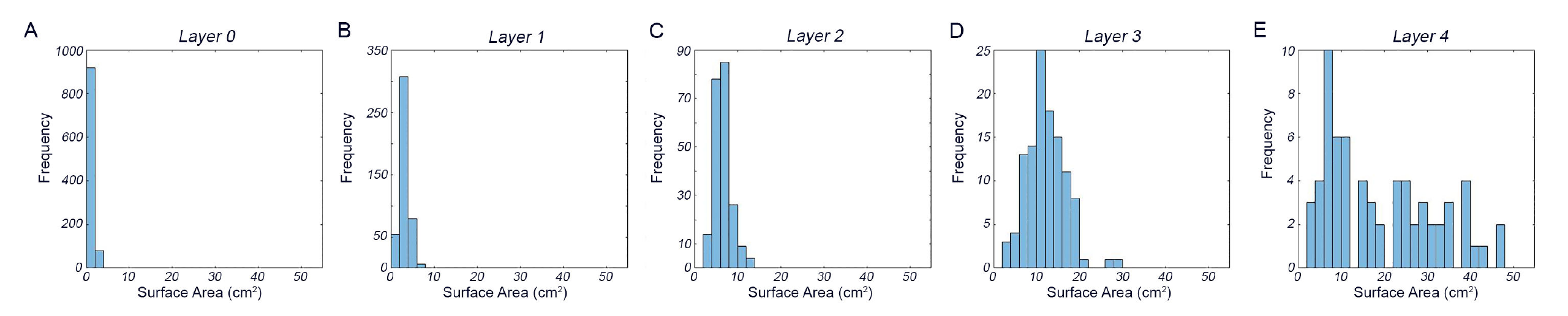}
	\caption{\textbf{Histograms of surface areas of regions at each connectome scale.} }
\end{figure*}

\newpage
\subsection{Average fiber length at different resolutions}
\begin{figure*}[h]
	\centering
	\includegraphics[width=0.85\linewidth]{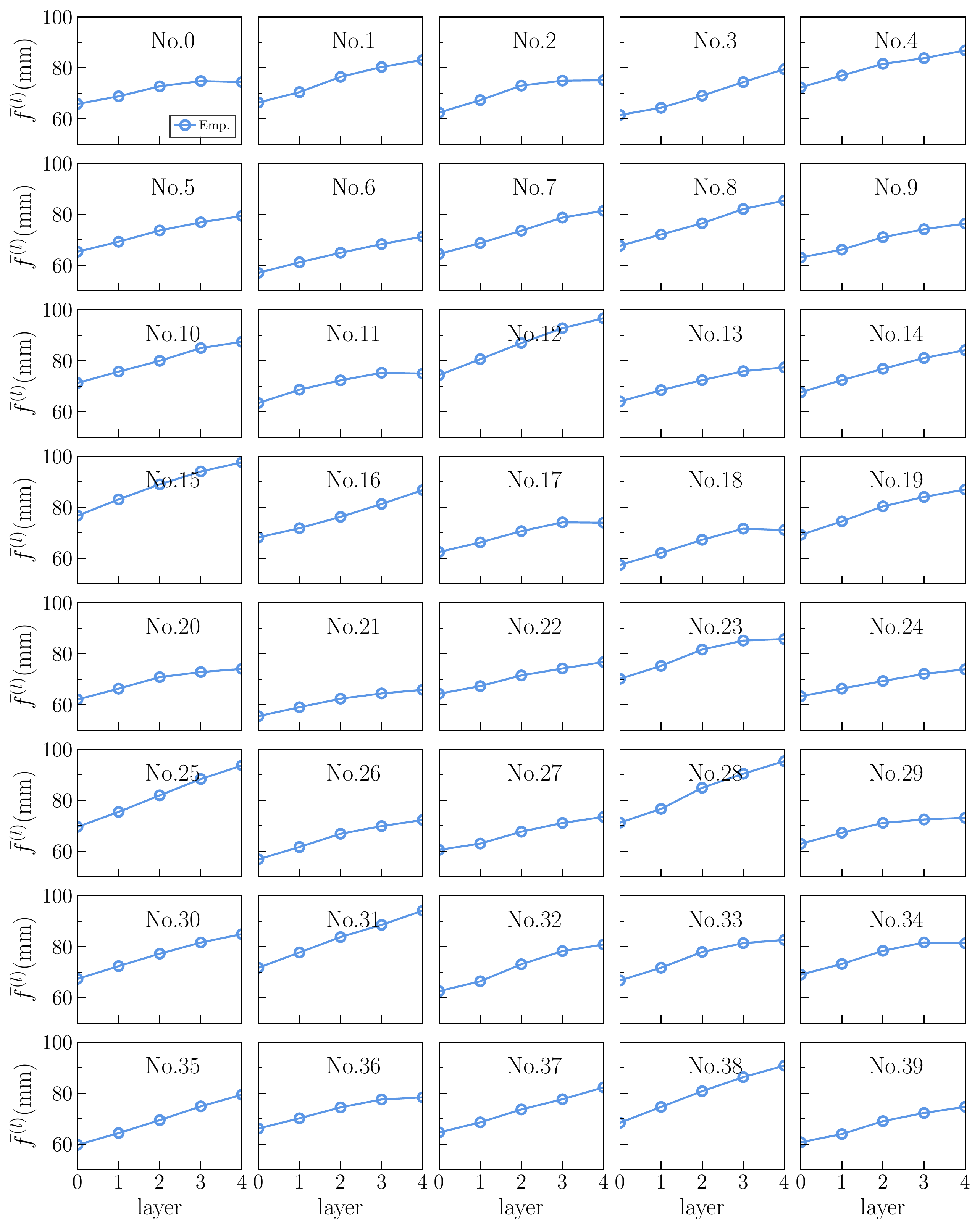}%
	\caption{
		Average fiber length $\bar{f}^{(l)}$(mm) for each subject in UL dataset. 
	}
\end{figure*}
\newpage
\subsection{MH connectomes vs GR flows for all subjects}

\begin{figure*}[th]
	\centering
	\includegraphics[width=0.8\linewidth]{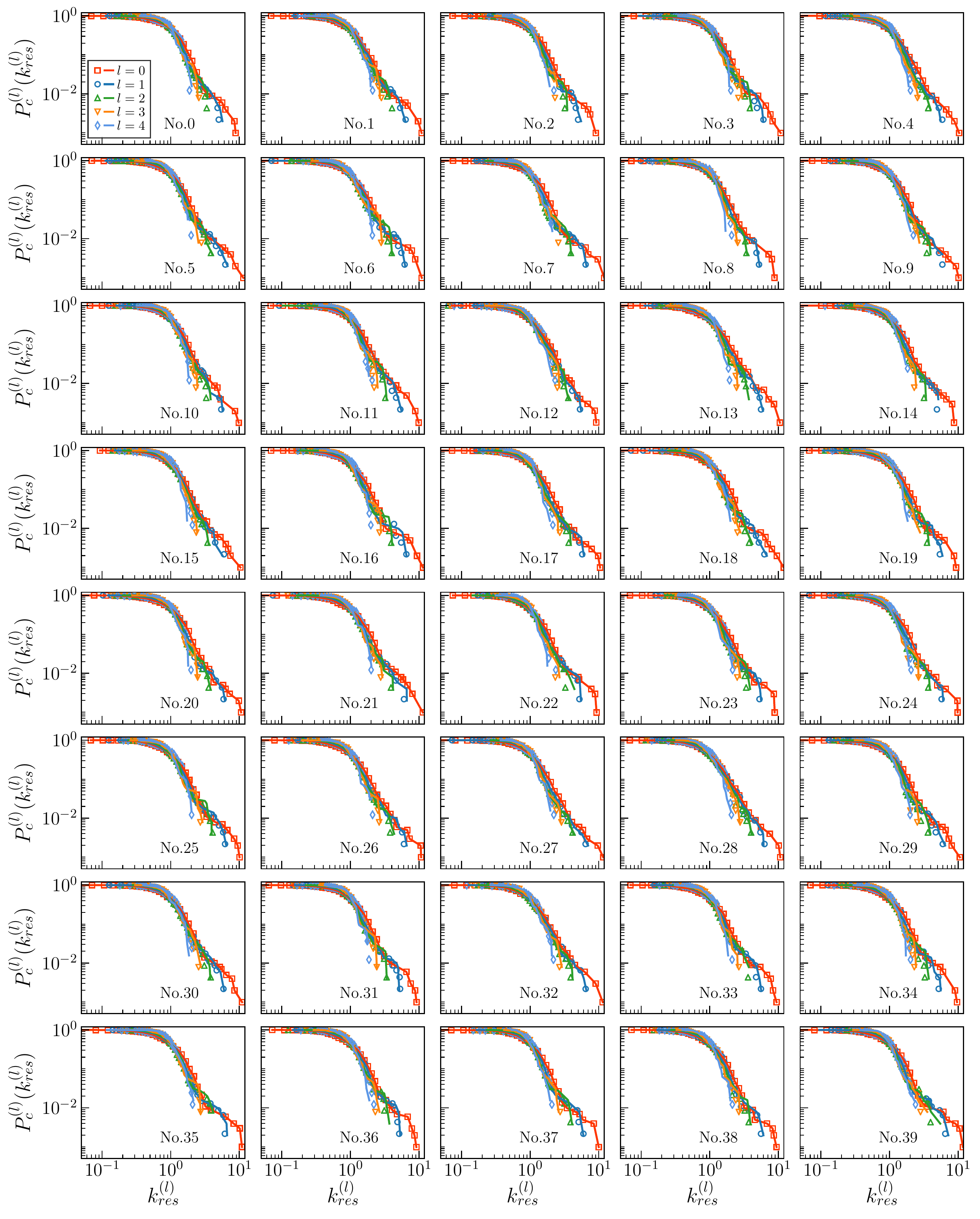}%
	\caption{\label{fig:FigA1}Complementary cumulative degree distribution $P_c^{(l)}(k_{res}^{(l)})$ of rescaled degrees $k^{(l)}_{res}$ for different layers $l$ in each subject as compared to the multiscale GR unfolding, where the symbols correspond to the empirical multiscale connectome and the line to the GR flow.}
\end{figure*}

\begin{figure*}[t]
	\centering
	\includegraphics[width=1\linewidth]{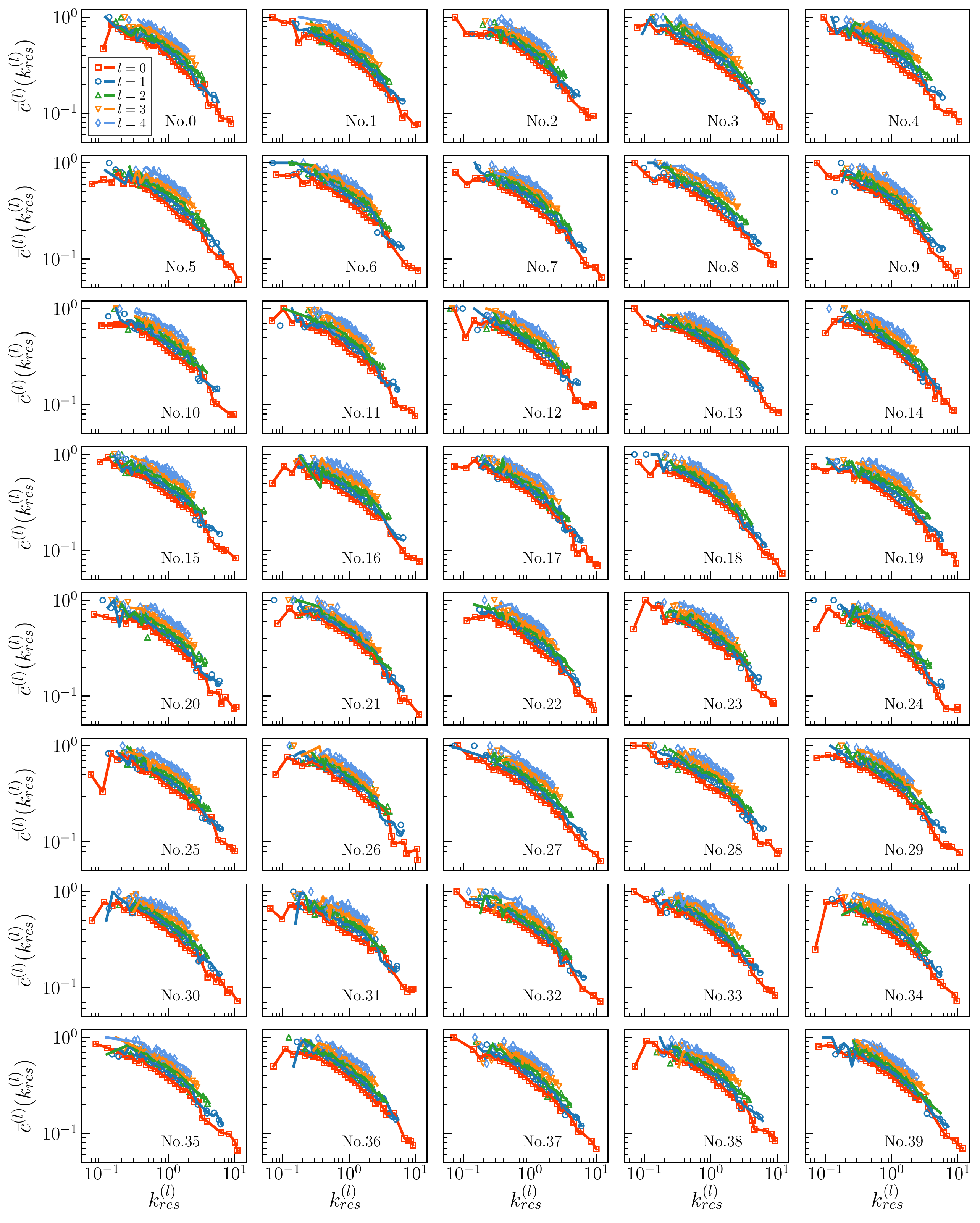}%
	\caption{\label{fig:FigA2} The degree-dependent clustering coefficient $\bar{c}^{(l)}(k_{res}^{(l)})$ of rescaled degrees $k^{(l)}_{res}$ for different layers $l$ in each subject as compared to the multiscale GR shell, where the symbols correspond to the empirical multiscale connectome and the line to the GR flow.}
\end{figure*}

\begin{figure*}[t]
	\centering
	\includegraphics[width=1\linewidth]{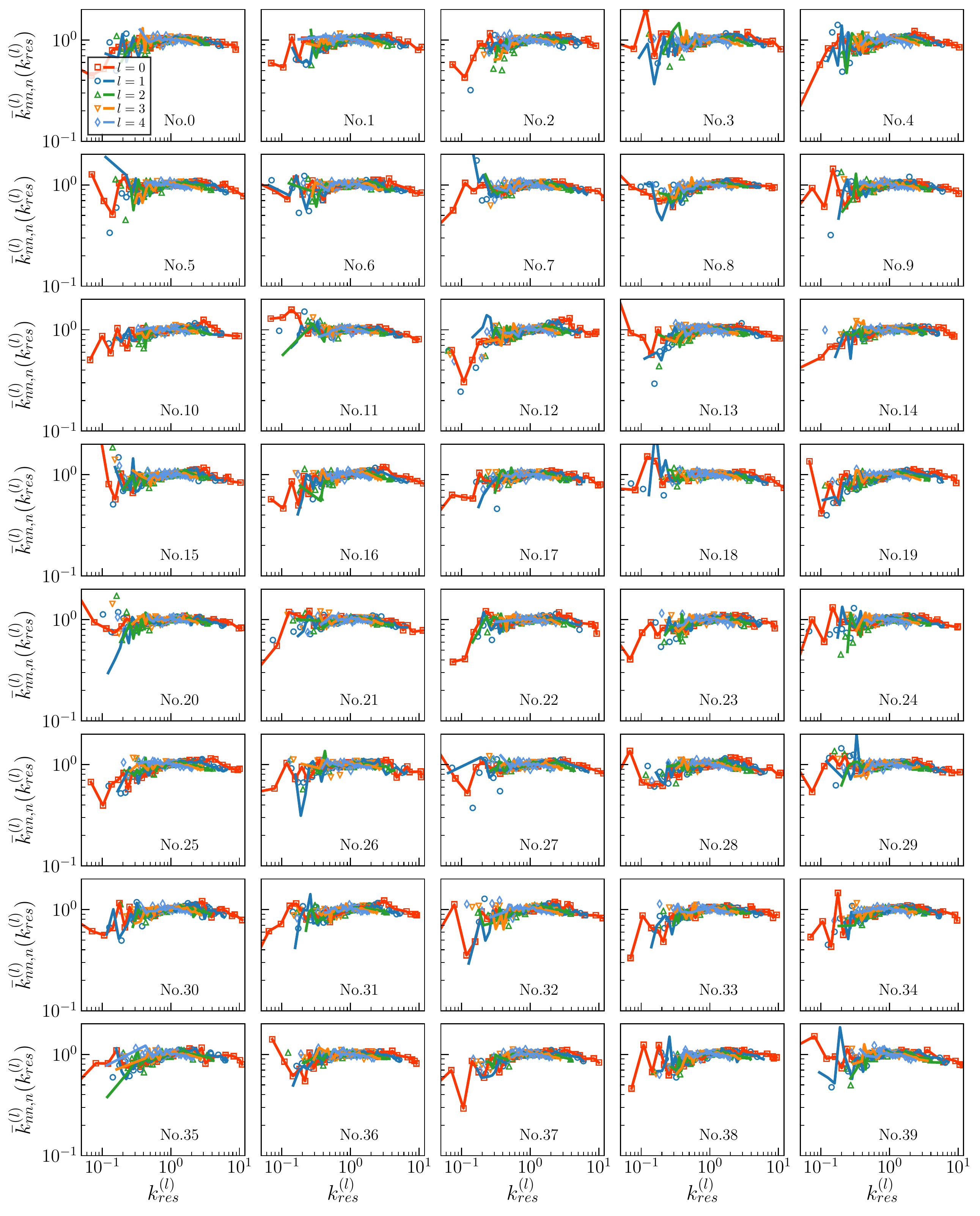}%
	\caption{\label{fig:FigA3} Normalized average nearest-neighbour degree $\bar{k}_{nn,n}^{(l)} (k_{res}^{(l)}) = \bar{k}_{nn}^{(l)} (k_{res}^{(l)}) \langle  k^{(l)}\rangle/\langle(k^{(l)})^2\rangle$ versus rescaled degrees $k^{(l)}_{res}$ for different layers $l$ in each subject as compared to the multiscale GR unfolding, where the symbols correspond to the empirical multiscale connectome and the line to the GR flow.}
\end{figure*}

\begin{figure*}[t]
	\centering
	\includegraphics[width=1\linewidth]{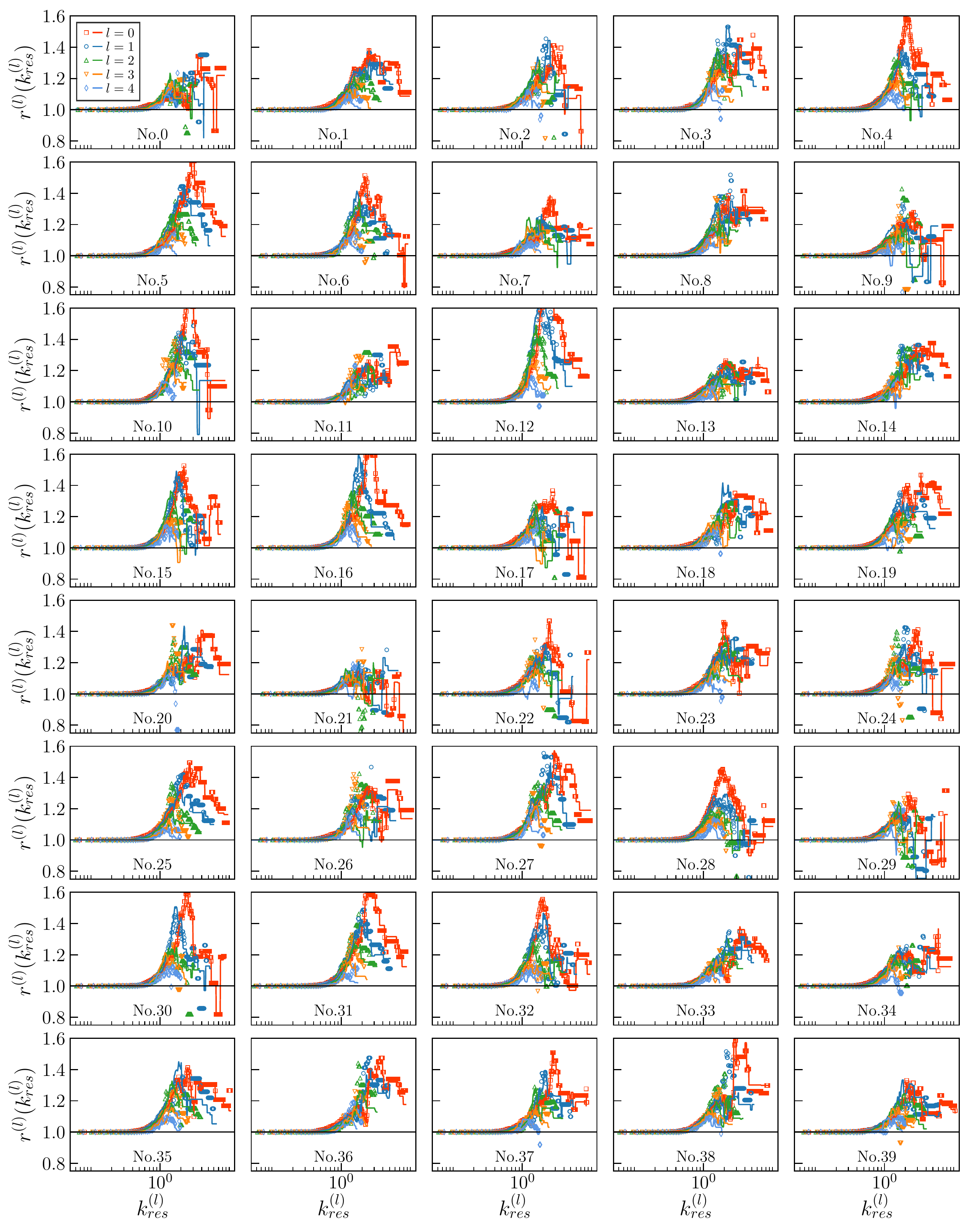}%
	\caption{\label{fig:FigA4} Rich club coefficient $r^{(l)}(k_{res}^{(l)})$ versus rescaled degrees $k^{(l)}_{res}$ for different layers $l$ in each subject as compared to the multiscale GR unfolding, where the symbols correspond to the empirical multiscale connectome and the line to the GR flow. The two largest hubs in subjects 2 and 21 are disconnected, giving two outlier values $0$. }
\end{figure*}

\begin{figure*}[t]
	\centering
	\includegraphics[width=1\linewidth]{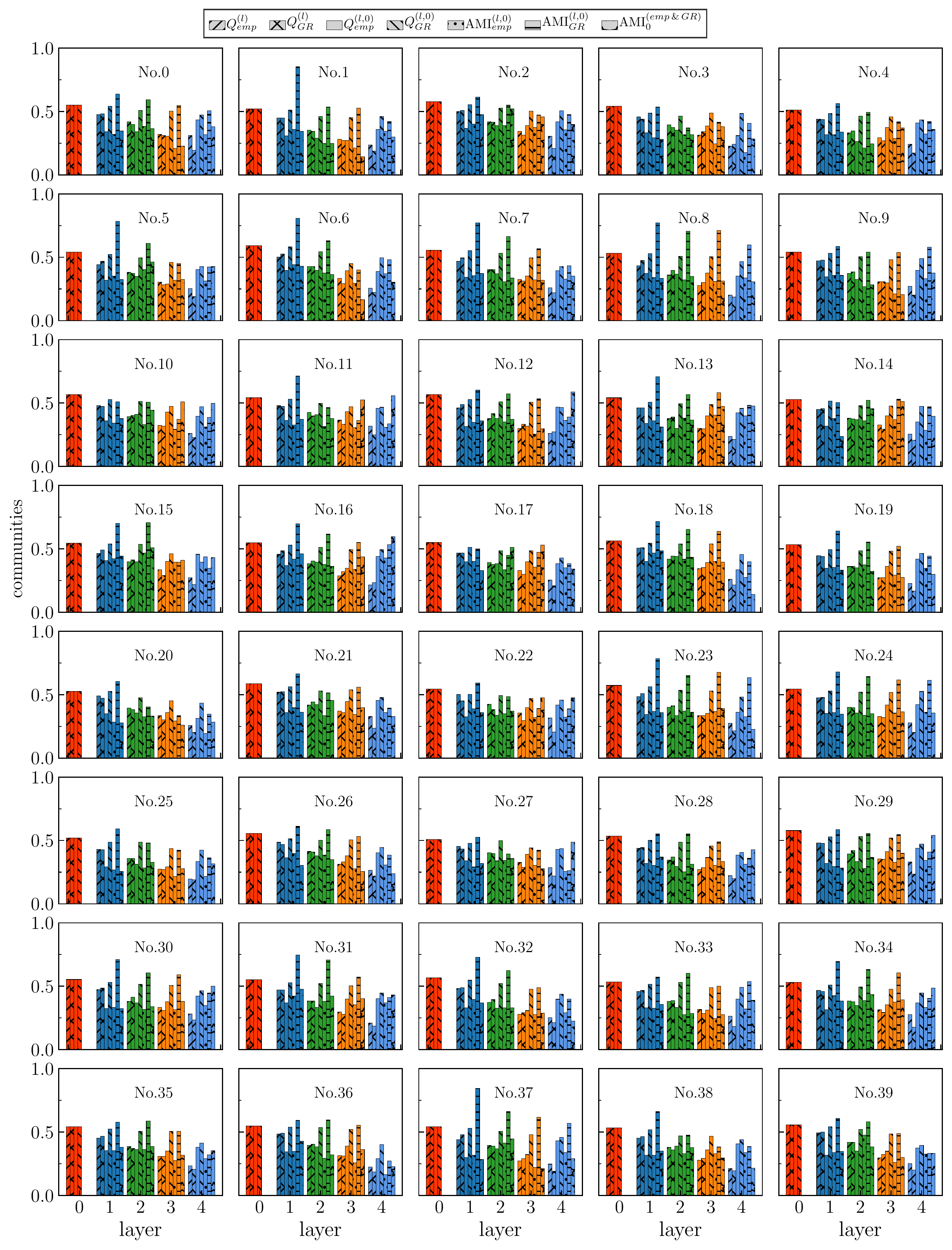}%
	\caption{\label{fig:FigA5} Community structure of the empirical multiscale connectomes and the GR unfolding.}
\end{figure*}

\begin{figure*}[t]
	\centering
	\includegraphics[width=1\linewidth]{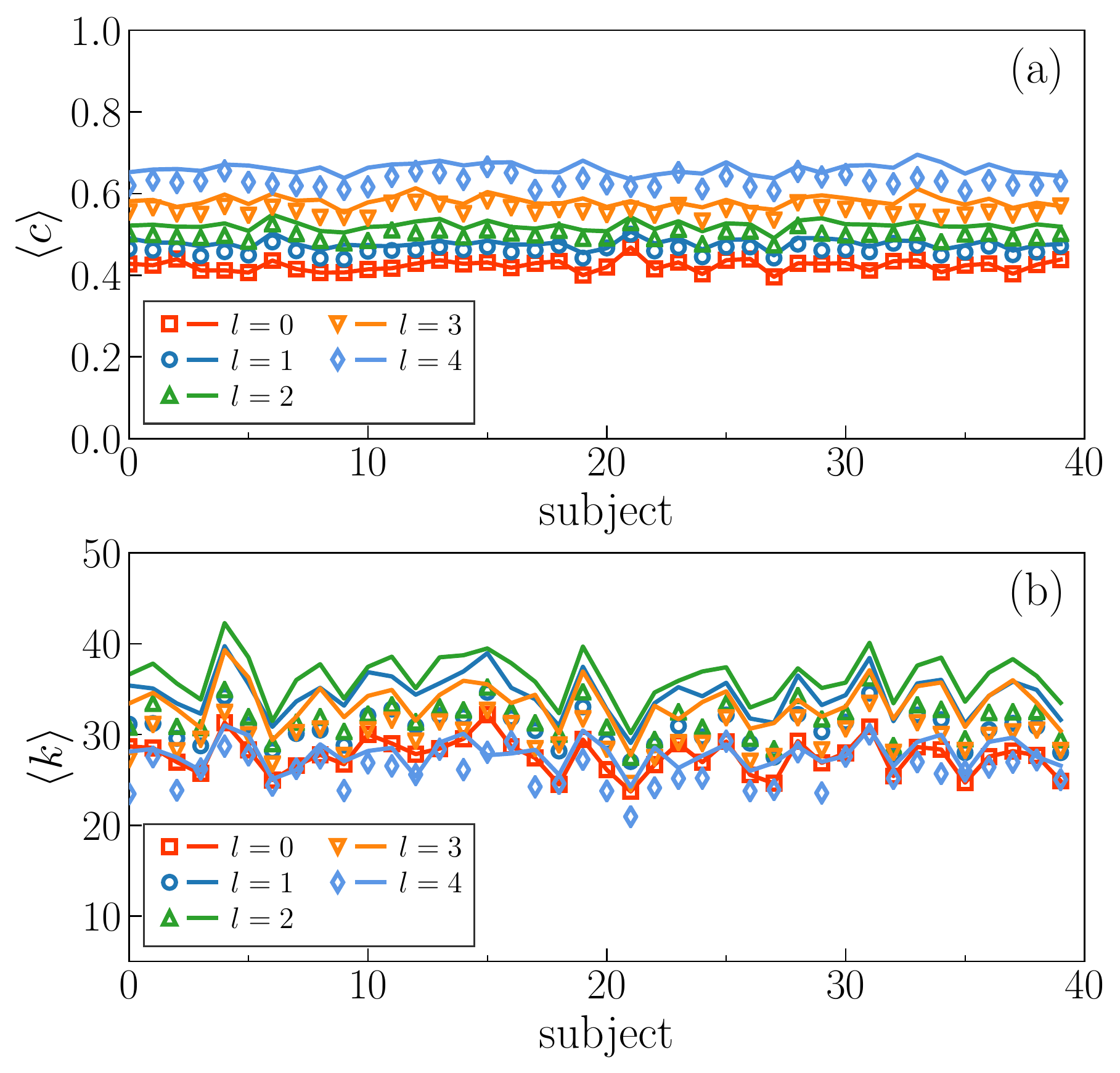}%
	\caption{\label{fig:FigA6}Average clustering coefficient and mean degree for all the layers in each subject as compared to the multiscale GR unfolding, where the symbols correspond to the empirical multiscale connectome and the lines to the GR flow.}
\end{figure*}

\begin{figure*}
	\centering
	\includegraphics[width=1\linewidth]{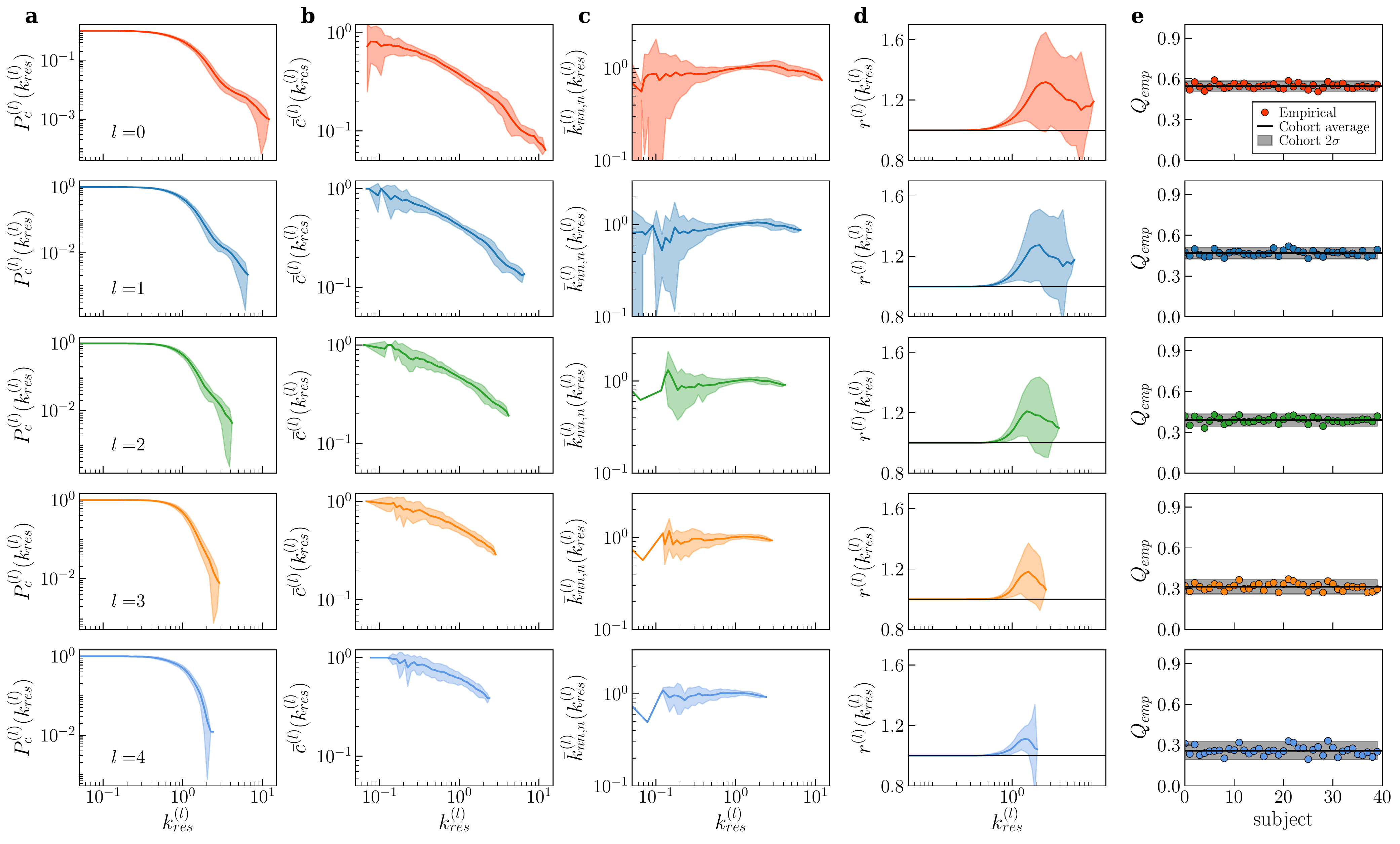}%
	\caption{\textbf{Network properties across 40 subjects for all layers in the UL dataset.} Each column shows the complementary cumulative degree distribution, degree-dependent clustering coefficient, degree-degree correlations, rich club coefficient and modularity. The degrees have been rescaled by the internal average degree of the corresponding layer $k^{(l)}_{res} = k^{(l)}/ \langle k^{(l)}\rangle$. The solid lines show the corresponding average values across 40 subjects in the cohort and the shadows indicate 2$\sigma$ deviations. }	
\end{figure*}
\begin{figure*}
	\centering
	\includegraphics[width=1\linewidth]{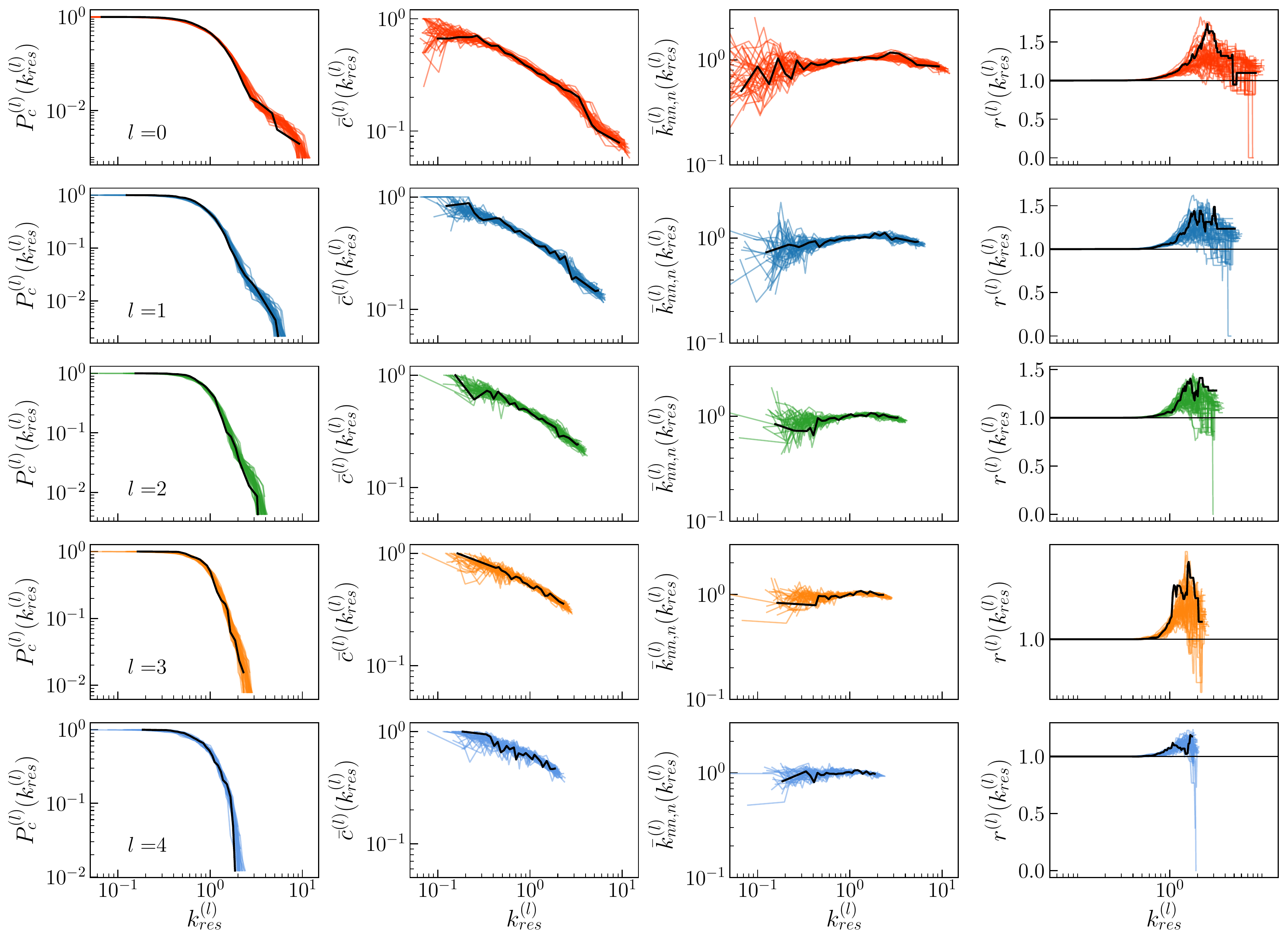}%
	\caption{\textbf{Subject No.~10 is a typical subject in the UL dataset.} Each column shows the complementary cumulative degree distribution, degree dependent clustering coefficient, degree-degree correlations and rich club coefficient. The degrees have been rescaled by the internal average degree of the corresponding layer $k^{(l)}_{res} = k^{(l)}/ \langle k^{(l)}\rangle$. Different lines correspond to different subject in each cohort. The results for subject No.~10 have been highlighted in black color.}	
\end{figure*}

\begin{figure*}[h]
	\centering
	\includegraphics[width=0.9\linewidth]{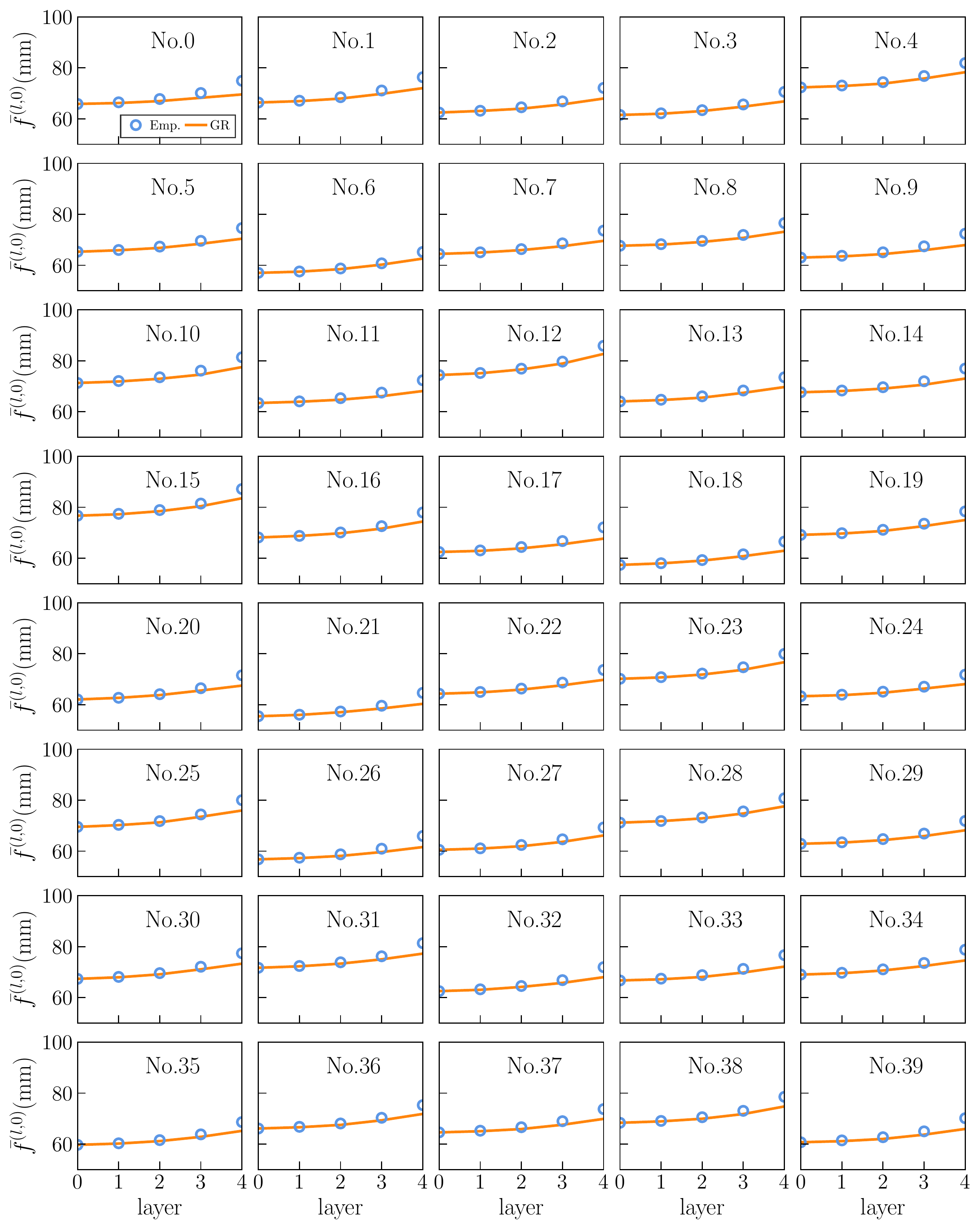}%
	\caption{
		Average fiber length $\bar{f}^{(l,0)}$ in layer $0$ of links outside supernodes in layer $l$, where supernodes are defined by the anatomical coarse-graining in the empirical curve (symbols) or the similarity coarse-graining in the GR case (lines). 
	}
\end{figure*}


\clearpage
\newpage

\clearpage
\newpage
\subsection{Behavior of the connection probabilities}
\begin{figure*}[!h]
	\centering
	\includegraphics[width=0.89\linewidth]{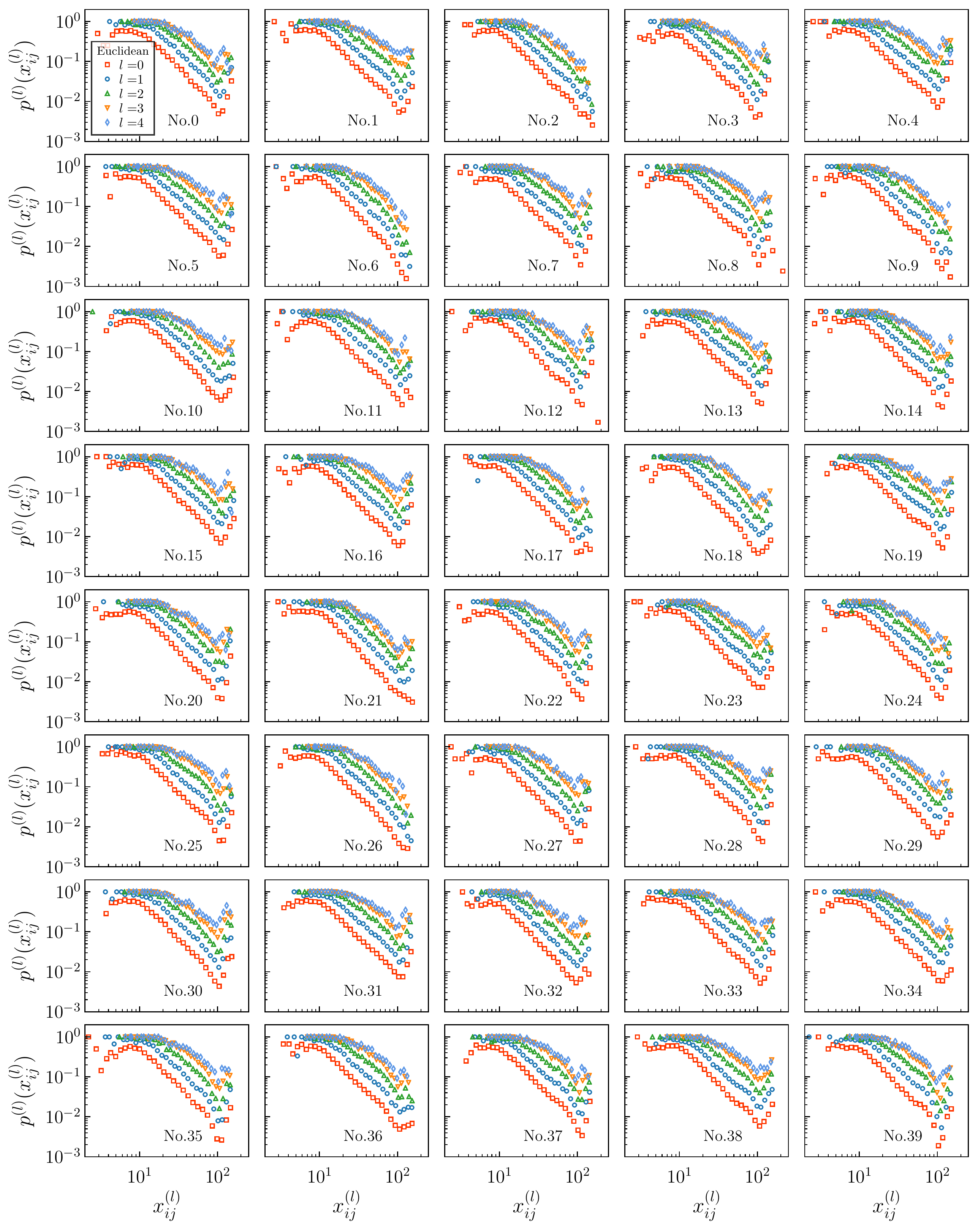}%
	\caption{Empirical connection probabilities $p^{(l)}(x_{ij}^{(l)})$ for each subject in the Euclidean space. The whole range of Euclidean distances $x_{ij}$ is binned, and for each bin the ratio of the number of connected connectome pairs to the total number of connectome pairs falling within this bin is shown.
	}
\end{figure*}

%

%
\begin{figure*}[t]
	\centering
	\includegraphics[width=0.95\linewidth]{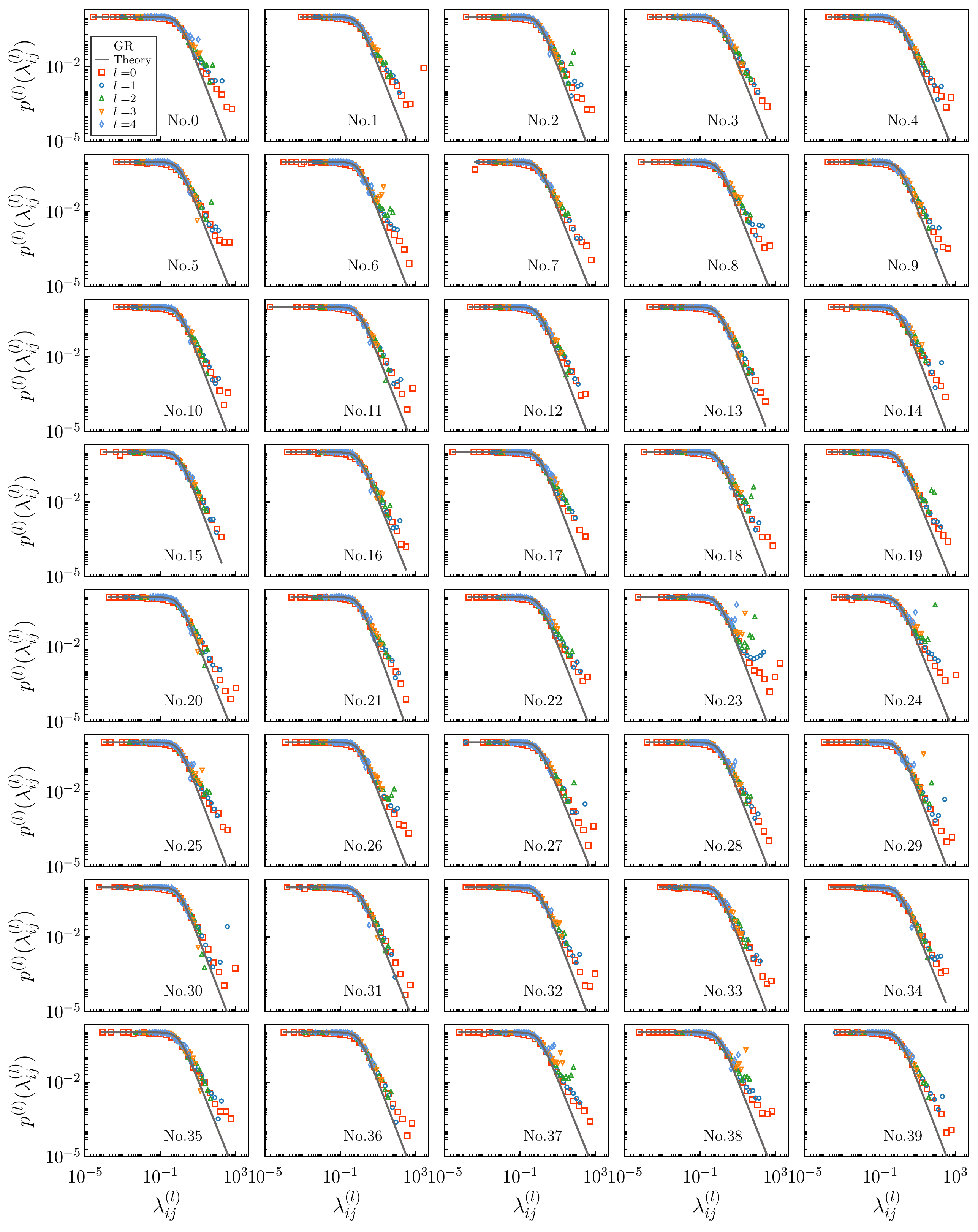}%
	\caption{
		Empirical versus theoretical connection probability  $p^{(l)}(\lambda_{ij}^{(l)})$ within a given range of $\lambda_{ij}^{(l)}$ in GR shell for each subject. Open symbols are the connection probability of GR networks within a given range of $\lambda_{ij}^{(l)}$ and the gray lines shows the theoretical curves. 
	}
\end{figure*}

\clearpage
\newpage
\begin{figure*}[!h]
	\centering
	\includegraphics[width=0.7\linewidth]{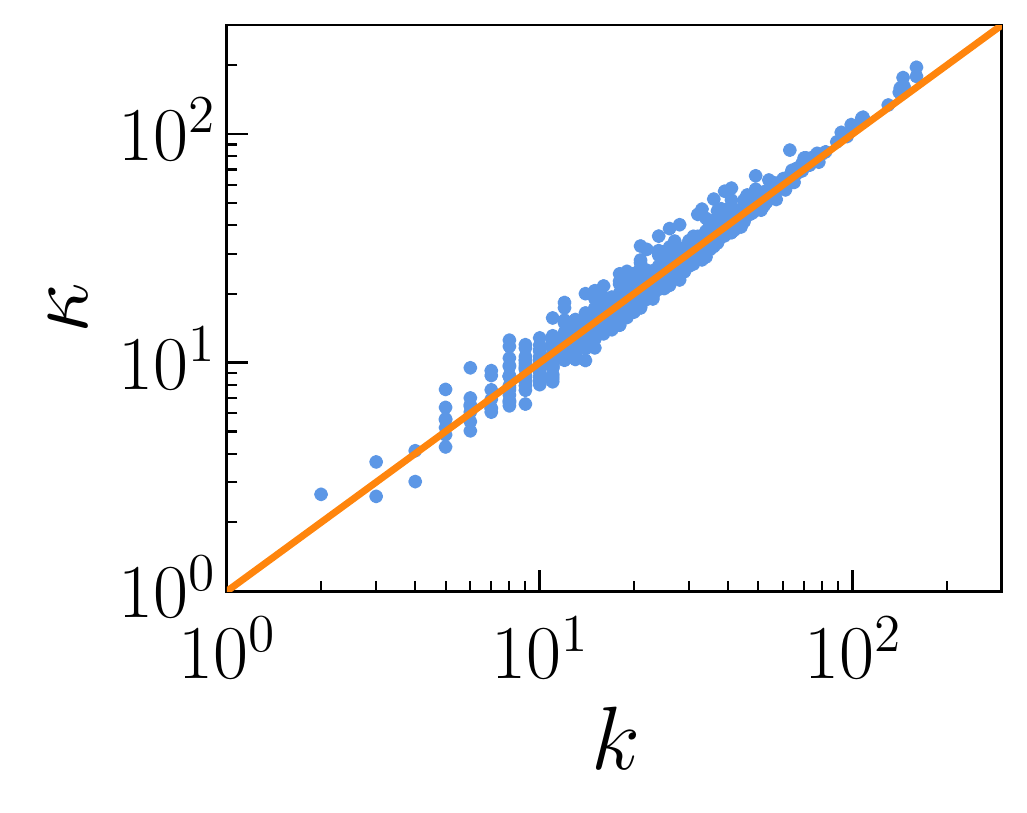}%
	\caption{\label{fig:FigA7}\textbf{Hidden degrees $\kappa$ versus observed degree $k$ of highest resolution layer in subject No.~10.}}
\end{figure*}

\clearpage
\newpage
\subsection{Navigability on the independent MH connectome layers and GR shell}
\begin{figure*}[htb]
	\centering
	\includegraphics[width=1\linewidth]{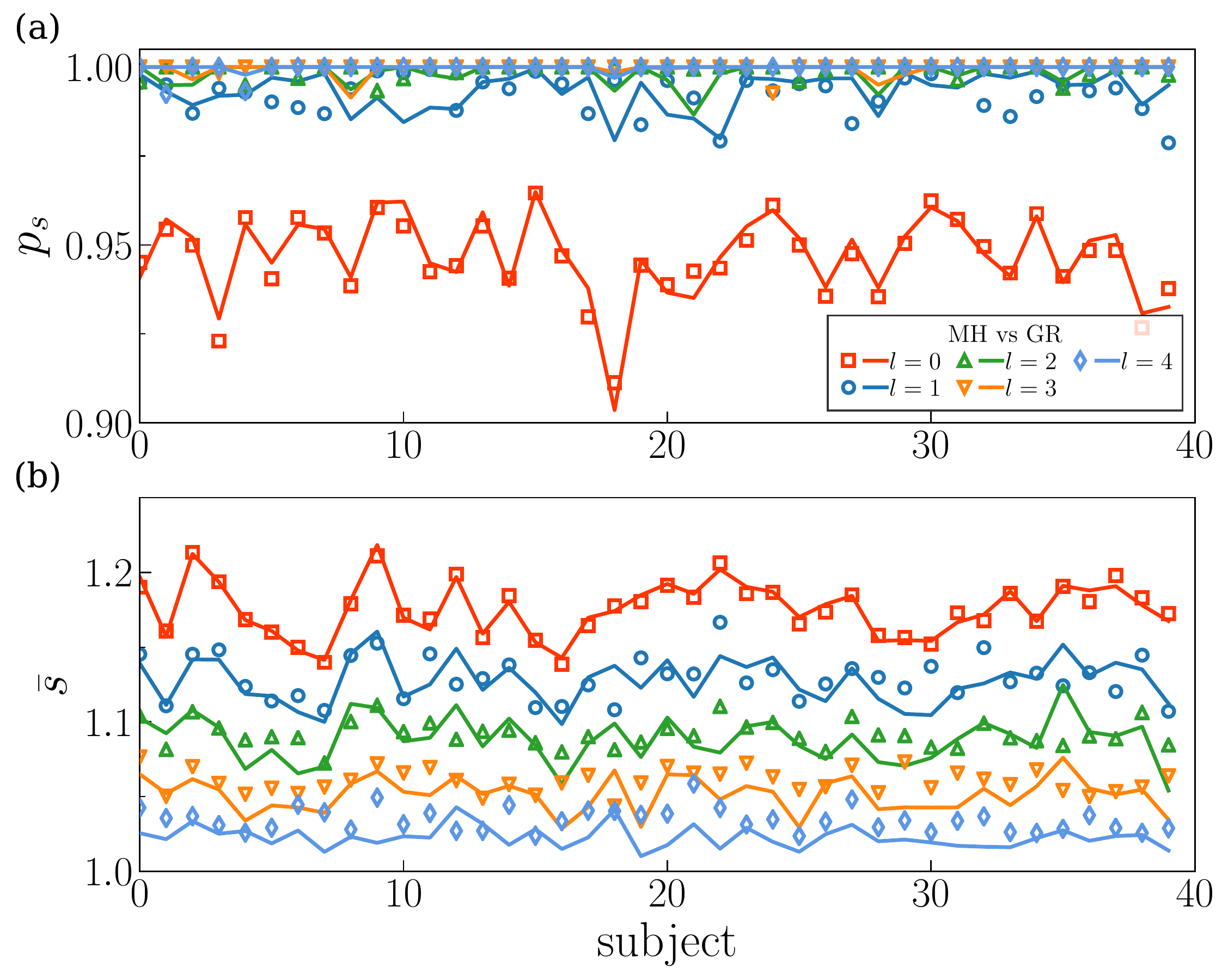}%
	\caption{\label{fig:FigA8} \textbf{Navigability in the hyperbolic space on the independent MH connectome layers and GR shell.} open symbols correspond to the independent hyperbolic embeddings of the MH connectome layers and the solid lines to the GR shell. }
\end{figure*}

\clearpage
\newpage
\subsection{Multiscale navigation protocol.} 

In this section, we study the performance of the multiscale GR protocol\cite{Garcia2018} in the GR shell of the $40$ MH connectomes. Notice that when a packet can get stuck into a loop in single layer greedy routing, the multiscale navigation protocol can find alternative paths by taking advantage of the increased efficiency of greedy forwarding in the coarse-grained layers. When node $i$ needs to send a packet to a destination node $j$, node $i$ performs a virtual greedy forwarding step in the highest possible layer to find which supernode should be next in the greedy path. Based on this, node $i$ then forwards the packet to its physical neighbour in the real network, which guarantees that it will eventually reach such supernode. The full details of this process is described as follows.

To guarantee navigation inside supernodes, we require an extra condition in the renormalization process and only consider blocks of connected consecutive nodes (a single node can be left alone forming a supernode by itself) to produce the GR shell. Notice that the new requirement does not alter the self-similarity of the renormalized networks forming the multiscale shell nor the congruency with the hidden metric space \cite{Garcia2018}. 

With respect to standard greedy routing in single layered networks, the multiscale navigation protocol requires adding the following information about the supernodes and their neighbors. 
\begin{itemize}
	
	\item[(i)] The coordinates $(r_i^{(l)},\theta_i^{(l)})$ of node $i$ in every layer $l$.
	
	\item[(ii)] For each node $i$, she should know her (super)neighbours list and their coordinates in each layer. 
	
	\item[(iii)] Let SuperN$(i, l)$ be the supernode to which $i$ belongs in layer $l$.
	Supposed that SuperN$(i, l)$ contains (super)nodes $\{i,i_1,i_2\ldots\}$ and SuperN$(k, l)$ has (super)nodes $\{k,k_1,k_2\ldots\}$ in layer $l$. If SuperN$(i, l)$ is connected to SuperN$(k, l)$, 
	there is at least one edge between (super)nodes $\{i,i_1,i_2\ldots\}$ and $\{k,k_1,k_2\ldots\}$ in layer $l$, and the connected (super)nodes of $\{i,i_1,i_2\ldots\}$ and $\{k,k_1,k_2\ldots\}$ are called ``gateway'. So, for every superneighbour of node SuperN$(i, l)$ in layer $l$, node $i$ knows which (super)node or (super)nodes in layer $l-1$ are gateways reaching it. 
	
	\item[(iv)] If SuperN$(i,l-1)$ is a gateway reaching some supernode $s$, at least one of its (super)neighbours in layer $l-1$ belongs to $s$; node $i$ knows which.
\end{itemize}

\begin{figure}[t]
	\centering
	\includegraphics[width=1\linewidth]{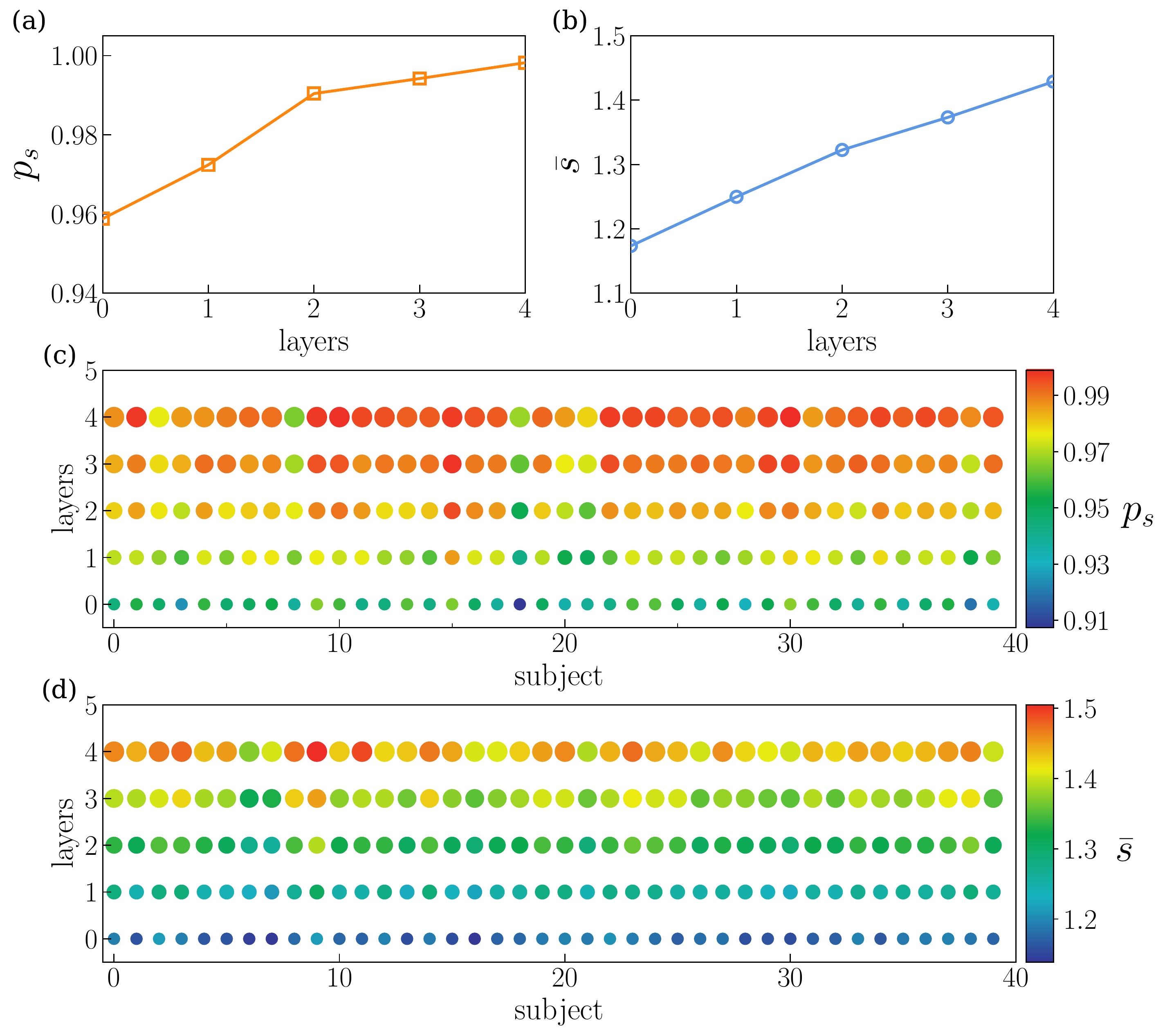}%
	\caption{\label{fig:Fig6}\textbf{Performance of the GR multiscale navigation protocol in the GR shells of the human connectomes.} (a) Success rate $p_s$ and (b) average stretch $\bar{s}$ as a function of the number of GR shell layers used in the routing process for subject No.~10, computed for $10^4$ randomly selected pairs of nodes in layer $l=0$. (c) and (d) The same for the 40 subjects in the cohort. The colors of the dots represent the magnitude of the corresponding property and their sizes are proportional to the number of layers used in the routing process.}
\end{figure}

This information allows us to navigate the network as follows.   
If node $i$ wants to send a packet to a destination node $j$, node $i$ 
should know $j'$s coordinates in all $L$ layers  $(r_i^{(l)},\theta_i^{(l)})$ and then node $i$ 
will first check if it is connected to $j$; in that case, the decision is clear. If it is not, it 
will performs a virtual greedy forwarding step in the highest possible layer to find which supernode should be next in the greedy path. The detailed steps are provided as following:
\begin{itemize}
	
	\item[1.]Find the highest layer $l_{max}$ in which SuperN$(i, l_{max})$ and SuperN$(j, l_{max})$ still have different coordinates. Set $l= l_{max}$.
	
	\item[2.] Perform a standard step of greedy routing in layer $l$: find the closest
	neighbour of SuperN$(i, l)$ to SuperN$(j, l)$. This is the current target SuperT$(l)$.
	
	\item[3.] While $l>0$, look into layer $l-1$:
	
	Set $l=l-1$.
	
	If SuperN$(i, l)$ is a gateway connecting to some (super)node within
	SuperT$(l+1)$, node $i$ sets as new current target SuperT$(l)$ its (super)neighbour belonging to SuperT$(l+1)$ closest to SuperN$(j, l)$.
	Else node $i$ sets as new target SuperT$(l)$ the gateway in SuperN$(i, l+ 1)$
	connecting to SuperT$(l+ 1)$ (its (super)neighbour belonging to SuperN$(i, l+ 1))$.
	
	\item[4.] In layer $l=0$, SuperT$(0$) belongs to the real network and she is a neighbour of $i$, so node $i$ forwards the message to SuperT$(0)$.
\end{itemize}
Fig.~\ref{fig:Fig6} (a) shows the gain in success rate as the number of renormalized layers used in the multiscale navigation process is increased, for the representative subject in the cohort. Interestingly, the navigability properties of every GR brain representation are very similar. For all subjects, the success rate increases significantly with the number of navigated layers in the shell, in fact it becomes very close to $100\%$ with the inclusion of just two renormalized layers (Fig.~\ref{fig:Fig6} (c)), and this improvement comes at the expense of only a mild increase of the stretch of successful paths (Fig.~\ref{fig:Fig6} (b) and (d)). 

\clearpage
\newpage
\subsection{Network properties of null models}
\begin{figure*}[!h]
	\centering
	\includegraphics[width=1\linewidth]{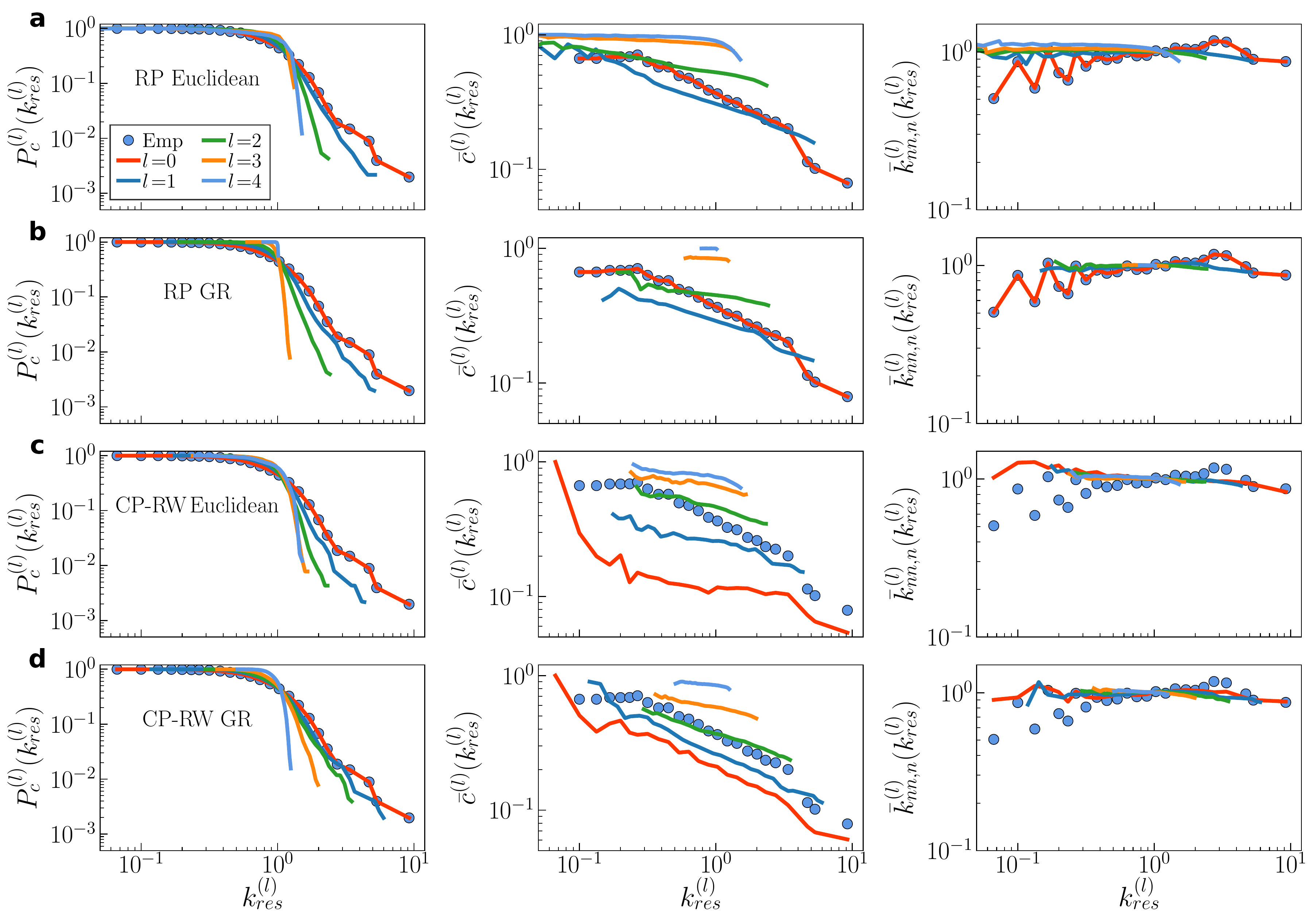}%
	\caption{\textbf{Loss the self-similarity in null networks}. Each column shows complementary cumulative degree distribution $P_c^{(l)}(k_{res}^{(l)})$, degree-dependent clustering coefficient  $\bar{c}^{(l)}(k_{res}^{(l)})$, and degree-degree correlations $\bar{k}_{nn,n}^{(l)} (k_{res}^{(l)})$. 
	}
\end{figure*}
\begin{figure*}[!h]
	\centering
	\includegraphics[width=1\linewidth]{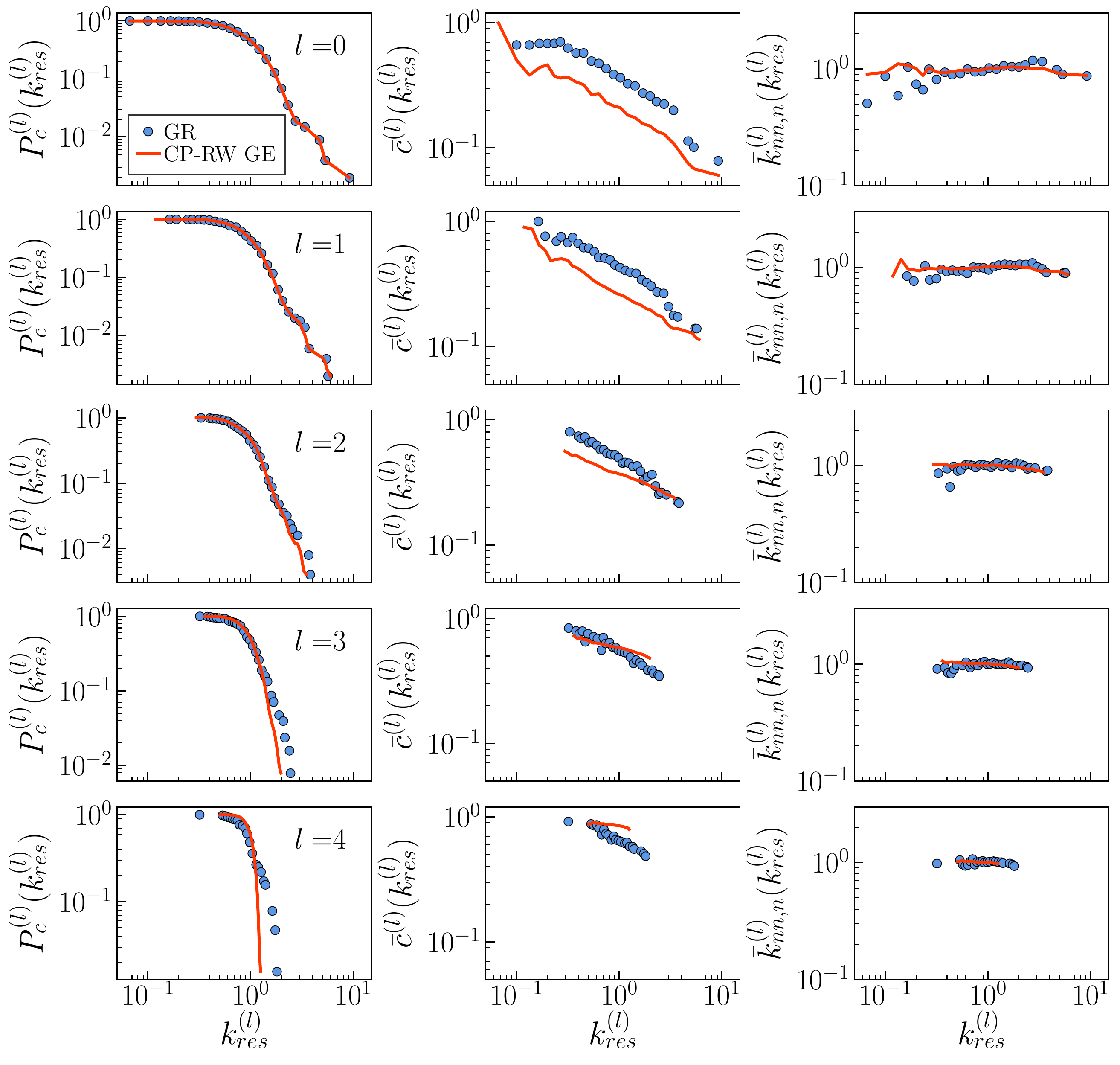}%
	\caption{\textbf{Comparision the network properties between standard GR and CP-RW GE model. } Each column shows complementary cumulative distribution $P_c^{(l)}(k_{res}^{(l)})$, degree dependent clustering coefficient  $\bar{c}^{(l)}(k_{res}^{(l)})$, and degree-degree correlations $\bar{k}_{nn,n}^{(l)} (k_{res}^{(l)})$. We found that the self-similarity was still preserved to some extents in layer 0 to 2.
	}
\end{figure*}

\clearpage
\newpage

\section{Cross-validating results in the HCP dataset}
\begin{figure*}[h]
	\centering
	\includegraphics[width=0.9\linewidth]{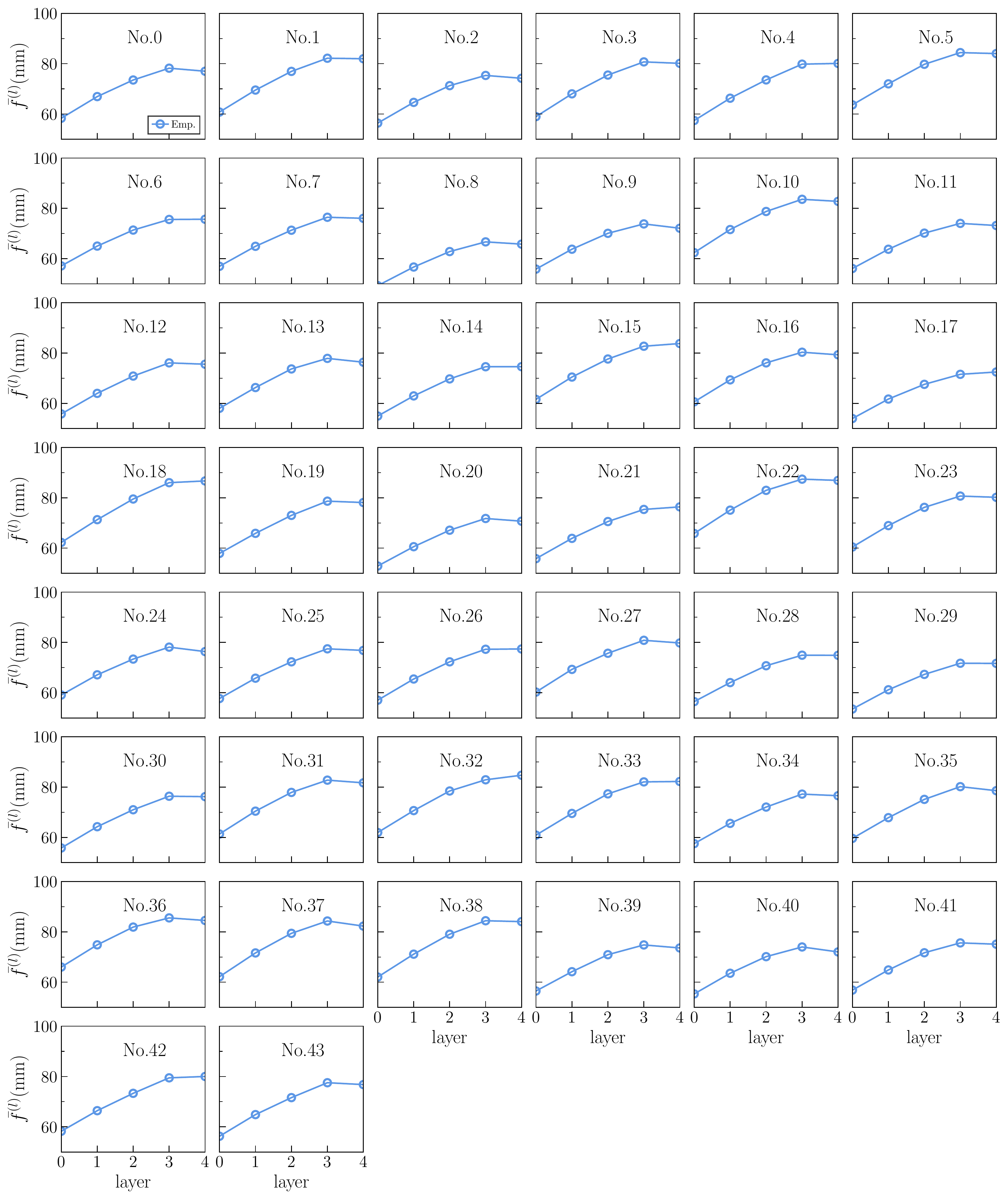}%
	\caption{
		Average fiber length $\bar{f}^{(l)}$ for each subject in HCP dataset. 
	}
\end{figure*}

\begin{figure*}[!h]
	\centering
	\includegraphics[width=1\linewidth]{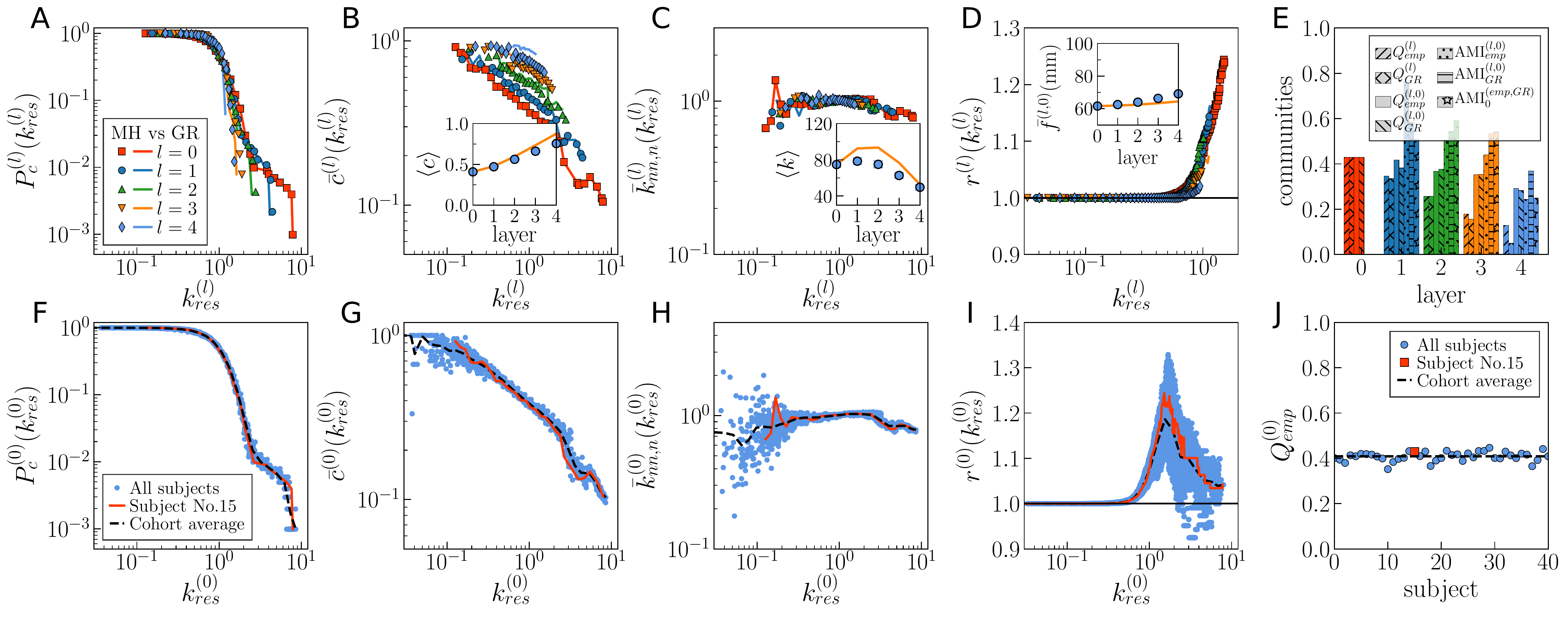}
	\caption{\textbf{Self-similarity of the MH connectome at different resolutions}. (\textit{A-E}) Results for HCP subject No.~15. Filled symbols correspond to the empirical MH connectome and the lines to the GR shell. 
		(\textit{A}) Complementary cumulative degree distribution $P_c^{(l)}(k_{res}^{(l)})$. 
		(\textit{B}) Degree dependent clustering coefficient $\bar{c}^{(l)}(k_{res}^{(l)})$. Inset: flow of the average clustering coefficient $\langle c \rangle$. 
		(\textit{C}) Degree-degree correlations $\bar{k}_{nn,n}^{(l)} (k_{res}^{(l)})$. Inset: flow of the average degree $\langle k\rangle$. 
		(\textit{D}) Rich club coefficient $r^{(l)}(k_{res}^{(l)})$ for low and intermediate values of the rescaled threshold degrees. Inset: average fiber length $\bar{f}^{(l,0)}$ in layer $0$ of links outside supernodes in layer $l$, where supernodes are defined by the anatomical coarse-graining in the empirical curve or the similarity coarse-graining in the GR case. In the three insets in (\textit{B})-(\textit{D}), error bars show the $\pm 2$ standard error interval around the mean; when not visible the bars are within symbol size. 
		(\textit{E}) Community structure of the multiscale connectomes. $Q^{(l)}$ is the modularity in layer $l$, $Q^{(l,0)}$ is the modularity that the community structure of layer $l$ induces in layer $0$, and $\textrm{AMI}^{(l)}$ is the adjusted mutual information between the latter and the community partition directly detected in layer $0$ (see Materials and Methods). The subindices $\{emp,GR\}$ indicate the empirical MH connectomes and the GR shell, respectively. $\textrm{AMI}_{0}^{(emp,GR)}$ is the adjusted mutual information between topological communities in the empirical MH connectomes at each layer and the GR flow measured in their projection over layer 0.
		(\textit{F-J}) Variability of topological properties in the HCP dataset. Blue symbols correspond to the properties of layer $0$ in all subjects. The red line correspond to HCP subject No.~10. The black dashed line represents the average value across the 44 subjects in the cohort. In the plots, degrees have been rescaled by the average degree of the corresponding layer $k^{(l)}_{res} = k^{(l)}/ \langle k^{(l)}\rangle$.} 
\end{figure*}
\clearpage
\newpage

\begin{figure*}[!h]
	\centering
	\includegraphics[width=1\linewidth]{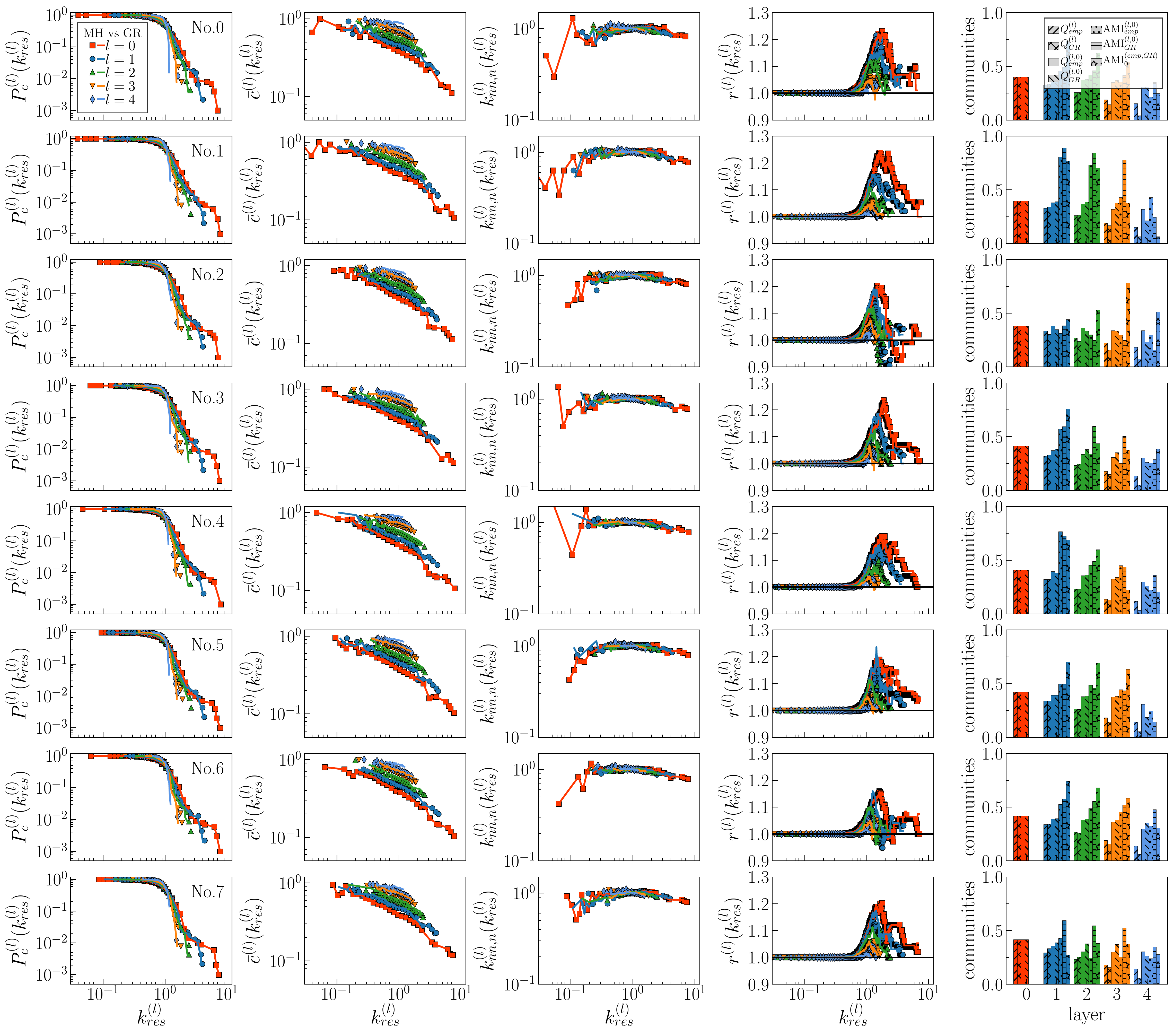}%
	\caption{\textbf{Self-similarity of the MH connectome at different resolutions}. We show results for subject No.~0-7 in HCP dataset. Filled symbols correspond to the empirical MH connectome and the lines to the GR shell. Each column shows complementary cumulative distribution $P_c^{(l)}(k_{res}^{(l)})$, degree dependent clustering coefficient  $\bar{c}^{(l)}(k_{res}^{(l)})$, degree-degree correlations $\bar{k}_{nn,n}^{(l)} (k_{res}^{(l)})$, rich club coefficient $r^{(l)}(k_{res}^{(l)})$, and community structure of the multiscale connectomes. 
	}
\end{figure*}
\clearpage
\newpage

\begin{figure*}[!h]
	\centering
	\includegraphics[width=1\linewidth]{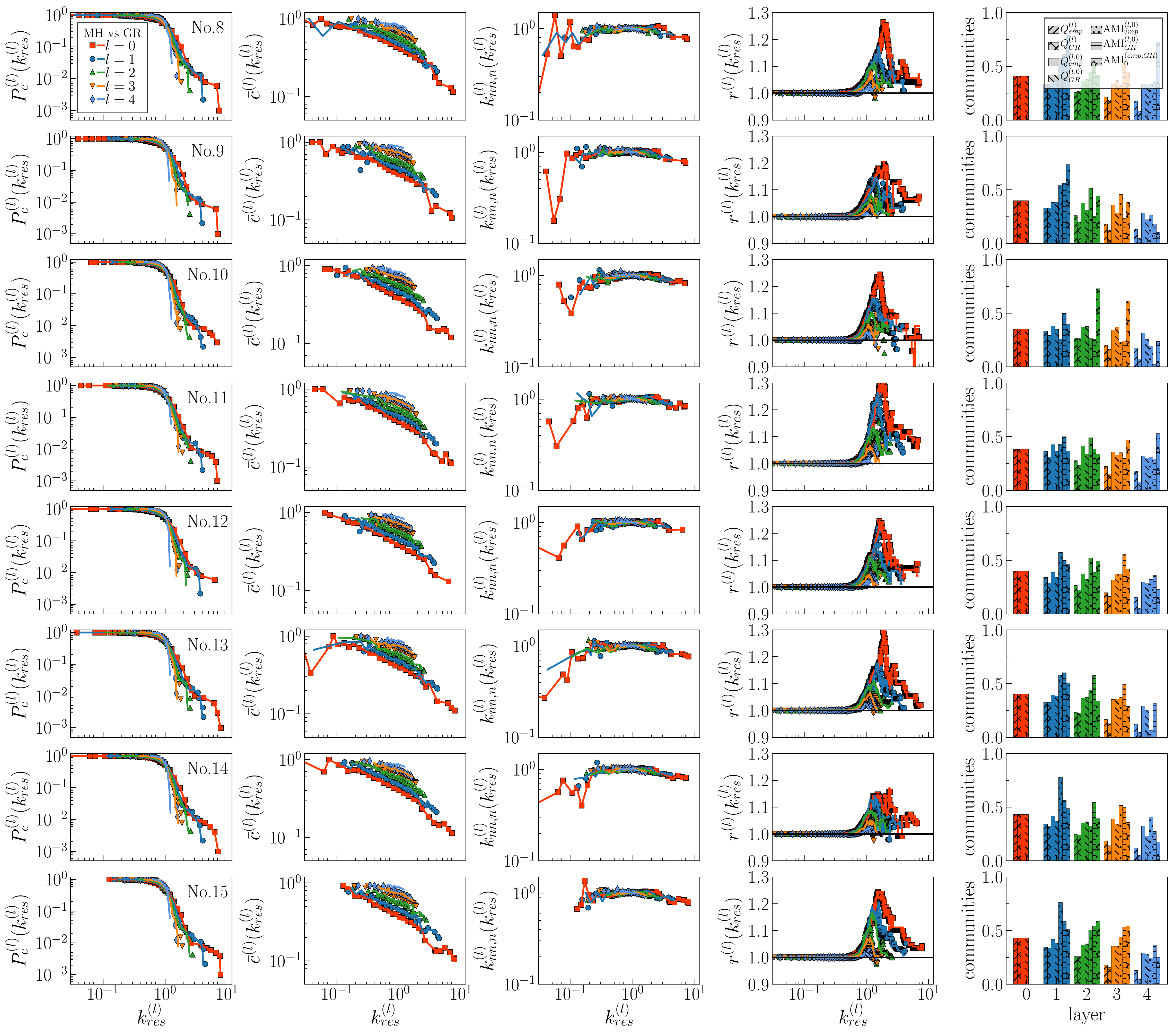}%
	\caption{\textbf{Self-similarity of the MH connectome at different resolutions}. We show results for subject No.~8-15 in HCP dataset. Filled symbols correspond to the empirical MH connectome and the lines to the GR shell. Each column shows complementary cumulative distribution $P_c^{(l)}(k_{res}^{(l)})$, degree dependent clustering coefficient  $\bar{c}^{(l)}(k_{res}^{(l)})$, degree-degree correlations $\bar{k}_{nn,n}^{(l)} (k_{res}^{(l)})$, rich club coefficient $r^{(l)}(k_{res}^{(l)})$, and community structure of the multiscale connectomes. 
	}
\end{figure*}

\begin{figure*}[!h]
	\centering
	\includegraphics[width=1\linewidth]{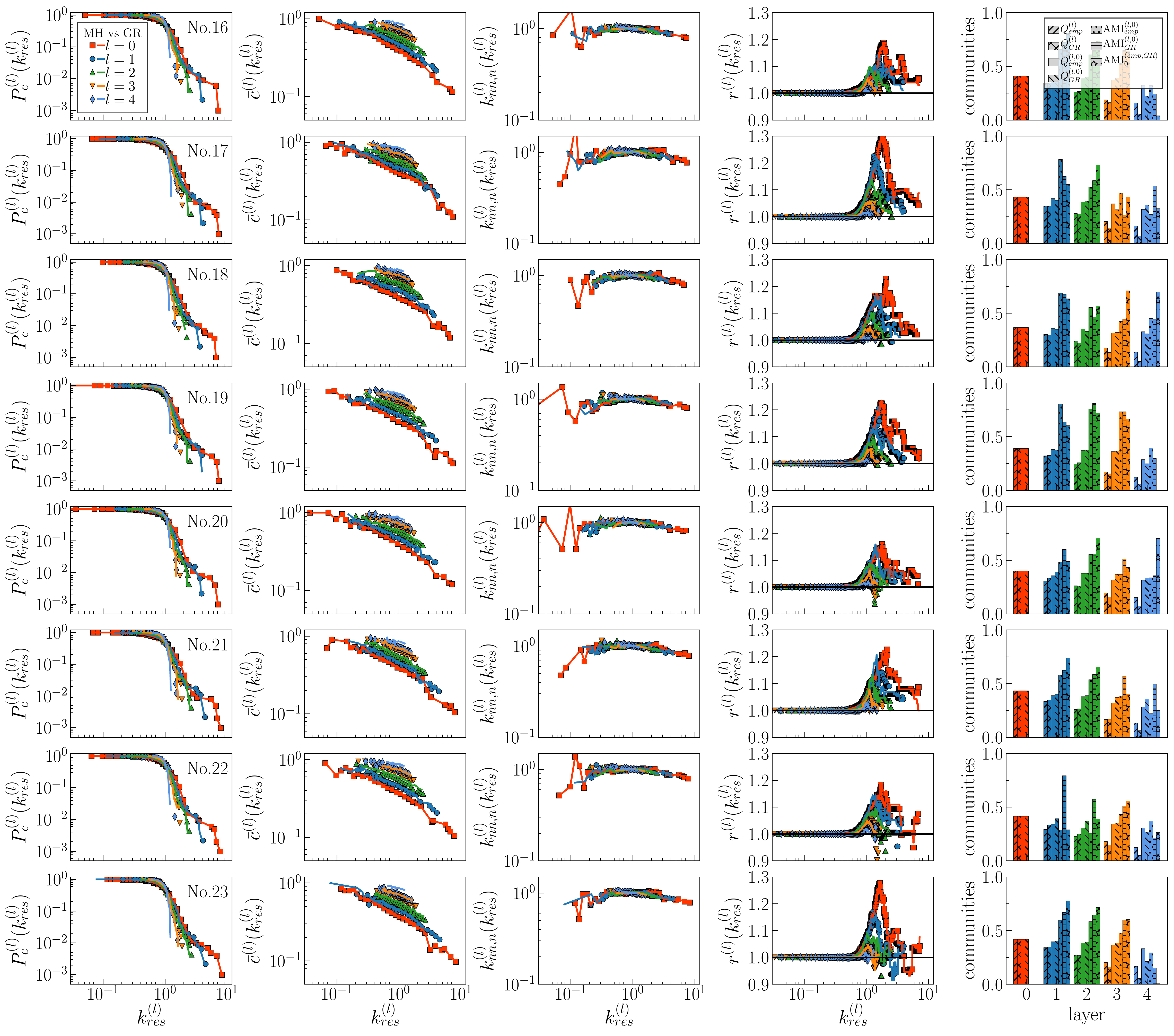}%
	\caption{\textbf{Self-similarity of the MH connectome at different resolutions}. We show results for subject No.~16-23 in HCP dataset. Filled symbols correspond to the empirical MH connectome and the lines to the GR shell. Each column shows complementary cumulative distribution $P_c^{(l)}(k_{res}^{(l)})$, degree dependent clustering coefficient  $\bar{c}^{(l)}(k_{res}^{(l)})$, degree-degree correlations $\bar{k}_{nn,n}^{(l)} (k_{res}^{(l)})$, rich club coefficient $r^{(l)}(k_{res}^{(l)})$, and community structure of the multiscale connectomes. 
	}
\end{figure*}
\begin{figure*}[!h]
	\centering
	\includegraphics[width=1\linewidth]{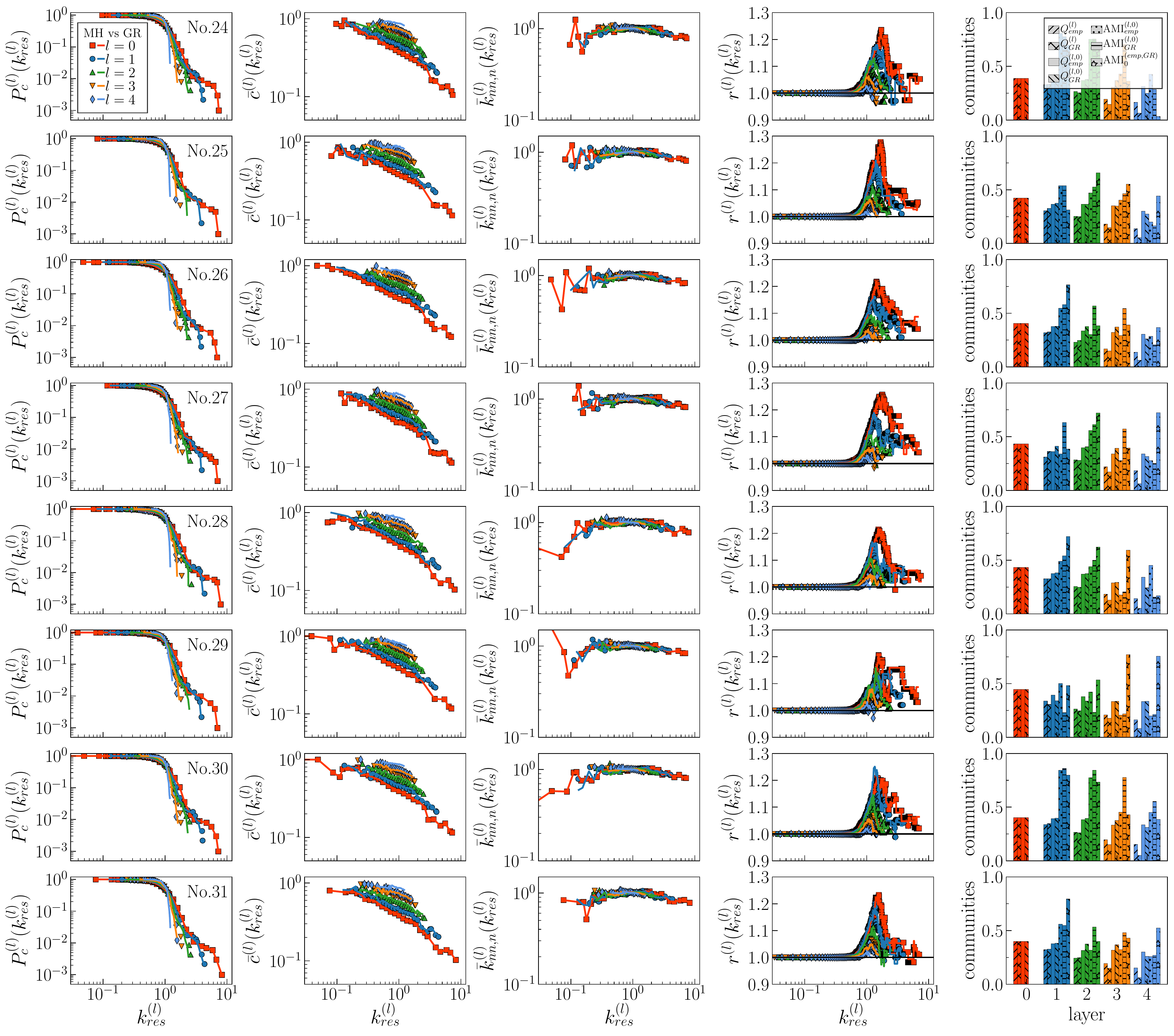}%
	\caption{\textbf{Self-similarity of the MH connectome at different resolutions}. We show results for subject No.~14-31 in HCP dataset. Filled symbols correspond to the empirical MH connectome and the lines to the GR shell. Each column shows complementary cumulative distribution $P_c^{(l)}(k_{res}^{(l)})$, degree dependent clustering coefficient  $\bar{c}^{(l)}(k_{res}^{(l)})$, degree-degree correlations $\bar{k}_{nn,n}^{(l)} (k_{res}^{(l)})$, rich club coefficient $r^{(l)}(k_{res}^{(l)})$, and community structure of the multiscale connectomes. 
	}
\end{figure*}
\begin{figure*}[!h]
	\centering
	\includegraphics[width=1\linewidth]{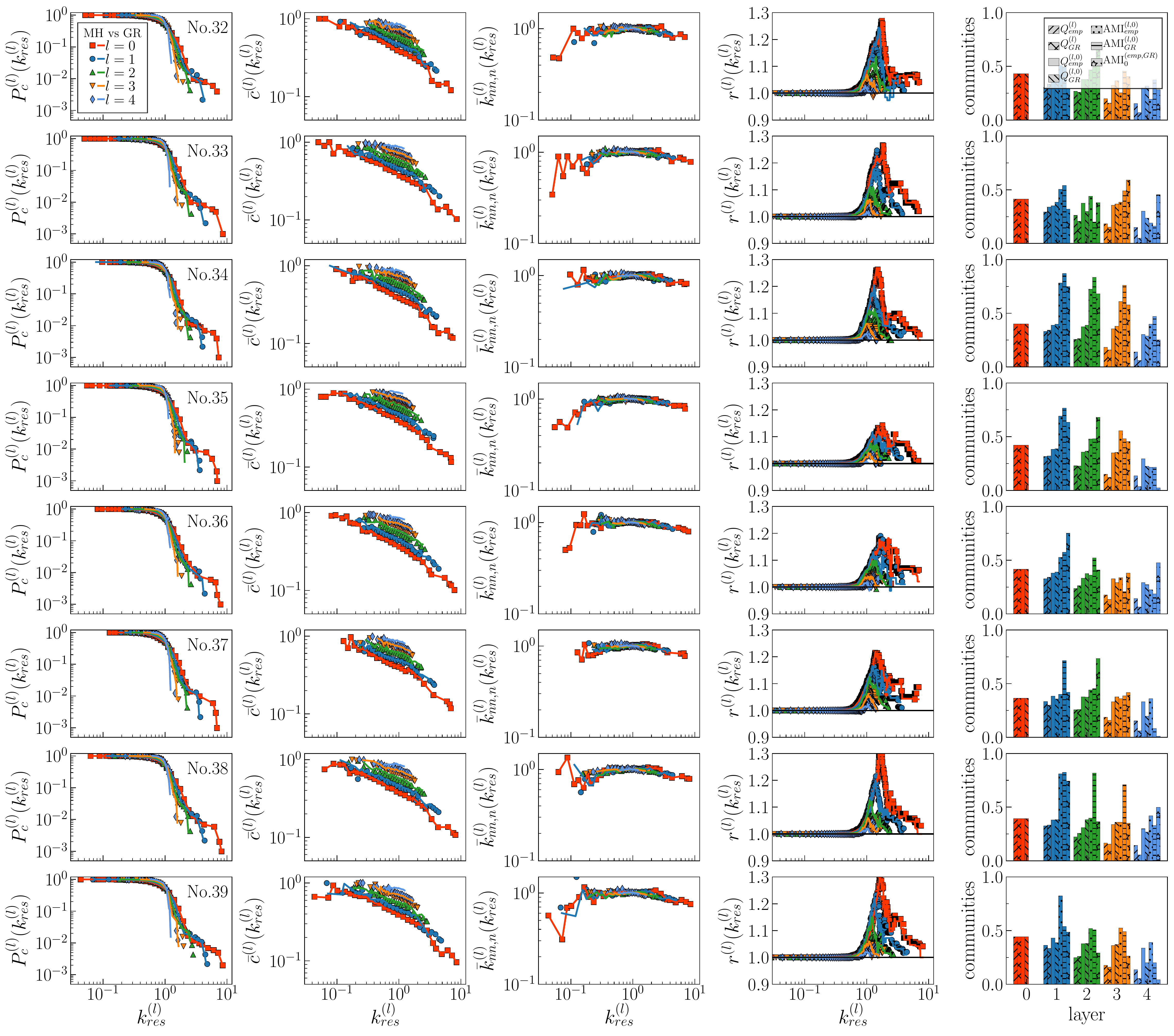}%
	\caption{\textbf{Self-similarity of the MH connectome at different resolutions}. We show results for subject No.~32-39 in HCP dataset. Filled symbols correspond to the empirical MH connectome and the lines to the GR shell. Each column shows complementary cumulative distribution $P_c^{(l)}(k_{res}^{(l)})$, degree dependent clustering coefficient  $\bar{c}^{(l)}(k_{res}^{(l)})$, degree-degree correlations $\bar{k}_{nn,n}^{(l)} (k_{res}^{(l)})$, rich club coefficient $r^{(l)}(k_{res}^{(l)})$, and community structure of the multiscale connectomes. 
	}
\end{figure*}
\begin{figure*}[!h]
	\centering
	\includegraphics[width=1\linewidth]{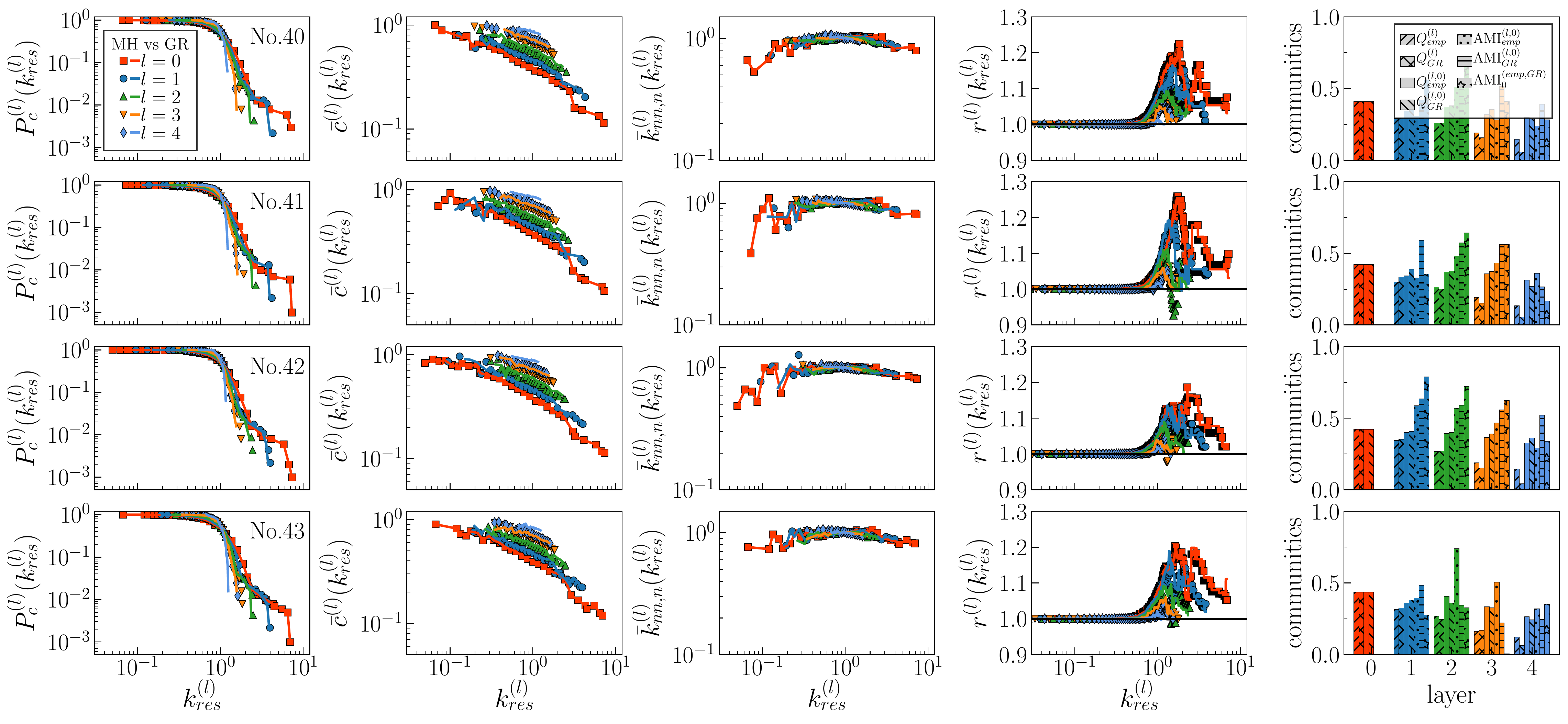}%
	\caption{\textbf{Self-similarity of the MH connectome at different resolutions}. We show results for subject No.~40-43 in HCP dataset. Filled symbols correspond to the empirical MH connectome and the lines to the GR shell. Each column shows complementary cumulative distribution $P_c^{(l)}(k_{res}^{(l)})$, degree dependent clustering coefficient  $\bar{c}^{(l)}(k_{res}^{(l)})$, degree-degree correlations $\bar{k}_{nn,n}^{(l)} (k_{res}^{(l)})$, rich club coefficient $r^{(l)}(k_{res}^{(l)})$, and community structure of the multiscale connectomes. 
	}
\end{figure*}
\begin{figure*}[!h]
	\centering
	\includegraphics[width=1\linewidth]{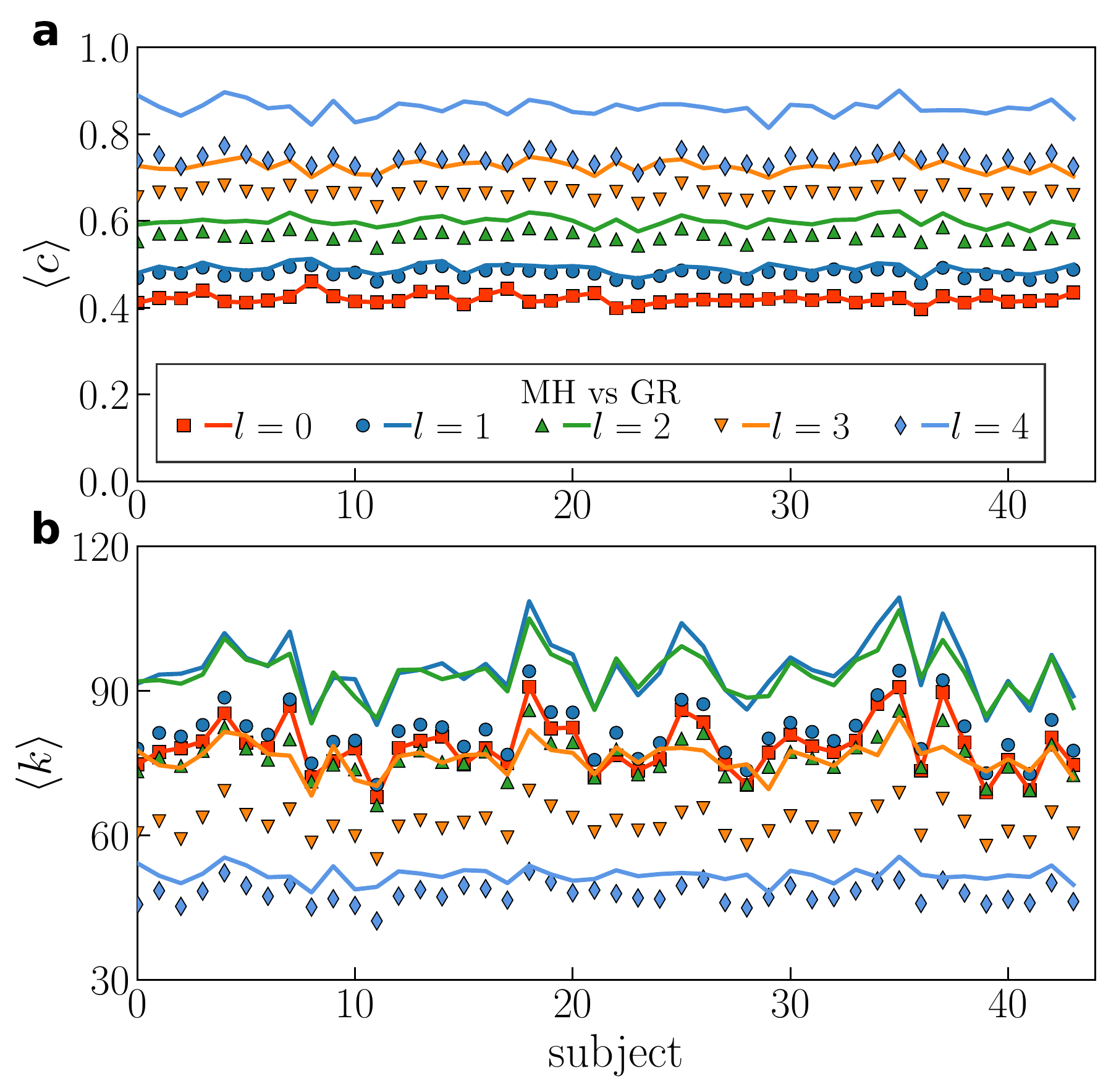}%
	\caption{ Average clustering coefficient (a) and mean degree (b) for all the layers in each subject as compared to the multiscale GR unfolding, where the symbols correspond to the empirical multiscale connectome and the lines to the GR flow.
	}
\end{figure*}
\begin{figure*}
	\centering
	\includegraphics[width=1\linewidth]{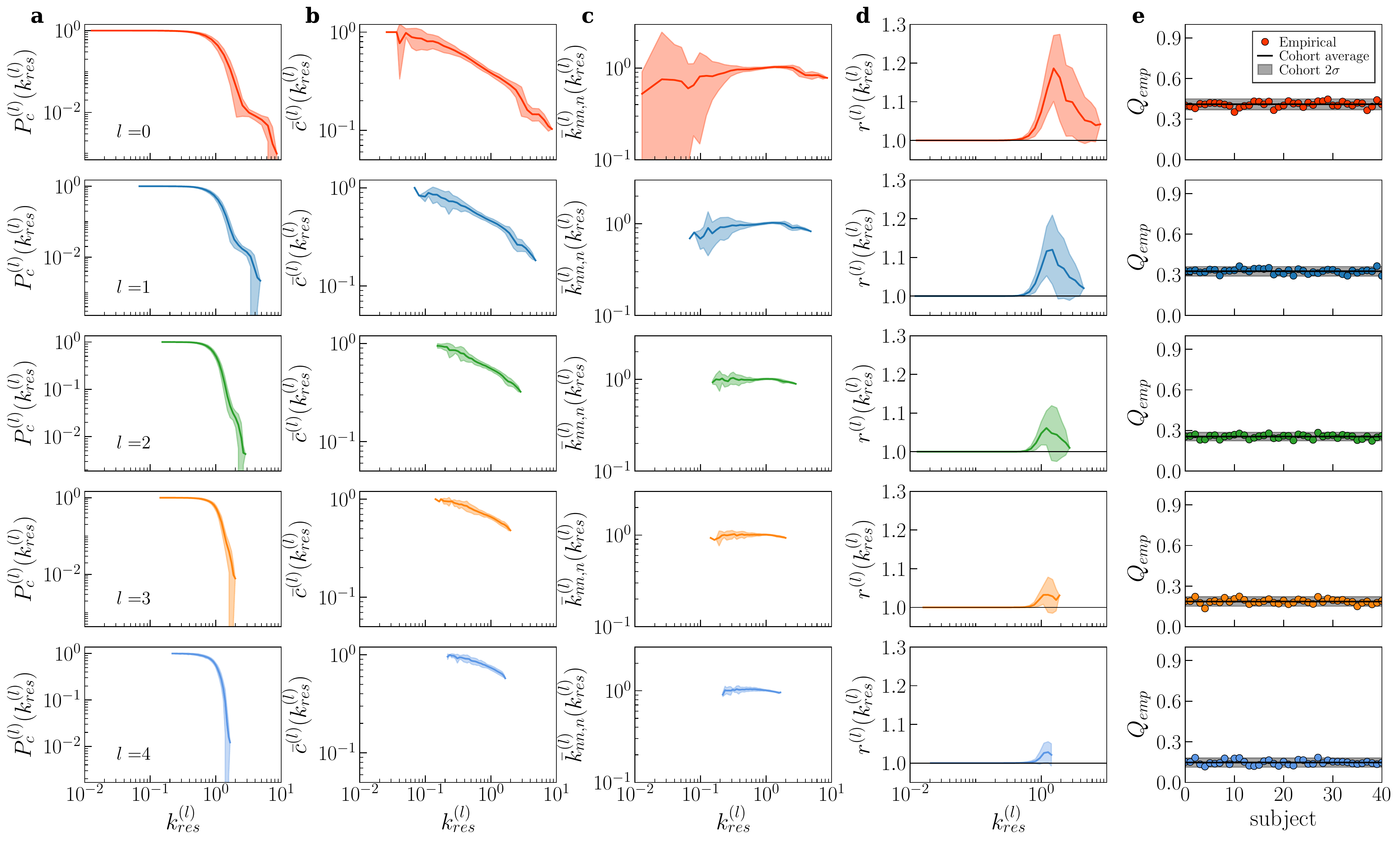}%
	\caption{\textbf{Network properties across 44 subjects for all layers in the HCP dataset.} Each column shows the complementary cumulative degree distribution, degree-dependent clustering coefficient, degree-degree correlations, rich club coefficient and modularity. The degrees have been rescaled by the internal average degree of the corresponding layer $k^{(l)}_{res} = k^{(l)}/ \langle k^{(l)}\rangle$. The solid lines show the corresponding average values across 44 subjects in the cohort and the shadows indicate 2$\sigma$ deviations. }
	\label{fig:Fig_group}
\end{figure*}
\begin{figure*}
	\centering
	\includegraphics[width=1\linewidth]{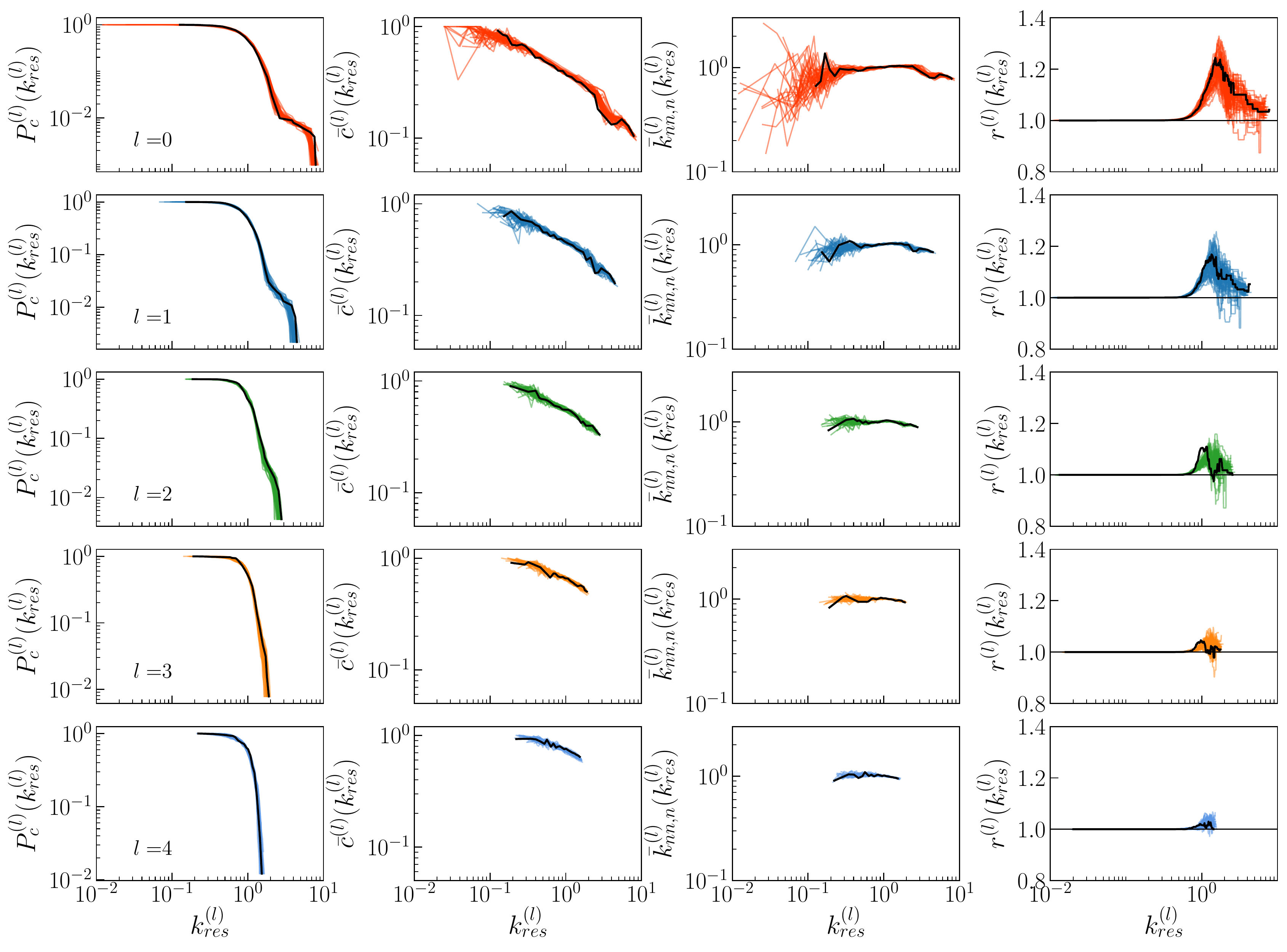}%
	\caption{\textbf{Subject No.~15 is a typical subject in HCP dataset.} Each column shows the complementary cumulative degree distribution, degree dependent clustering coefficient, degree-degree correlations and rich club coefficient. The degrees have been rescaled by the internal average degree of the corresponding layer $k^{(l)}_{res} = k^{(l)}/ \langle k^{(l)}\rangle$. Different lines correspond to different subject in each cohort. The results for subject No.~15 have been highlighted in black color.}	
\end{figure*}

\clearpage
\newpage
\begin{figure*}[ht]
	\centering
	\includegraphics[width=0.8\linewidth]{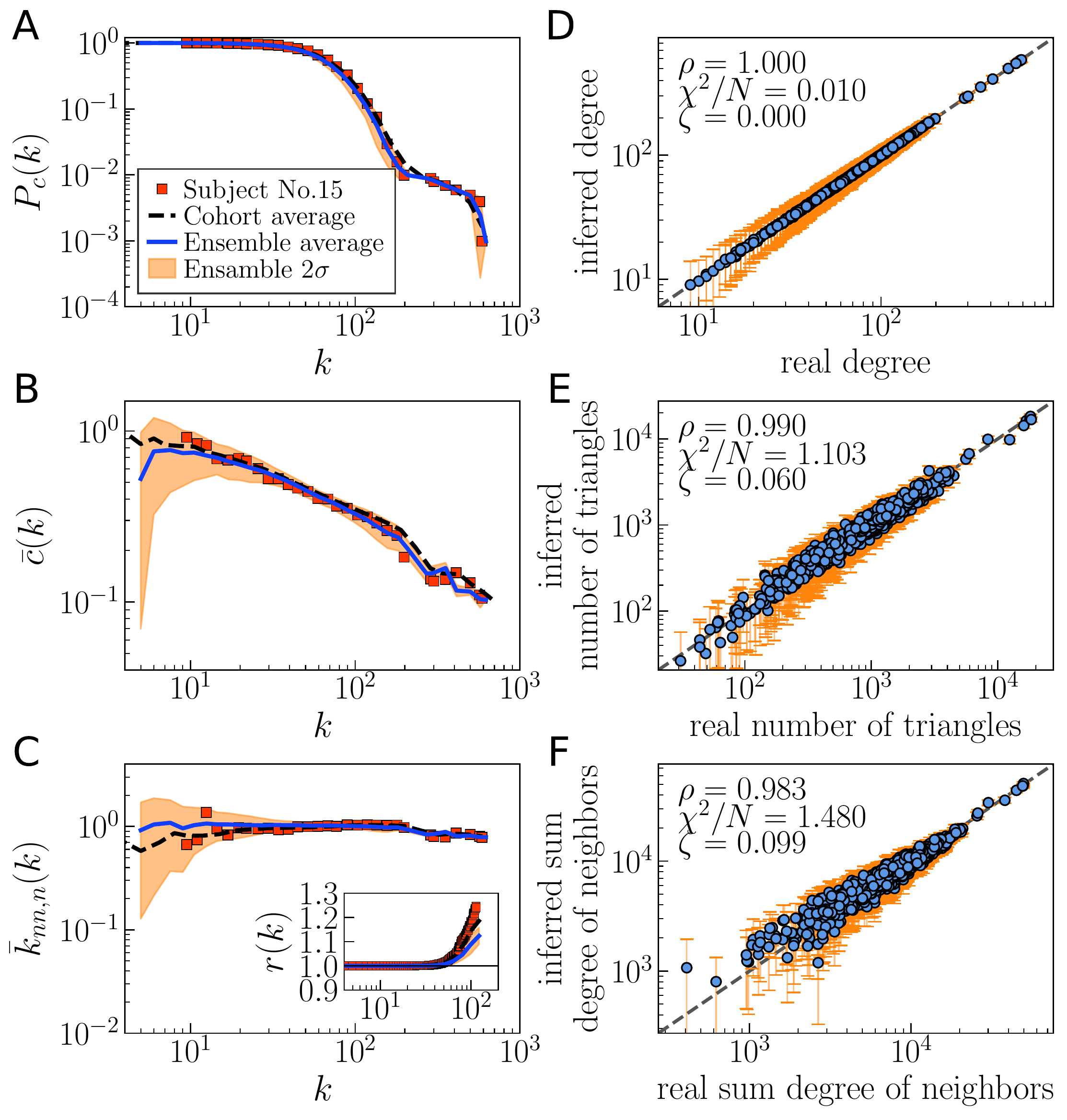}
	\caption{\textbf{Network properties of $l=0$ as compared to model predictions for HCP subject No.~15}. 
		(\textit{A}) Complementary cumulative degree distribution, 
		(\textit{B}) custering spectrum,
		(\textit{C}) average nearest neighbors degree, and 
		(\textit{C})-inset rich club coefficient. Red symbols correspond to subject No.~15, and black dashed lines to the group average across the 44 subjects in the sample. Blue lines correspond to the ensemble average over $100$ synthetic networks generated with the $\mathbb{S}^1$ model using the coordinates and parameters inferred by Mercator~\cite{Guille2019}, and the orange regions to the $2\sigma$ confidence interval around the expected value.
		(\textit{D})-(\textit{F}) Comparison of predictions in our model (average over the ensemble of $100$ synthetic networks) with real values for (\textit{D}) degrees,
		(\textit{E}) number of triangles, and (\textit{F}) sum of degrees of neighbors. Error bars show the $2\sigma$ confidence interval around the expected values. Statistical tests for the goodness of fit --- Pearson correlation coefficient $\rho$, $\chi^2$ test is normalized by the number of nodes, and $\zeta$ score--- are reported in each graph.}
\end{figure*}

\begin{figure*}[t]
	\centering
	\includegraphics[width=1\linewidth]{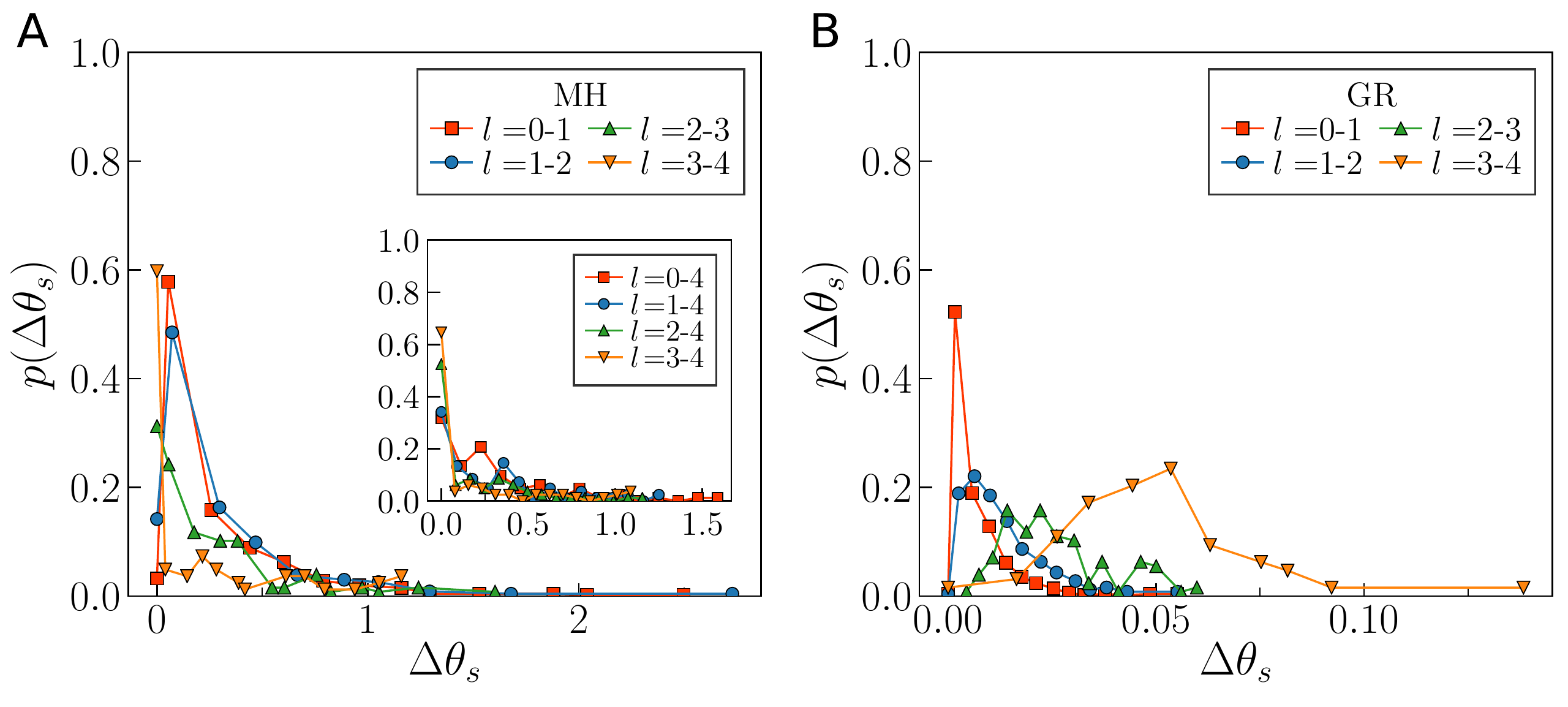}
	\caption{Results for HCP subject No.~15. (\textit{A}) and (\textit{B}). Distribution $p(\Delta \theta_s)$ of average angular separation between subnodes of coarse-grained nodes from one layer to the next in MH and GR, respectively. The inset in (\textit{A}) shows the distribution $p(\Delta \theta_s)$ from layer $l$ to $4$ in MH.}
\end{figure*}
\clearpage
\newpage

\begin{figure*}[h]
	\centering
	\includegraphics[width=0.9\linewidth]{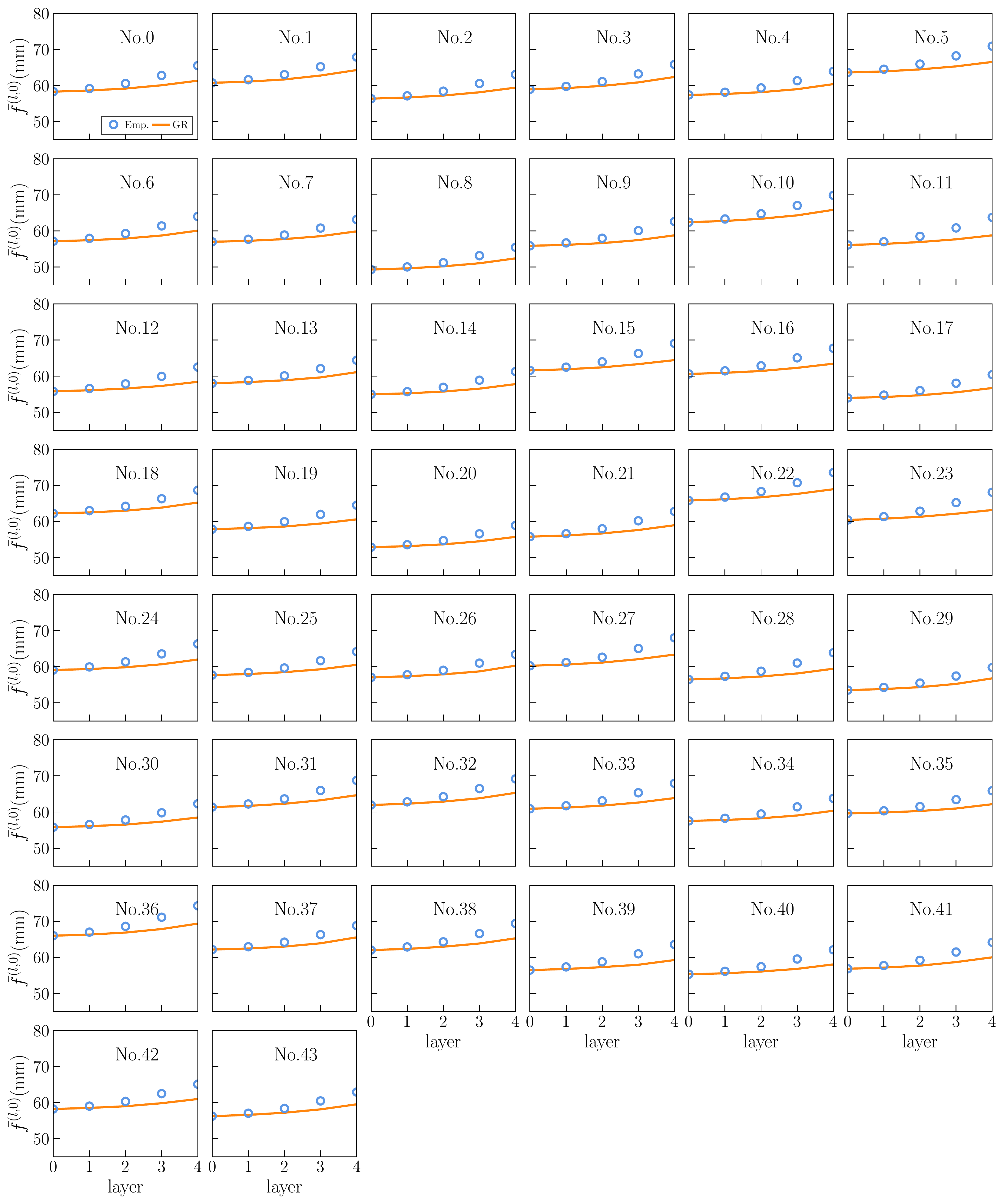}%
	\caption{
		Average fiber length $\bar{f}^{(l,0)}$ in layer $0$ of links outside supernodes in layer $l$, where supernodes are defined by the anatomical coarse-graining in the empirical curve (symbols) or the similarity coarse-graining in the GR case (lines). 
	}
\end{figure*}

\begin{figure}[t]
	\centering
	\includegraphics[width=0.8\linewidth]{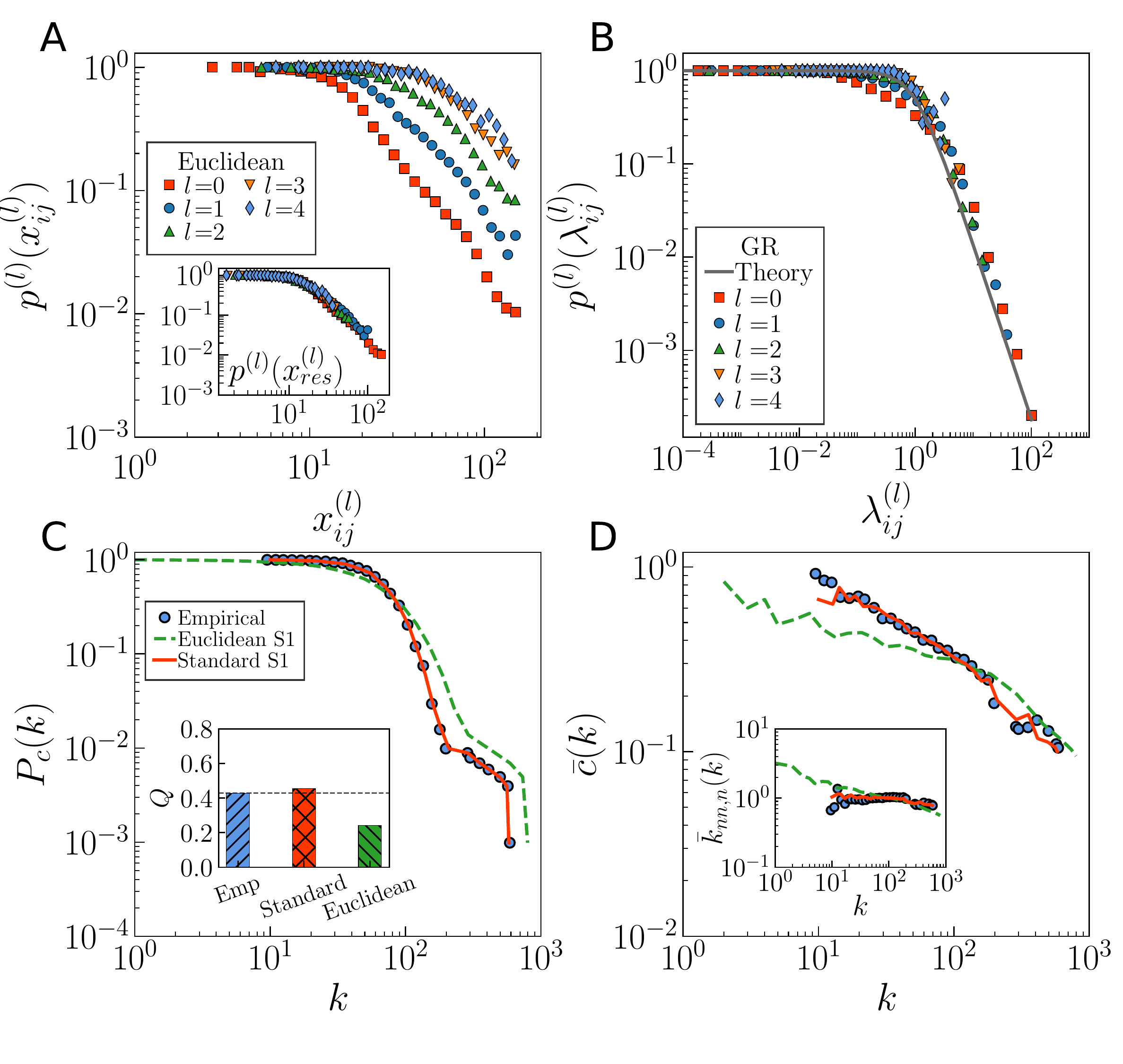}
	\caption{\textbf{Empirical vs theoretical probability of connection.} Results for HCP subject No. 15. 
		(\textit{A}) Empirical connection probabilities $p^{(l)}(x_{ij}^{(l)})$ in Euclidean space. Euclidean distances $x_{ij}$ are binned, and for each bin the ratio of the number of connected node pairs to the total number of pairs falling within the bin is shown. Inset shows the empirical connection probabilities $p^{(l)}(x_{res}^{(l)})$ as a function of rescale distances $x_{res}^{(l)}=x^{(l)}/a^{(l)}$ in the MH connectome, where the $a^{(l)}=[1.0, 1.5, 2.6, 3.8, 4.0]$ for different layer $l$. 
		(\textit{B}) Empirical versus theoretical connection probability  $p^{(l)}(\lambda_{ij}^{(l)})$  in the GR shell as a function of hyperbolic distance $\lambda_{ij}^{(l)}$. 
		(\textit{C}) Complementary cumulative degree distribution $P_c(k)$. Modularity $Q$, as measured by the Louvain method, is shown in the inset. 
		(\textit{D})Degree dependent clustering coefficient  $\bar{c}(k)$. Inset: degree-degree correlations $\bar{k}_{nn,n}(k)$. The filled symbols correspond to the empirical connectome of Subject No.~15. Green dashed lines are generated using the $\mathbb{S}^1$ model with Euclidean distances($\beta=2.31, \mu=0.0030$), and red lines correspond to the standard $\mathbb{S}^1$ model ($\beta=1.87, \mu=0.0039$).}
\end{figure}

\begin{figure*}[t]
	\centering
	\includegraphics[width=1\linewidth]{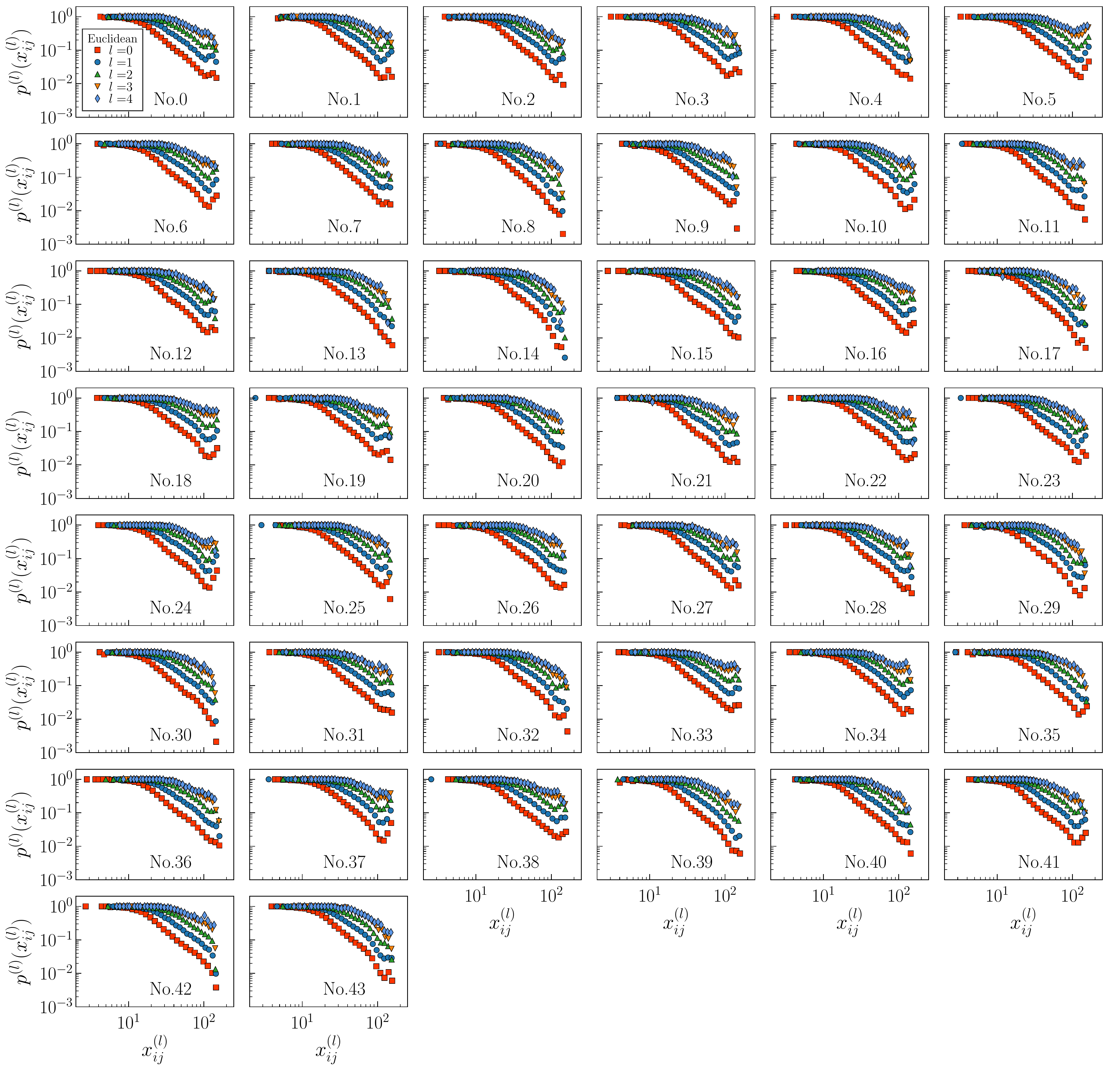}%
	\caption{\label{fig:FigA9} Empirical connection probabilities $p^{(l)}(x_{ij}^{(l)})$ for each subject in HCP dataset. The whole range of Euclidean distances $x_{ij}$ is binned, and for each bin the ratio of the number of connected connectome pairs to the total number of connectome pairs falling within this bin is shown.
	}
\end{figure*}

\begin{figure*}[t]
	\centering
	\includegraphics[width=1\linewidth]{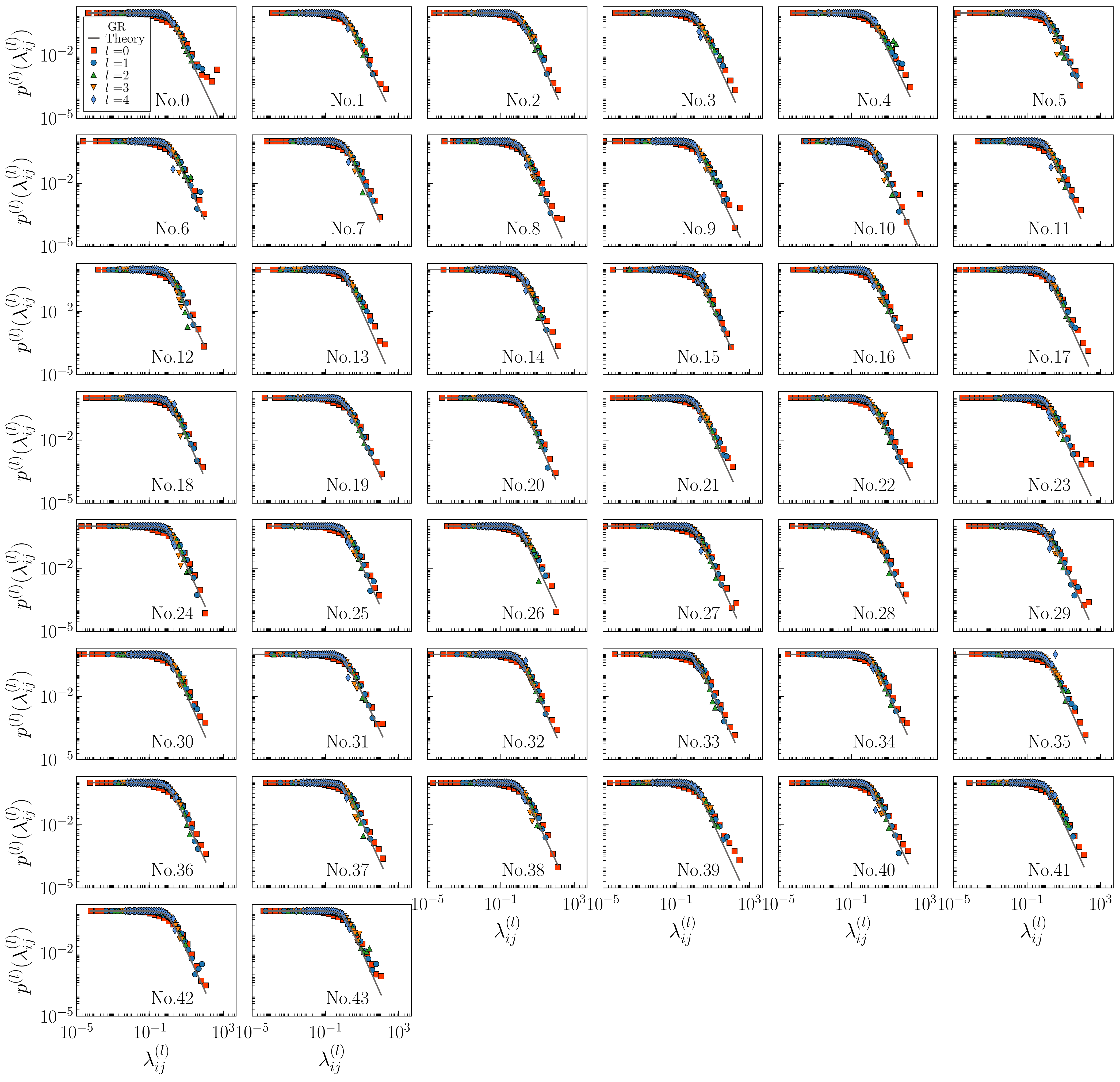}%
	\caption{
		Empirical versus theoretical connection probability  $p^{(l)}(\lambda_{ij}^{(l)})$ within a given range of $\lambda_{ij}^{(l)}$ on GR shell for each subject in HCP dataset. Symbols are the connection probability of GR networks within a given range of $\lambda_{ij}^{(l)}$ and the gray lines shows the theoretical curves. 
	}
\end{figure*}
\begin{figure}[!t]
	\centering
	\includegraphics[width=1\linewidth]{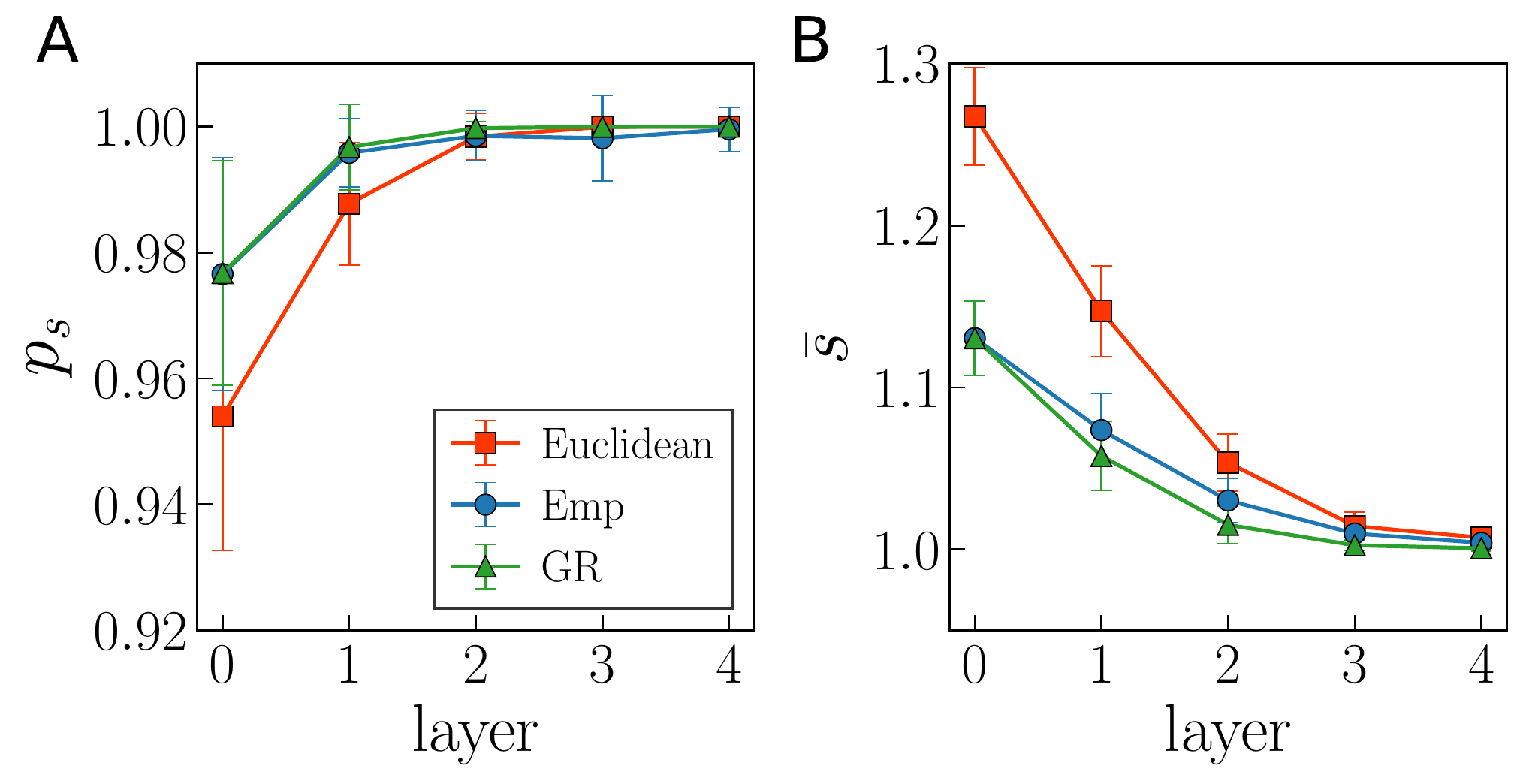}
	\caption{
		\textbf{Navigability of MH and GR maps at different resolutions.} (\textit{A}) average success rate (\textit{B}) and average stretch for all HCP subjects. The error bars show the $2\sigma$ confidence interval around the expected values. }
\end{figure}

\clearpage
\section{Comparison of similarity distances with Euclidean distances and homophily}
\begin{figure*}[!h]
	\centering
	\includegraphics[width=1\linewidth]{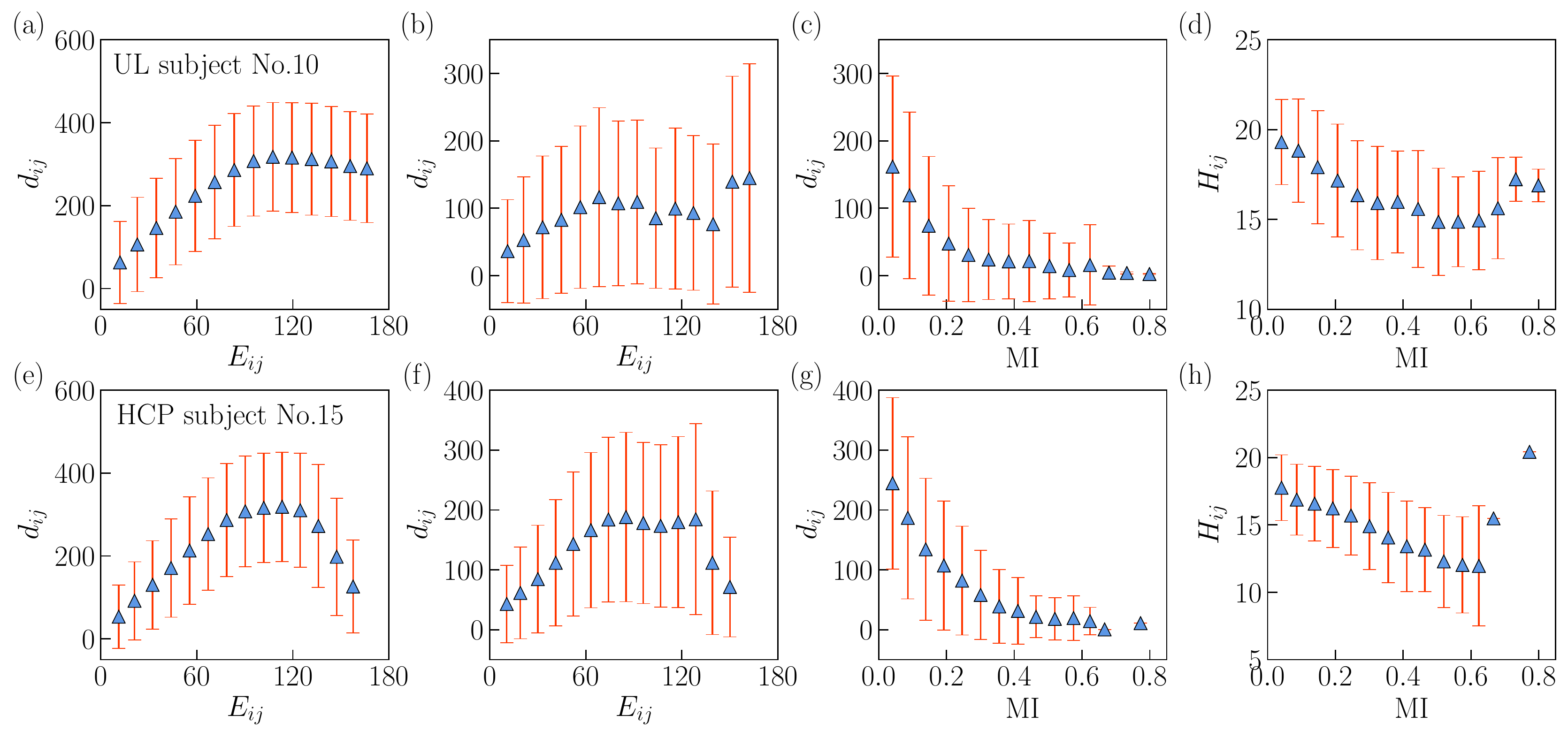}%
	\caption{\textbf{Similarity distance vs Euclidean distance and homophily.} (a)-(d) show results for subject No.~10 in the UL dataset and (e)-(h) for subject No.~15 in the HCP dataset. (a) and (e), Similarity distance ($d_{ij}=R\Delta \theta_{ij}$) vs Euclidean distance ($E_{ij}$) for all pairs of nodes in the connectome. (b) and (f), Similarity distance($d_{ij}$) vs Euclidean($E_{ij}$) for connected pairs of nodes. (c) and (g), similarity distance ($d_{ij}$) vs matching index (MI). (d) and (h) hyperbolic distance ($H_{ij}$) vs matching index (MI). The matching index is a normalized measure of overlap in two nodes' neighborhoods (ratio between the number of common neighbors and the total number of neighbors of the two discounting the pair), as defined in Ref.~\cite{betzel2016generative}. In each plot, we show a binned statistic with 15 equal-width bins. Each symbol shows the mean value in the bin and error bar indicates $1$ standard deviation around the average value. }	
\end{figure*}

\clearpage
\section{Statistics for subjects in the UL and the HCP datasets}
\renewcommand{\arraystretch}{0.55} 
\begin{table*}[!h] \small
	\begin{ruledtabular}
		\centering	
		\caption{Overview of the 40 connectomes in the UL dataeset. The number of nodes ($N$), the number of links ($L$), the density of links 
			($\rho_l = 2L/N(N-1)$), its average degree ($\langle k \rangle = 2L/N$), the average local clustering coefficient ($\langle c \rangle$), the average fiber length ($f^{(l)}$ (mm)), and corresponding $\pm 1$ standard error interval around the mean (SEM), the assortativity coefficient ($r_c$), the modularity ($Q$), the number of the communities ($Nc$), and the hyperbolic embedding parameter $\beta$ and $\mu$.}
		\begin{tabular}{*{13}{c}} 
			\thd{Subject}    & \thd{Layer} & \thd{$N$} & \thd{$L$} & \thd{$\rho_l$} & \thd{$\langle k\rangle\pm \text{SEM}$}  &  \thd{$\langle c\rangle\pm \text{SEM}$}  & \thd{$f^{(l)}\pm \text{SEM}$} &\thd{$r_c$} &\thd{$Q$} &\thd{$N_c$} & \thd{$\beta$} & \thd{$\mu$}\\
			\hline
			\multirow{5}{*}{No. 0}  & 0      & 1009    & 14470       & 0.03        & 28.68$\pm$0.669   & 0.43$\pm$0.005    & 65.82$\pm$43.675  & -0.016     & 0.55       & 5          & 1.96       & 0.011     \\ 
			& 1      & 461     & 7177        & 0.07        & 31.14$\pm$0.825   & 0.47$\pm$0.006    & 68.82$\pm$44.082  & 0.005      & 0.48       & 5          & 2.22        & 0.011     \\ 
			& 2      & 233     & 3586        & 0.13        & 30.78$\pm$0.937   & 0.50$\pm$0.008    & 72.74$\pm$45.021  & 0.031      & 0.42       & 5          & 2.41        & 0.012     \\ 
			& 3      & 128     & 1744        & 0.21        & 27.25$\pm$1.031   & 0.56$\pm$0.012    & 74.80$\pm$46.285  & 0.040      & 0.32       & 3          & 2.63        & 0.014     \\ 
			& 4      & 82      & 962         & 0.29        & 23.46$\pm$1.102   & 0.62$\pm$0.016    & 74.39$\pm$47.996  & 0.021      & 0.31       & 3          & 3.08        & 0.018     \\ 
			\hline 
			\multirow{5}{*}{No. 1}  & 0      & 1010    & 14406       & 0.03        & 28.53$\pm$0.729   & 0.42$\pm$0.005    & 66.38$\pm$44.958  & -0.003     & 0.52       & 5          & 1.93       & 0.011     \\ 
			& 1      & 461     & 7195        & 0.07        & 31.21$\pm$0.944   & 0.46$\pm$0.006    & 70.46$\pm$45.278  & 0.003      & 0.45       & 6          & 2.11        & 0.011     \\ 
			& 2      & 233     & 3893        & 0.14        & 33.42$\pm$1.158   & 0.50$\pm$0.008    & 76.44$\pm$46.113  & -0.012     & 0.35       & 3          & 2.26        & 0.011     \\ 
			& 3      & 128     & 1992        & 0.25        & 31.12$\pm$1.257   & 0.57$\pm$0.010    & 80.33$\pm$47.629  & -0.039     & 0.28       & 3          & 2.48        & 0.012     \\ 
			& 4      & 82      & 1129        & 0.34        & 27.54$\pm$1.328   & 0.63$\pm$0.013    & 83.11$\pm$50.287  & -0.080     & 0.23       & 4          & 2.78        & 0.015     \\ 
			\hline 
			\multirow{5}{*}{No. 2}  & 0      & 1014    & 13671       & 0.03        & 26.96$\pm$0.638   & 0.44$\pm$0.005    & 62.45$\pm$42.759  & 0.008      & 0.58       & 6          & 2.00       & 0.012     \\ 
			& 1      & 462     & 6833        & 0.06        & 29.58$\pm$0.812   & 0.46$\pm$0.006    & 67.32$\pm$43.889  & 0.027      & 0.50       & 5          & 2.19        & 0.012     \\ 
			& 2      & 233     & 3599        & 0.13        & 30.89$\pm$0.954   & 0.49$\pm$0.008    & 73.01$\pm$45.070  & 0.056      & 0.42       & 4          & 2.41        & 0.012     \\ 
			& 3      & 128     & 1806        & 0.22        & 28.22$\pm$1.031   & 0.55$\pm$0.011    & 74.95$\pm$45.415  & 0.043      & 0.34       & 4          & 2.63        & 0.014     \\ 
			& 4      & 82      & 978         & 0.29        & 23.85$\pm$1.131   & 0.63$\pm$0.015    & 75.12$\pm$47.046  & -0.035     & 0.30       & 3          & 3.23        & 0.018     \\ 
			\hline 
			\multirow{5}{*}{No. 3}  & 0      & 1011    & 12991       & 0.03        & 25.70$\pm$0.648   & 0.41$\pm$0.005    & 61.51$\pm$43.188  & 0.003      & 0.54       & 7          & 1.89       & 0.012     \\ 
			& 1      & 462     & 6642        & 0.06        & 28.75$\pm$0.842   & 0.45$\pm$0.006    & 64.32$\pm$43.813  & 0.029      & 0.46       & 4          & 2.08        & 0.011     \\ 
			& 2      & 233     & 3581        & 0.13        & 30.74$\pm$1.016   & 0.49$\pm$0.008    & 69.07$\pm$45.923  & 0.046      & 0.40       & 5          & 2.26        & 0.012     \\ 
			& 3      & 128     & 1888        & 0.23        & 29.50$\pm$1.135   & 0.55$\pm$0.011    & 74.38$\pm$49.242  & 0.049      & 0.31       & 3          & 2.48        & 0.013     \\ 
			& 4      & 82      & 1074        & 0.32        & 26.20$\pm$1.302   & 0.63$\pm$0.015    & 79.49$\pm$53.949  & 0.005      & 0.23       & 2          & 2.78        & 0.015     \\ 
			\hline 
			\multirow{5}{*}{No. 4}  & 0      & 1014    & 15879       & 0.03        & 31.32$\pm$0.778   & 0.41$\pm$0.005    & 72.32$\pm$50.544  & 0.014      & 0.51       & 6          & 1.85       & 0.009     \\ 
			& 1      & 462     & 7896        & 0.07        & 34.18$\pm$0.999   & 0.46$\pm$0.006    & 76.98$\pm$52.410  & 0.036      & 0.44       & 4          & 2.08        & 0.010     \\ 
			& 2      & 233     & 4069        & 0.15        & 34.93$\pm$1.165   & 0.50$\pm$0.008    & 81.56$\pm$53.710  & 0.057      & 0.33       & 4          & 2.26        & 0.010     \\ 
			& 3      & 128     & 2077        & 0.26        & 32.45$\pm$1.254   & 0.57$\pm$0.010    & 83.79$\pm$54.812  & 0.042      & 0.29       & 3          & 2.63        & 0.012     \\ 
			& 4      & 82      & 1177        & 0.35        & 28.71$\pm$1.408   & 0.66$\pm$0.013    & 86.85$\pm$58.033  & -0.027     & 0.24       & 3          & 2.93        & 0.014     \\ 
			\hline 
			\multirow{5}{*}{No. 5}  & 0      & 1014    & 14340       & 0.03        & 28.28$\pm$0.698   & 0.41$\pm$0.005    & 65.30$\pm$42.579  & -0.013     & 0.54       & 5          & 1.89       & 0.011     \\ 
			& 1      & 462     & 7175        & 0.07        & 31.06$\pm$0.900   & 0.45$\pm$0.006    & 69.23$\pm$43.346  & -0.003     & 0.44       & 5          & 2.11        & 0.011     \\ 
			& 2      & 233     & 3719        & 0.14        & 31.92$\pm$1.043   & 0.48$\pm$0.007    & 73.64$\pm$43.925  & 0.008      & 0.38       & 4          & 2.22        & 0.011     \\ 
			& 3      & 128     & 1919        & 0.24        & 29.98$\pm$1.141   & 0.55$\pm$0.009    & 76.85$\pm$45.317  & 0.004      & 0.30       & 4          & 2.48        & 0.013     \\ 
			& 4      & 82      & 1137        & 0.34        & 27.73$\pm$1.260   & 0.63$\pm$0.012    & 79.36$\pm$47.597  & -0.021     & 0.25       & 3          & 2.78        & 0.014     \\ 
			\hline 
			\multirow{5}{*}{No. 6}  & 0      & 1013    & 12660       & 0.02        & 25.00$\pm$0.653   & 0.44$\pm$0.005    & 57.01$\pm$38.389  & -0.012     & 0.59       & 6          & 1.96       & 0.013     \\ 
			& 1      & 462     & 6519        & 0.06        & 28.22$\pm$0.865   & 0.48$\pm$0.007    & 61.16$\pm$39.025  & -0.006     & 0.50       & 5          & 2.26        & 0.013     \\ 
			& 2      & 233     & 3375        & 0.12        & 28.97$\pm$1.009   & 0.53$\pm$0.009    & 64.87$\pm$39.810  & -0.014     & 0.43       & 4          & 2.52        & 0.013     \\ 
			& 3      & 128     & 1712        & 0.21        & 26.75$\pm$1.080   & 0.57$\pm$0.012    & 68.31$\pm$41.896  & -0.014     & 0.33       & 3          & 2.78        & 0.015     \\ 
			& 4      & 82      & 1000        & 0.30        & 24.39$\pm$1.196   & 0.62$\pm$0.015    & 71.27$\pm$45.710  & -0.017     & 0.26       & 4          & 3.00        & 0.017     \\ 
			\hline 
			\multirow{5}{*}{No. 7}  & 0      & 1014    & 13474       & 0.03        & 26.58$\pm$0.702   & 0.42$\pm$0.005    & 64.45$\pm$43.876  & -0.023     & 0.56       & 6          & 1.89       & 0.011     \\ 
			& 1      & 462     & 6955        & 0.07        & 30.11$\pm$0.902   & 0.46$\pm$0.006    & 68.69$\pm$45.530  & -0.012     & 0.47       & 4          & 2.11        & 0.011     \\ 
			& 2      & 233     & 3651        & 0.14        & 31.34$\pm$1.019   & 0.50$\pm$0.007    & 73.55$\pm$47.353  & -0.006     & 0.40       & 3          & 2.41        & 0.012     \\ 
			& 3      & 128     & 1934        & 0.24        & 30.22$\pm$1.079   & 0.56$\pm$0.010    & 78.72$\pm$50.132  & 0.000      & 0.32       & 4          & 2.63        & 0.013     \\ 
			& 4      & 82      & 1076        & 0.32        & 26.24$\pm$1.201   & 0.62$\pm$0.014    & 81.38$\pm$53.156  & -0.041     & 0.26       & 3          & 2.86        & 0.015     \\ 
		\end{tabular}
	\end{ruledtabular}
\end{table*}

\renewcommand{\arraystretch}{0.54} 
\begin{table*}[b] \footnotesize
	\begin{ruledtabular}
		\centering	
		\label{tab:A2}		
		\begin{tabular}{*{13}{c}} 
			\thd{Subject}    & \thd{Layer} & \thd{$N$} & \thd{$L$} & \thd{$\rho_l$} & \thd{$\langle k\rangle\pm \text{SEM}$}  &  \thd{$\langle c\rangle\pm \text{SEM}$}  & \thd{$f^{(l)}\pm \text{SEM}$} &\thd{$r_c$} &\thd{$Q$} &\thd{$N_c$} & \thd{$\beta$} & \thd{$\mu$}\\
			\hline 
			\multirow{5}{*}{No. 8}  & 0      & 1002    & 13910       & 0.03        & 27.76$\pm$0.652   & 0.41$\pm$0.005    & 67.65$\pm$45.004  & 0.017      & 0.53       & 5          & 1.89       & 0.011     \\ 
			& 1      & 462     & 7041        & 0.07        & 30.48$\pm$0.845   & 0.44$\pm$0.007    & 72.07$\pm$46.547  & 0.051      & 0.43       & 4          & 2.04        & 0.011     \\ 
			& 2      & 233     & 3723        & 0.14        & 31.96$\pm$1.053   & 0.49$\pm$0.008    & 76.47$\pm$47.376  & 0.067      & 0.36       & 4          & 2.19        & 0.011     \\ 
			& 3      & 128     & 1960        & 0.24        & 30.62$\pm$1.177   & 0.54$\pm$0.010    & 82.08$\pm$50.685  & 0.056      & 0.28       & 3          & 2.41        & 0.012     \\ 
			& 4      & 82      & 1124        & 0.34        & 27.41$\pm$1.317   & 0.62$\pm$0.013    & 85.38$\pm$54.262  & 0.005      & 0.20       & 4          & 2.48        & 0.014     \\ 
			\hline 
			\multirow{5}{*}{No. 9}  & 0      & 1010    & 13496       & 0.03        & 26.72$\pm$0.632   & 0.41$\pm$0.005    & 63.02$\pm$41.763  & -0.007     & 0.54       & 5          & 1.89       & 0.011     \\ 
			& 1      & 462     & 6658        & 0.06        & 28.82$\pm$0.802   & 0.44$\pm$0.006    & 66.13$\pm$42.236  & 0.000      & 0.47       & 4          & 2.08        & 0.011     \\ 
			& 2      & 233     & 3532        & 0.13        & 30.32$\pm$0.938   & 0.48$\pm$0.008    & 71.00$\pm$43.525  & 0.010      & 0.37       & 3          & 2.26        & 0.012     \\ 
			& 3      & 128     & 1774        & 0.22        & 27.72$\pm$1.022   & 0.54$\pm$0.011    & 74.12$\pm$45.548  & 0.004      & 0.31       & 3          & 2.56        & 0.014     \\ 
			& 4      & 82      & 977         & 0.29        & 23.83$\pm$1.147   & 0.61$\pm$0.016    & 76.36$\pm$47.999  & -0.052     & 0.27       & 4          & 2.78        & 0.017     \\ 
			\hline 
			\multirow{5}{*}{No. 10}  & 0      & 1014    & 15222       & 0.03        & 30.02$\pm$0.675   & 0.41$\pm$0.005    & 71.29$\pm$48.238  & 0.004      & 0.57       & 6          & 1.96       & 0.010     \\ 
			& 1      & 462     & 7414        & 0.07        & 32.10$\pm$0.855   & 0.46$\pm$0.006    & 75.74$\pm$49.299  & 0.019      & 0.48       & 4          & 2.19        & 0.011     \\ 
			& 2      & 233     & 3754        & 0.14        & 32.22$\pm$0.943   & 0.49$\pm$0.008    & 79.97$\pm$50.254  & 0.043      & 0.39       & 4          & 2.34        & 0.011     \\ 
			& 3      & 128     & 1954        & 0.24        & 30.53$\pm$1.037   & 0.54$\pm$0.010    & 85.00$\pm$52.882  & 0.060      & 0.32       & 3          & 2.60        & 0.013     \\ 
			& 4      & 82      & 1102        & 0.33        & 26.88$\pm$1.213   & 0.62$\pm$0.013    & 87.45$\pm$57.162  & 0.015      & 0.26       & 3          & 2.78        & 0.015     \\ 
			\hline 
			\multirow{5}{*}{No. 11}  & 0      & 1013    & 14695       & 0.03        & 29.01$\pm$0.702   & 0.42$\pm$0.005    & 63.42$\pm$42.951  & -0.030     & 0.54       & 6          & 1.93       & 0.011     \\ 
			& 1      & 462     & 7577        & 0.07        & 32.80$\pm$0.891   & 0.46$\pm$0.007    & 68.64$\pm$44.197  & -0.021     & 0.48       & 4          & 2.19        & 0.011     \\ 
			& 2      & 233     & 3872        & 0.14        & 33.24$\pm$1.000   & 0.51$\pm$0.009    & 72.30$\pm$44.861  & -0.010     & 0.43       & 3          & 2.48        & 0.011     \\ 
			& 3      & 128     & 2023        & 0.25        & 31.61$\pm$1.053   & 0.58$\pm$0.011    & 75.29$\pm$46.142  & 0.013      & 0.36       & 3          & 2.93        & 0.013     \\ 
			& 4      & 82      & 1088        & 0.33        & 26.54$\pm$1.146   & 0.64$\pm$0.015    & 74.97$\pm$47.933  & 0.002      & 0.32       & 3          & 3.45        & 0.016     \\ 
			\hline 
			\multirow{5}{*}{No. 12}  & 0      & 1001    & 13933       & 0.03        & 27.84$\pm$0.663   & 0.43$\pm$0.005    & 74.40$\pm$57.807  & 0.044      & 0.57       & 7          & 1.96       & 0.011     \\ 
			& 1      & 460     & 7112        & 0.07        & 30.92$\pm$0.870   & 0.46$\pm$0.006    & 80.62$\pm$60.153  & 0.080      & 0.46       & 5          & 2.19        & 0.011     \\ 
			& 2      & 233     & 3655        & 0.14        & 31.37$\pm$1.031   & 0.50$\pm$0.008    & 86.96$\pm$62.239  & 0.081      & 0.38       & 4          & 2.34        & 0.012     \\ 
			& 3      & 128     & 1875        & 0.23        & 29.30$\pm$1.174   & 0.58$\pm$0.011    & 92.78$\pm$66.547  & 0.071      & 0.30       & 3          & 2.78        & 0.014     \\ 
			& 4      & 82      & 1049        & 0.32        & 25.59$\pm$1.368   & 0.65$\pm$0.014    & 96.71$\pm$71.598  & -0.014     & 0.26       & 3          & 3.08        & 0.016     \\ 
			\hline 
			\multirow{5}{*}{No. 13}  & 0      & 1013    & 14409       & 0.03        & 28.45$\pm$0.771   & 0.44$\pm$0.005    & 64.03$\pm$43.296  & -0.011     & 0.54       & 6          & 1.93       & 0.011     \\ 
			& 1      & 462     & 7409        & 0.07        & 32.07$\pm$0.994   & 0.47$\pm$0.006    & 68.46$\pm$43.863  & 0.002      & 0.46       & 4          & 2.11        & 0.010     \\ 
			& 2      & 233     & 3845        & 0.14        & 33.00$\pm$1.143   & 0.51$\pm$0.008    & 72.39$\pm$43.825  & 0.009      & 0.38       & 5          & 2.34        & 0.011     \\ 
			& 3      & 128     & 2009        & 0.25        & 31.39$\pm$1.275   & 0.57$\pm$0.010    & 75.90$\pm$45.009  & -0.027     & 0.30       & 4          & 2.63        & 0.012     \\ 
			& 4      & 82      & 1163        & 0.35        & 28.37$\pm$1.400   & 0.65$\pm$0.014    & 77.41$\pm$46.731  & -0.087     & 0.24       & 4          & 2.78        & 0.014     \\ 
			\hline 
			\multirow{5}{*}{No. 14}  & 0      & 1013    & 14959       & 0.03        & 29.53$\pm$0.713   & 0.43$\pm$0.005    & 67.63$\pm$45.728  & 0.001      & 0.53       & 5          & 1.96       & 0.011     \\ 
			& 1      & 462     & 7382        & 0.07        & 31.96$\pm$0.890   & 0.46$\pm$0.007    & 72.39$\pm$47.018  & 0.011      & 0.45       & 4          & 2.19        & 0.011     \\ 
			& 2      & 233     & 3808        & 0.14        & 32.69$\pm$1.025   & 0.49$\pm$0.008    & 76.85$\pm$47.985  & 0.013      & 0.38       & 3          & 2.34        & 0.011     \\ 
			& 3      & 128     & 1952        & 0.24        & 30.50$\pm$1.104   & 0.55$\pm$0.010    & 81.09$\pm$50.685  & 0.027      & 0.33       & 3          & 2.63        & 0.013     \\ 
			& 4      & 82      & 1071        & 0.32        & 26.12$\pm$1.264   & 0.63$\pm$0.014    & 84.18$\pm$54.978  & 0.014      & 0.26       & 3          & 3.08        & 0.016     \\ 
			\hline 
			\multirow{5}{*}{No. 15}  & 0      & 1007    & 16208       & 0.03        & 32.19$\pm$0.749   & 0.43$\pm$0.005    & 76.68$\pm$53.253  & 0.001      & 0.54       & 5          & 1.96       & 0.010     \\ 
			& 1      & 462     & 8003        & 0.08        & 34.65$\pm$0.966   & 0.47$\pm$0.006    & 83.08$\pm$55.373  & 0.022      & 0.46       & 3          & 2.19        & 0.010     \\ 
			& 2      & 233     & 4102        & 0.15        & 35.21$\pm$1.119   & 0.51$\pm$0.008    & 88.97$\pm$57.472  & 0.043      & 0.40       & 4          & 2.41        & 0.011     \\ 
			& 3      & 128     & 2092        & 0.26        & 32.69$\pm$1.202   & 0.58$\pm$0.011    & 94.04$\pm$60.183  & 0.037      & 0.34       & 3          & 2.78        & 0.012     \\ 
			& 4      & 82      & 1150        & 0.35        & 28.05$\pm$1.384   & 0.67$\pm$0.014    & 97.63$\pm$65.524  & -0.025     & 0.27       & 3          & 3.26        & 0.015     \\ 
			\hline 
			\multirow{5}{*}{No. 16}  & 0      & 1014    & 14539       & 0.03        & 28.68$\pm$0.738   & 0.42$\pm$0.005    & 68.14$\pm$46.715  & -0.004     & 0.55       & 5          & 1.89       & 0.010     \\ 
			& 1      & 462     & 7352        & 0.07        & 31.83$\pm$0.958   & 0.46$\pm$0.006    & 71.78$\pm$47.023  & 0.004      & 0.46       & 4          & 2.11        & 0.011     \\ 
			& 2      & 233     & 3801        & 0.14        & 32.63$\pm$1.124   & 0.50$\pm$0.008    & 76.22$\pm$48.008  & 0.014      & 0.38       & 3          & 2.34        & 0.011     \\ 
			& 3      & 128     & 1999        & 0.25        & 31.23$\pm$1.259   & 0.57$\pm$0.010    & 81.28$\pm$50.587  & 0.029      & 0.29       & 3          & 2.48        & 0.012     \\ 
			& 4      & 82      & 1198        & 0.36        & 29.22$\pm$1.448   & 0.65$\pm$0.012    & 86.73$\pm$55.434  & 0.015      & 0.22       & 3          & 2.78        & 0.014     \\ 
			\hline 
			\multirow{5}{*}{No. 17}  & 0      & 1014    & 13901       & 0.03        & 27.42$\pm$0.683   & 0.43$\pm$0.005    & 62.44$\pm$43.569  & -0.011     & 0.55       & 6          & 1.96       & 0.011     \\ 
			& 1      & 462     & 7030        & 0.07        & 30.43$\pm$0.861   & 0.46$\pm$0.006    & 66.20$\pm$43.898  & -0.005     & 0.47       & 4          & 2.19        & 0.011     \\ 
			& 2      & 233     & 3642        & 0.13        & 31.26$\pm$0.987   & 0.50$\pm$0.009    & 70.61$\pm$44.926  & 0.001      & 0.39       & 3          & 2.34        & 0.012     \\ 
			& 3      & 128     & 1821        & 0.22        & 28.45$\pm$1.061   & 0.55$\pm$0.011    & 74.07$\pm$46.623  & 0.017      & 0.33       & 3          & 2.63        & 0.014     \\ 
			& 4      & 82      & 994         & 0.30        & 24.24$\pm$1.166   & 0.61$\pm$0.015    & 73.95$\pm$47.219  & -0.015     & 0.26       & 4          & 2.78        & 0.017     \\ 
			\hline 
			\multirow{5}{*}{No. 18}  & 0      & 1014    & 12418       & 0.02        & 24.49$\pm$0.653   & 0.43$\pm$0.005    & 57.36$\pm$40.798  & -0.029     & 0.56       & 5          & 1.96       & 0.013     \\ 
			& 1      & 462     & 6510        & 0.06        & 28.18$\pm$0.837   & 0.47$\pm$0.007    & 62.05$\pm$41.924  & -0.024     & 0.51       & 4          & 2.22        & 0.012     \\ 
			& 2      & 233     & 3497        & 0.13        & 30.02$\pm$0.969   & 0.51$\pm$0.008    & 67.22$\pm$42.933  & -0.014     & 0.42       & 3          & 2.48        & 0.013     \\ 
			& 3      & 128     & 1847        & 0.23        & 28.86$\pm$1.052   & 0.56$\pm$0.011    & 71.59$\pm$44.098  & -0.025     & 0.34       & 4          & 2.78        & 0.014     \\ 
			& 4      & 82      & 1008        & 0.30        & 24.59$\pm$1.140   & 0.62$\pm$0.015    & 71.06$\pm$45.854  & -0.044     & 0.26       & 2          & 3.08        & 0.017     \\ 			 
			
		\end{tabular}
	\end{ruledtabular}
\end{table*}

\renewcommand{\arraystretch}{0.54} 
\begin{table*}[t] \footnotesize
	\begin{ruledtabular}
		\centering	
		\label{tab:A3}		
		\begin{tabular}{*{13}{c}} 
			\thd{Subject}    & \thd{Layer} & \thd{$N$} & \thd{$L$} & \thd{$\rho_l$} & \thd{$\langle k\rangle\pm \text{SEM}$}  &  \thd{$\langle c\rangle\pm \text{SEM}$}  & \thd{$f^{(l)}\pm \text{SEM}$} &\thd{$r_c$} &\thd{$Q$} &\thd{$N_c$} & \thd{$\beta$} & \thd{$\mu$}\\
			\hline
			\multirow{5}{*}{No. 19}  & 0      & 1011    & 14883       & 0.03        & 29.44$\pm$0.708   & 0.40$\pm$0.005    & 69.15$\pm$46.830  & 0.017      & 0.53       & 6          & 1.85       & 0.010     \\ 
			& 1      & 462     & 7633        & 0.07        & 33.04$\pm$0.932   & 0.44$\pm$0.007    & 74.46$\pm$49.017  & 0.033      & 0.45       & 4          & 2.04        & 0.010     \\ 
			& 2      & 233     & 4042        & 0.15        & 34.70$\pm$1.130   & 0.49$\pm$0.008    & 80.35$\pm$51.407  & 0.034      & 0.36       & 4          & 2.22        & 0.010     \\ 
			& 3      & 128     & 2032        & 0.25        & 31.75$\pm$1.246   & 0.56$\pm$0.010    & 84.05$\pm$54.276  & 0.033      & 0.27       & 3          & 2.48        & 0.012     \\ 
			& 4      & 82      & 1117        & 0.34        & 27.24$\pm$1.412   & 0.64$\pm$0.013    & 86.97$\pm$58.517  & -0.000     & 0.23       & 3          & 2.48        & 0.014     \\ 
			\hline 
			\multirow{5}{*}{No. 20}  & 0      & 1011    & 13201       & 0.03        & 26.11$\pm$0.654   & 0.42$\pm$0.005    & 62.04$\pm$42.881  & -0.013     & 0.53       & 5          & 1.89       & 0.011     \\ 
			& 1      & 462     & 6749        & 0.06        & 29.22$\pm$0.852   & 0.47$\pm$0.007    & 66.28$\pm$43.375  & -0.008     & 0.49       & 4          & 2.19        & 0.012     \\ 
			& 2      & 233     & 3589        & 0.13        & 30.81$\pm$0.967   & 0.49$\pm$0.008    & 70.82$\pm$44.369  & 0.020      & 0.40       & 4          & 2.34        & 0.012     \\ 
			& 3      & 128     & 1816        & 0.22        & 28.38$\pm$1.037   & 0.55$\pm$0.012    & 72.81$\pm$44.735  & 0.021      & 0.33       & 4          & 2.63        & 0.014     \\ 
			& 4      & 82      & 975         & 0.29        & 23.78$\pm$1.170   & 0.62$\pm$0.016    & 74.04$\pm$46.391  & 0.002      & 0.26       & 2          & 2.93        & 0.017     \\ 
			\hline 
			\multirow{5}{*}{No. 21}  & 0      & 1012    & 12004       & 0.02        & 23.72$\pm$0.636   & 0.47$\pm$0.006    & 55.45$\pm$38.946  & -0.044     & 0.59       & 6          & 2.08       & 0.014     \\ 
			& 1      & 462     & 6241        & 0.06        & 27.02$\pm$0.806   & 0.50$\pm$0.007    & 59.01$\pm$38.968  & -0.022     & 0.52       & 6          & 2.37        & 0.014     \\ 
			& 2      & 233     & 3200        & 0.12        & 27.47$\pm$0.923   & 0.53$\pm$0.009    & 62.35$\pm$39.676  & -0.020     & 0.42       & 3          & 2.63        & 0.014     \\ 
			& 3      & 128     & 1576        & 0.19        & 24.62$\pm$0.967   & 0.56$\pm$0.013    & 64.41$\pm$41.296  & 0.008      & 0.37       & 3          & 2.78        & 0.016     \\ 
			& 4      & 82      & 859         & 0.26        & 20.95$\pm$1.068   & 0.62$\pm$0.017    & 65.83$\pm$43.853  & 0.022      & 0.33       & 3          & 3.04        & 0.020     \\ 
			\hline 
			\multirow{5}{*}{No. 22}  & 0      & 1014    & 13519       & 0.03        & 26.66$\pm$0.657   & 0.42$\pm$0.005    & 64.29$\pm$44.784  & -0.013     & 0.54       & 6          & 1.93       & 0.011     \\ 
			& 1      & 462     & 6669        & 0.06        & 28.87$\pm$0.819   & 0.46$\pm$0.006    & 67.28$\pm$45.278  & 0.002      & 0.50       & 5          & 2.17        & 0.012     \\ 
			& 2      & 233     & 3429        & 0.13        & 29.43$\pm$0.941   & 0.49$\pm$0.008    & 71.49$\pm$45.832  & 0.015      & 0.43       & 4          & 2.34        & 0.012     \\ 
			& 3      & 128     & 1758        & 0.22        & 27.47$\pm$0.982   & 0.55$\pm$0.011    & 74.21$\pm$47.045  & 0.023      & 0.36       & 3          & 2.63        & 0.014     \\ 
			& 4      & 82      & 989         & 0.30        & 24.12$\pm$1.079   & 0.62$\pm$0.014    & 76.73$\pm$49.833  & -0.028     & 0.32       & 3          & 3.15        & 0.017     \\ 
			\hline 
			\multirow{5}{*}{No. 23}  & 0      & 1014    & 14709       & 0.03        & 29.01$\pm$0.701   & 0.43$\pm$0.005    & 70.15$\pm$50.763  & 0.006      & 0.57       & 5          & 1.96       & 0.011     \\ 
			& 1      & 462     & 7147        & 0.07        & 30.94$\pm$0.887   & 0.47$\pm$0.006    & 75.22$\pm$52.027  & 0.017      & 0.48       & 4          & 2.19        & 0.011     \\ 
			& 2      & 233     & 3783        & 0.14        & 32.47$\pm$1.059   & 0.51$\pm$0.008    & 81.64$\pm$53.650  & 0.036      & 0.40       & 4          & 2.41        & 0.011     \\ 
			& 3      & 128     & 1864        & 0.23        & 29.12$\pm$1.148   & 0.57$\pm$0.012    & 85.15$\pm$56.933  & 0.022      & 0.33       & 4          & 2.71        & 0.014     \\ 
			& 4      & 82      & 1031        & 0.31        & 25.15$\pm$1.288   & 0.65$\pm$0.015    & 85.76$\pm$59.914  & -0.047     & 0.28       & 3          & 3.23        & 0.017     \\ 
			\hline 
			\multirow{5}{*}{No. 24}  & 0      & 1014    & 13661       & 0.03        & 26.94$\pm$0.660   & 0.40$\pm$0.005    & 63.30$\pm$40.449  & -0.005     & 0.54       & 5          & 1.85       & 0.011     \\ 
			& 1      & 462     & 6868        & 0.06        & 29.73$\pm$0.849   & 0.45$\pm$0.006    & 66.30$\pm$40.190  & 0.009      & 0.47       & 4          & 2.08        & 0.011     \\ 
			& 2      & 233     & 3587        & 0.13        & 30.79$\pm$0.970   & 0.48$\pm$0.008    & 69.28$\pm$40.086  & 0.031      & 0.40       & 4          & 2.22        & 0.011     \\ 
			& 3      & 128     & 1858        & 0.23        & 29.03$\pm$1.072   & 0.53$\pm$0.010    & 72.08$\pm$42.551  & 0.029      & 0.33       & 3          & 2.41        & 0.013     \\ 
			& 4      & 82      & 1033        & 0.31        & 25.20$\pm$1.216   & 0.61$\pm$0.014    & 73.88$\pm$45.514  & -0.036     & 0.28       & 3          & 2.78        & 0.016     \\ 
			\hline 
			\multirow{5}{*}{No. 25}  & 0      & 1013    & 14802       & 0.03        & 29.22$\pm$0.737   & 0.44$\pm$0.005    & 69.53$\pm$50.198  & 0.015      & 0.52       & 6          & 1.96       & 0.011     \\ 
			& 1      & 462     & 7428        & 0.07        & 32.16$\pm$0.979   & 0.47$\pm$0.006    & 75.38$\pm$52.684  & 0.012      & 0.43       & 5          & 2.15        & 0.011     \\ 
			& 2      & 233     & 3893        & 0.14        & 33.42$\pm$1.185   & 0.51$\pm$0.008    & 81.90$\pm$55.150  & -0.002     & 0.36       & 5          & 2.26        & 0.011     \\ 
			& 3      & 128     & 2039        & 0.25        & 31.86$\pm$1.297   & 0.56$\pm$0.011    & 88.28$\pm$59.237  & -0.010     & 0.27       & 4          & 2.48        & 0.012     \\ 
			& 4      & 82      & 1200        & 0.36        & 29.27$\pm$1.422   & 0.64$\pm$0.014    & 93.61$\pm$63.824  & -0.050     & 0.20       & 3          & 2.63        & 0.013     \\ 
			\hline 
			\multirow{5}{*}{No. 26}  & 0      & 1013    & 12942       & 0.03        & 25.55$\pm$0.673   & 0.44$\pm$0.005    & 56.78$\pm$40.265  & -0.020     & 0.55       & 5          & 1.96       & 0.012     \\ 
			& 1      & 462     & 6709        & 0.06        & 29.04$\pm$0.857   & 0.47$\pm$0.007    & 61.63$\pm$41.598  & 0.000      & 0.49       & 5          & 2.19        & 0.012     \\ 
			& 2      & 233     & 3445        & 0.13        & 29.57$\pm$0.995   & 0.51$\pm$0.009    & 66.80$\pm$43.586  & 0.009      & 0.41       & 4          & 2.41        & 0.013     \\ 
			& 3      & 128     & 1732        & 0.21        & 27.06$\pm$1.056   & 0.55$\pm$0.012    & 69.83$\pm$46.121  & 0.004      & 0.31       & 3          & 2.63        & 0.014     \\ 
			& 4      & 82      & 975         & 0.29        & 23.78$\pm$1.176   & 0.62$\pm$0.015    & 72.24$\pm$50.272  & -0.008     & 0.26       & 3          & 2.93        & 0.017     \\ 
			\hline 
			\multirow{5}{*}{No. 27}  & 0      & 1014    & 12483       & 0.02        & 24.62$\pm$0.619   & 0.40$\pm$0.005    & 60.51$\pm$41.126  & -0.004     & 0.51       & 5          & 1.85       & 0.012     \\ 
			& 1      & 462     & 6352        & 0.06        & 27.50$\pm$0.798   & 0.44$\pm$0.006    & 62.96$\pm$41.281  & 0.009      & 0.46       & 4          & 2.08        & 0.012     \\ 
			& 2      & 233     & 3308        & 0.12        & 28.39$\pm$0.946   & 0.47$\pm$0.008    & 67.63$\pm$42.449  & 0.018      & 0.40       & 5          & 2.22        & 0.012     \\ 
			& 3      & 128     & 1766        & 0.22        & 27.59$\pm$1.051   & 0.54$\pm$0.011    & 71.06$\pm$44.524  & 0.034      & 0.33       & 4          & 2.48        & 0.014     \\ 
			& 4      & 82      & 981         & 0.30        & 23.93$\pm$1.179   & 0.61$\pm$0.015    & 73.43$\pm$47.249  & -0.021     & 0.29       & 3          & 2.78        & 0.017     \\ 
			\hline 
			\multirow{5}{*}{No. 28}  & 0      & 1012    & 14812       & 0.03        & 29.27$\pm$0.774   & 0.43$\pm$0.005    & 71.18$\pm$52.646  & 0.000      & 0.53       & 6          & 1.93       & 0.010     \\ 
			& 1      & 462     & 7440        & 0.07        & 32.21$\pm$1.024   & 0.48$\pm$0.007    & 76.57$\pm$54.983  & -0.001     & 0.44       & 5          & 2.19        & 0.011     \\ 
			& 2      & 233     & 3995        & 0.15        & 34.29$\pm$1.243   & 0.52$\pm$0.009    & 84.86$\pm$59.155  & -0.012     & 0.35       & 4          & 2.34        & 0.011     \\ 
			& 3      & 128     & 2039        & 0.25        & 31.86$\pm$1.333   & 0.58$\pm$0.012    & 90.38$\pm$61.600  & -0.026     & 0.27       & 3          & 2.56        & 0.012     \\ 
			& 4      & 82      & 1151        & 0.35        & 28.07$\pm$1.442   & 0.65$\pm$0.015    & 95.31$\pm$65.566  & -0.039     & 0.22       & 3          & 2.78        & 0.014     \\ 
			\hline 
			\multirow{5}{*}{No. 29}  & 0      & 1012    & 13601       & 0.03        & 26.88$\pm$0.670   & 0.43$\pm$0.005    & 62.88$\pm$41.839  & -0.016     & 0.58       & 6          & 1.96       & 0.012     \\ 
			& 1      & 462     & 6998        & 0.07        & 30.29$\pm$0.852   & 0.46$\pm$0.006    & 67.22$\pm$42.522  & -0.010     & 0.48       & 4          & 2.19        & 0.011     \\ 
			& 2      & 233     & 3694        & 0.14        & 31.71$\pm$0.952   & 0.50$\pm$0.008    & 71.09$\pm$42.622  & -0.000     & 0.39       & 3          & 2.41        & 0.012     \\ 
			& 3      & 128     & 1813        & 0.22        & 28.33$\pm$1.011   & 0.57$\pm$0.011    & 72.42$\pm$43.307  & 0.021      & 0.36       & 3          & 2.78        & 0.014     \\ 
			& 4      & 82      & 966         & 0.29        & 23.56$\pm$1.116   & 0.64$\pm$0.016    & 73.09$\pm$45.577  & -0.002     & 0.33       & 3          & 3.38        & 0.018     \\ 
			
			
		\end{tabular}
	\end{ruledtabular}
\end{table*}

\renewcommand{\arraystretch}{0.55} 
\begin{table*}[t] \footnotesize
	\begin{ruledtabular}
		\centering
		\begin{tabular}{*{13}{c}} 
			\thd{Subject}    & \thd{Layer} & \thd{$N$} & \thd{$L$} & \thd{$\rho_l$} & \thd{$\langle k\rangle\pm \text{SEM}$}  &  \thd{$\langle c\rangle\pm \text{SEM}$}  & \thd{$f^{(l)}\pm \text{SEM}$} &\thd{$r_c$} &\thd{$Q$} &\thd{$N_c$} & \thd{$\beta$} & \thd{$\mu$}\\
			\hline 
			\hline 
			\multirow{5}{*}{No. 30}  & 0      & 1013    & 14177       & 0.03        & 27.99$\pm$0.711   & 0.43$\pm$0.005    & 67.36$\pm$47.975  & -0.008     & 0.55       & 5          & 1.96       & 0.011     \\ 
			& 1      & 462     & 7246        & 0.07        & 31.37$\pm$0.911   & 0.46$\pm$0.006    & 72.35$\pm$49.696  & 0.022      & 0.47       & 5          & 2.15        & 0.011     \\ 
			& 2      & 233     & 3778        & 0.14        & 32.43$\pm$1.067   & 0.50$\pm$0.008    & 77.23$\pm$50.352  & 0.016      & 0.38       & 3          & 2.32        & 0.011     \\ 
			& 3      & 128     & 1960        & 0.24        & 30.62$\pm$1.130   & 0.56$\pm$0.011    & 81.62$\pm$52.396  & 0.028      & 0.33       & 3          & 2.63        & 0.013     \\ 
			& 4      & 82      & 1131        & 0.34        & 27.59$\pm$1.287   & 0.65$\pm$0.014    & 84.91$\pm$56.072  & -0.004     & 0.28       & 3          & 3.08        & 0.015     \\ 		
			\multirow{5}{*}{No. 31}  & 0      & 1014    & 15641       & 0.03        & 30.85$\pm$0.760   & 0.41$\pm$0.005    & 71.71$\pm$51.033  & 0.014      & 0.55       & 6          & 1.85       & 0.009     \\ 
			& 1      & 462     & 7983        & 0.07        & 34.56$\pm$0.994   & 0.46$\pm$0.006    & 77.72$\pm$54.036  & 0.026      & 0.47       & 4          & 2.11        & 0.010     \\ 
			& 2      & 233     & 4220        & 0.16        & 36.22$\pm$1.132   & 0.50$\pm$0.007    & 83.75$\pm$57.298  & 0.051      & 0.38       & 4          & 2.34        & 0.010     \\ 
			& 3      & 128     & 2141        & 0.26        & 33.45$\pm$1.216   & 0.56$\pm$0.010    & 88.56$\pm$59.873  & 0.032      & 0.30       & 3          & 2.56        & 0.011     \\ 
			& 4      & 82      & 1231        & 0.37        & 30.02$\pm$1.386   & 0.63$\pm$0.012    & 94.13$\pm$65.808  & 0.002      & 0.21       & 3          & 2.48        & 0.013     \\ 
			\hline 
			\multirow{5}{*}{No. 32}  & 0      & 1013    & 12875       & 0.03        & 25.42$\pm$0.673   & 0.43$\pm$0.005    & 62.54$\pm$46.293  & -0.001     & 0.57       & 6          & 1.96       & 0.012     \\ 
			& 1      & 462     & 6409        & 0.06        & 27.74$\pm$0.867   & 0.48$\pm$0.007    & 66.40$\pm$47.918  & 0.009      & 0.48       & 6          & 2.22        & 0.013     \\ 
			& 2      & 233     & 3351        & 0.12        & 28.76$\pm$1.052   & 0.50$\pm$0.009    & 73.08$\pm$51.161  & 0.034      & 0.37       & 4          & 2.26        & 0.012     \\ 
			& 3      & 128     & 1795        & 0.22        & 28.05$\pm$1.174   & 0.55$\pm$0.011    & 78.25$\pm$53.290  & 0.028      & 0.28       & 4          & 2.41        & 0.013     \\ 
			& 4      & 82      & 1031        & 0.31        & 25.15$\pm$1.271   & 0.62$\pm$0.015    & 80.85$\pm$55.859  & 0.005      & 0.25       & 3          & 2.78        & 0.016     \\ 
			\hline 
			\multirow{5}{*}{No. 33}  & 0      & 1014    & 14510       & 0.03        & 28.62$\pm$0.714   & 0.44$\pm$0.005    & 66.73$\pm$46.049  & -0.008     & 0.53       & 5          & 1.96       & 0.011     \\ 
			& 1      & 462     & 7427        & 0.07        & 32.15$\pm$0.929   & 0.47$\pm$0.007    & 71.73$\pm$46.530  & 0.002      & 0.46       & 4          & 2.19        & 0.011     \\ 
			& 2      & 233     & 3876        & 0.14        & 33.27$\pm$1.066   & 0.50$\pm$0.008    & 77.93$\pm$47.755  & 0.009      & 0.38       & 3          & 2.34        & 0.011     \\ 
			& 3      & 128     & 2004        & 0.25        & 31.31$\pm$1.138   & 0.56$\pm$0.010    & 81.38$\pm$49.065  & 0.016      & 0.32       & 3          & 2.63        & 0.012     \\ 
			& 4      & 82      & 1106        & 0.33        & 26.98$\pm$1.292   & 0.64$\pm$0.015    & 82.64$\pm$51.067  & -0.043     & 0.26       & 3          & 3.08        & 0.015     \\ 
			\hline 
			\multirow{5}{*}{No. 34}  & 0      & 1012    & 14344       & 0.03        & 28.35$\pm$0.706   & 0.41$\pm$0.005    & 69.04$\pm$45.713  & -0.001     & 0.53       & 5          & 1.85       & 0.010     \\ 
			& 1      & 461     & 7292        & 0.07        & 31.64$\pm$0.904   & 0.45$\pm$0.006    & 73.20$\pm$45.988  & 0.020      & 0.47       & 5          & 2.04        & 0.010     \\ 
			& 2      & 233     & 3814        & 0.14        & 32.74$\pm$1.025   & 0.48$\pm$0.007    & 78.36$\pm$47.041  & 0.035      & 0.38       & 4          & 2.19        & 0.011     \\ 
			& 3      & 128     & 1921        & 0.24        & 30.02$\pm$1.047   & 0.54$\pm$0.011    & 81.64$\pm$48.906  & 0.024      & 0.31       & 4          & 2.56        & 0.013     \\ 
			& 4      & 82      & 1053        & 0.32        & 25.68$\pm$1.197   & 0.63$\pm$0.015    & 81.33$\pm$50.079  & -0.015     & 0.28       & 3          & 3.08        & 0.016     \\ 
			\hline 
			\multirow{5}{*}{No. 35}  & 0      & 1009    & 12460       & 0.02        & 24.70$\pm$0.665   & 0.42$\pm$0.005    & 59.69$\pm$41.052  & -0.001     & 0.54       & 6          & 1.89       & 0.012     \\ 
			& 1      & 462     & 6467        & 0.06        & 28.00$\pm$0.867   & 0.46$\pm$0.006    & 64.31$\pm$42.744  & 0.012      & 0.45       & 5          & 2.11        & 0.012     \\ 
			& 2      & 233     & 3439        & 0.13        & 29.52$\pm$1.035   & 0.50$\pm$0.008    & 69.39$\pm$43.860  & 0.012      & 0.39       & 5          & 2.34        & 0.012     \\ 
			& 3      & 128     & 1808        & 0.22        & 28.25$\pm$1.136   & 0.55$\pm$0.010    & 74.81$\pm$46.898  & 0.015      & 0.31       & 3          & 2.48        & 0.013     \\ 
			& 4      & 82      & 1064        & 0.32        & 25.95$\pm$1.276   & 0.61$\pm$0.014    & 79.37$\pm$50.431  & -0.040     & 0.23       & 3          & 2.48        & 0.015     \\ 
			\hline 
			\multirow{5}{*}{No. 36}  & 0      & 1014    & 13978       & 0.03        & 27.57$\pm$0.693   & 0.43$\pm$0.005    & 66.11$\pm$45.586  & -0.011     & 0.55       & 6          & 1.96       & 0.011     \\ 
			& 1      & 462     & 7053        & 0.07        & 30.53$\pm$0.891   & 0.47$\pm$0.007    & 70.12$\pm$46.290  & -0.006     & 0.49       & 4          & 2.19        & 0.011     \\ 
			& 2      & 233     & 3775        & 0.14        & 32.40$\pm$1.052   & 0.50$\pm$0.008    & 74.40$\pm$46.276  & -0.004     & 0.40       & 3          & 2.34        & 0.011     \\ 
			& 3      & 128     & 1915        & 0.24        & 29.92$\pm$1.157   & 0.55$\pm$0.011    & 77.52$\pm$48.562  & 0.009      & 0.32       & 3          & 2.48        & 0.013     \\ 
			& 4      & 82      & 1083        & 0.33        & 26.41$\pm$1.288   & 0.63$\pm$0.015    & 78.34$\pm$51.617  & -0.034     & 0.22       & 2          & 2.93        & 0.016     \\ 
			\hline 
			\multirow{5}{*}{No. 37}  & 0      & 1014    & 14282       & 0.03        & 28.17$\pm$0.693   & 0.40$\pm$0.005    & 64.59$\pm$43.099  & -0.004     & 0.54       & 5          & 1.88       & 0.011     \\ 
			& 1      & 462     & 7216        & 0.07        & 31.24$\pm$0.898   & 0.45$\pm$0.006    & 68.52$\pm$43.720  & -0.006     & 0.44       & 4          & 2.11        & 0.011     \\ 
			& 2      & 233     & 3783        & 0.14        & 32.47$\pm$1.036   & 0.49$\pm$0.008    & 73.59$\pm$44.947  & -0.009     & 0.39       & 4          & 2.30        & 0.011     \\ 
			& 3      & 128     & 1940        & 0.24        & 30.31$\pm$1.132   & 0.55$\pm$0.010    & 77.60$\pm$47.272  & -0.024     & 0.27       & 3          & 2.52        & 0.013     \\ 
			& 4      & 82      & 1105        & 0.33        & 26.95$\pm$1.282   & 0.62$\pm$0.013    & 82.25$\pm$52.116  & -0.044     & 0.25       & 3          & 2.71        & 0.015     \\ 
			\hline 
			\multirow{5}{*}{No. 38}  & 0      & 1011    & 14001       & 0.03        & 27.70$\pm$0.679   & 0.43$\pm$0.005    & 68.35$\pm$47.067  & 0.025      & 0.53       & 6          & 1.93       & 0.011     \\ 
			& 1      & 462     & 7129        & 0.07        & 30.86$\pm$0.909   & 0.46$\pm$0.006    & 74.59$\pm$49.171  & 0.040      & 0.45       & 5          & 2.08        & 0.011     \\ 
			& 2      & 233     & 3788        & 0.14        & 32.52$\pm$1.086   & 0.49$\pm$0.008    & 80.76$\pm$51.217  & 0.045      & 0.38       & 4          & 2.26        & 0.011     \\ 
			& 3      & 128     & 1952        & 0.24        & 30.50$\pm$1.235   & 0.55$\pm$0.010    & 86.29$\pm$54.329  & 0.033      & 0.28       & 4          & 2.34        & 0.012     \\ 
			& 4      & 82      & 1115        & 0.34        & 27.20$\pm$1.378   & 0.62$\pm$0.012    & 90.79$\pm$59.115  & -0.019     & 0.21       & 4          & 2.48        & 0.014     \\ 
			\hline 
			\multirow{5}{*}{No. 39}  & 0      & 1013    & 12599       & 0.02        & 24.87$\pm$0.695   & 0.44$\pm$0.005    & 60.71$\pm$40.883  & -0.031     & 0.56       & 5          & 1.96       & 0.013     \\ 
			& 1      & 462     & 6468        & 0.06        & 28.00$\pm$0.902   & 0.47$\pm$0.007    & 63.92$\pm$40.596  & -0.028     & 0.49       & 5          & 2.19        & 0.012     \\ 
			& 2      & 233     & 3433        & 0.13        & 29.47$\pm$1.053   & 0.50$\pm$0.008    & 68.99$\pm$41.314  & -0.027     & 0.42       & 5          & 2.34        & 0.012     \\ 
			& 3      & 128     & 1808        & 0.22        & 28.25$\pm$1.162   & 0.57$\pm$0.011    & 72.21$\pm$42.214  & -0.027     & 0.30       & 3          & 2.71        & 0.014     \\ 
			& 4      & 82      & 1026        & 0.31        & 25.02$\pm$1.262   & 0.63$\pm$0.015    & 74.63$\pm$44.861  & -0.075     & 0.25       & 3          & 2.93        & 0.016     \\ 		
		\end{tabular}
	\end{ruledtabular}
\end{table*}


\renewcommand{\arraystretch}{0.55} 
\begin{table*}[b] \footnotesize
	\begin{ruledtabular}
		\centering	
		\caption{Overview of the 44 connectomes in the HCP dataeset. The number of nodes ($N$), the number of links ($L$), the density of links 
			($\rho_l = 2L/N(N-1)$), its average degree ($\langle k \rangle = 2L/N$), the average local clustering coefficient ($\langle c \rangle$), the average fiber length ($f^{(l)}$ (mm)), and corresponding $\pm 1$ standard error interval around the mean (SEM), the assortativity coefficient ($r_c$), the modularity ($Q$), the number of the communities ($Nc$), and the hyperbolic embedding parameter $\beta$ and $\mu$.}
		\label{tab:A1}		
		\begin{tabular}{*{13}{c}} 
			\thd{Subject}    & \thd{Layer} & \thd{$N$} & \thd{$L$} & \thd{$\rho_l$} & \thd{$\langle k\rangle\pm \text{SEM}$}  &  \thd{$\langle c\rangle\pm \text{SEM}$}  & \thd{$f^{(l)}\pm \text{SEM}$} &\thd{$r_c$} &\thd{$Q$} &\thd{$N_c$} & \thd{$\beta$} & \thd{$\mu$}\\
			\hline
			\multirow{5}{*}{No. 0}  & 0      & 1014    & 37910       & 0.07        & 74.77$\pm$1.509   & 0.41$\pm$0.004    & 58.31$\pm$43.175  & -0.021     & 0.40       & 5          & 1.83       & 0.004     \\ 
			& 1      & 462     & 18024       & 0.17        & 78.03$\pm$1.758   & 0.47$\pm$0.004    & 66.92$\pm$44.813  & -0.005     & 0.33       & 3          & 2.02        & 0.004     \\ 
			& 2      & 233     & 8540        & 0.32        & 73.30$\pm$1.765   & 0.55$\pm$0.006    & 73.51$\pm$45.896  & -0.016     & 0.25       & 3          & 2.19        & 0.005     \\ 
			& 3      & 128     & 3871        & 0.48        & 60.48$\pm$1.617   & 0.65$\pm$0.007    & 78.20$\pm$47.528  & -0.052     & 0.19       & 3          & 3.02        & 0.007     \\ 
			& 4      & 82      & 1873        & 0.56        & 45.68$\pm$1.591   & 0.74$\pm$0.010    & 77.00$\pm$49.433  & -0.098     & 0.15       & 2          & 3.58        & 0.010     \\ 
			\hline 
			\multirow{5}{*}{No. 1}  & 0      & 1014    & 39210       & 0.08        & 77.34$\pm$1.661   & 0.42$\pm$0.004    & 60.77$\pm$44.773  & -0.036     & 0.39       & 3          & 1.83       & 0.004     \\ 
			& 1      & 462     & 18776       & 0.18        & 81.28$\pm$1.910   & 0.48$\pm$0.005    & 69.51$\pm$45.948  & -0.035     & 0.33       & 3          & 2.00        & 0.004     \\ 
			& 2      & 233     & 8875        & 0.33        & 76.18$\pm$1.851   & 0.57$\pm$0.006    & 76.93$\pm$47.307  & -0.041     & 0.26       & 3          & 2.37        & 0.005     \\ 
			& 3      & 128     & 4029        & 0.50        & 62.95$\pm$1.608   & 0.67$\pm$0.006    & 82.16$\pm$48.826  & -0.047     & 0.19       & 2          & 2.67        & 0.006     \\ 
			& 4      & 82      & 1990        & 0.60        & 48.54$\pm$1.531   & 0.75$\pm$0.008    & 81.97$\pm$50.412  & -0.090     & 0.15       & 2          & 3.61        & 0.009     \\ 
			\hline 
			\multirow{5}{*}{No. 2}  & 0      & 1014    & 39609       & 0.08        & 78.12$\pm$1.539   & 0.42$\pm$0.004    & 56.38$\pm$40.686  & -0.028     & 0.38       & 4          & 1.92       & 0.004     \\ 
			& 1      & 462     & 18603       & 0.17        & 80.53$\pm$1.754   & 0.48$\pm$0.005    & 64.62$\pm$41.902  & -0.025     & 0.33       & 3          & 2.16        & 0.004     \\ 
			& 2      & 233     & 8671        & 0.32        & 74.43$\pm$1.710   & 0.57$\pm$0.005    & 71.26$\pm$43.164  & -0.023     & 0.27       & 3          & 2.54        & 0.005     \\ 
			& 3      & 128     & 3791        & 0.47        & 59.23$\pm$1.545   & 0.66$\pm$0.007    & 75.33$\pm$44.626  & -0.036     & 0.22       & 2          & 3.60        & 0.007     \\ 
			& 4      & 82      & 1857        & 0.56        & 45.29$\pm$1.419   & 0.73$\pm$0.009    & 74.19$\pm$45.822  & -0.073     & 0.18       & 2          & 3.82        & 0.010     \\ 
			\hline 
			\multirow{5}{*}{No. 3}  & 0      & 1014    & 40309       & 0.08        & 79.50$\pm$1.711   & 0.44$\pm$0.004    & 58.96$\pm$43.097  & -0.041     & 0.41       & 4          & 1.91       & 0.004     \\ 
			& 1      & 462     & 19158       & 0.18        & 82.94$\pm$1.965   & 0.49$\pm$0.005    & 68.01$\pm$44.740  & -0.041     & 0.32       & 3          & 2.02        & 0.004     \\ 
			& 2      & 233     & 9033        & 0.33        & 77.54$\pm$1.903   & 0.58$\pm$0.006    & 75.47$\pm$46.003  & -0.054     & 0.23       & 3          & 2.39        & 0.005     \\ 
			& 3      & 128     & 4078        & 0.50        & 63.72$\pm$1.660   & 0.68$\pm$0.007    & 80.71$\pm$48.110  & -0.062     & 0.18       & 2          & 2.92        & 0.006     \\ 
			& 4      & 82      & 1986        & 0.60        & 48.44$\pm$1.512   & 0.75$\pm$0.009    & 80.13$\pm$49.525  & -0.097     & 0.13       & 2          & 3.70        & 0.009     \\ 
			\hline 
			\multirow{5}{*}{No. 4}  & 0      & 1014    & 43276       & 0.08        & 85.36$\pm$1.687   & 0.41$\pm$0.004    & 57.41$\pm$41.454  & -0.040     & 0.41       & 4          & 1.81       & 0.003     \\ 
			& 1      & 462     & 20470       & 0.19        & 88.61$\pm$1.886   & 0.47$\pm$0.004    & 66.27$\pm$43.139  & -0.033     & 0.32       & 4          & 1.97        & 0.004     \\ 
			& 2      & 233     & 9609        & 0.36        & 82.48$\pm$1.812   & 0.57$\pm$0.005    & 73.54$\pm$44.661  & -0.043     & 0.23       & 3          & 2.37        & 0.004     \\ 
			& 3      & 128     & 4431        & 0.55        & 69.23$\pm$1.575   & 0.68$\pm$0.005    & 79.81$\pm$46.430  & -0.042     & 0.14       & 3          & 3.08        & 0.006     \\ 
			& 4      & 82      & 2142        & 0.64        & 52.24$\pm$1.520   & 0.77$\pm$0.006    & 80.09$\pm$48.261  & -0.065     & 0.12       & 2          & 3.20        & 0.008     \\ 
			\hline 
			\multirow{5}{*}{No. 5}  & 0      & 1014    & 40165       & 0.08        & 79.22$\pm$1.546   & 0.41$\pm$0.004    & 63.63$\pm$45.504  & -0.032     & 0.42       & 4          & 1.85       & 0.004     \\ 
			& 1      & 462     & 19100       & 0.18        & 82.68$\pm$1.769   & 0.47$\pm$0.005    & 71.98$\pm$46.051  & -0.032     & 0.34       & 3          & 2.06        & 0.004     \\ 
			& 2      & 233     & 9093        & 0.34        & 78.05$\pm$1.721   & 0.56$\pm$0.005    & 79.75$\pm$47.253  & -0.041     & 0.26       & 3          & 2.52        & 0.005     \\ 
			& 3      & 128     & 4114        & 0.51        & 64.28$\pm$1.550   & 0.67$\pm$0.006    & 84.40$\pm$48.945  & -0.059     & 0.18       & 3          & 3.13        & 0.007     \\ 
			& 4      & 82      & 2033        & 0.61        & 49.59$\pm$1.454   & 0.75$\pm$0.008    & 84.01$\pm$51.240  & -0.095     & 0.14       & 2          & 4.31        & 0.009     \\ 
			\hline 
			\multirow{5}{*}{No. 6}  & 0      & 1014    & 39638       & 0.08        & 78.18$\pm$1.571   & 0.42$\pm$0.004    & 57.13$\pm$41.277  & -0.042     & 0.42       & 4          & 1.89       & 0.004     \\ 
			& 1      & 462     & 18692       & 0.18        & 80.92$\pm$1.784   & 0.48$\pm$0.005    & 65.01$\pm$42.186  & -0.042     & 0.34       & 3          & 2.10        & 0.004     \\ 
			& 2      & 233     & 8821        & 0.33        & 75.72$\pm$1.749   & 0.57$\pm$0.006    & 71.37$\pm$43.120  & -0.048     & 0.27       & 3          & 2.53        & 0.005     \\ 
			& 3      & 128     & 3958        & 0.49        & 61.84$\pm$1.562   & 0.66$\pm$0.007    & 75.59$\pm$44.708  & -0.054     & 0.19       & 3          & 2.84        & 0.007     \\ 
			& 4      & 82      & 1942        & 0.58        & 47.37$\pm$1.481   & 0.74$\pm$0.009    & 75.67$\pm$46.882  & -0.094     & 0.14       & 3          & 2.81        & 0.008     \\ 
			\hline 
			\multirow{5}{*}{No. 7}  & 0      & 1014    & 44089       & 0.09        & 86.96$\pm$1.696   & 0.43$\pm$0.004    & 56.98$\pm$39.373  & -0.033     & 0.42       & 4          & 1.94       & 0.004     \\ 
			& 1      & 462     & 20383       & 0.19        & 88.24$\pm$1.900   & 0.49$\pm$0.004    & 64.92$\pm$40.040  & -0.020     & 0.29       & 3          & 2.14        & 0.004     \\ 
			& 2      & 233     & 9312        & 0.34        & 79.93$\pm$1.822   & 0.58$\pm$0.005    & 71.32$\pm$40.703  & -0.023     & 0.23       & 2          & 2.45        & 0.005     \\ 
			& 3      & 128     & 4186        & 0.52        & 65.41$\pm$1.598   & 0.68$\pm$0.006    & 76.46$\pm$42.401  & -0.049     & 0.18       & 2          & 3.00        & 0.006     \\ 
			& 4      & 82      & 2044        & 0.62        & 49.85$\pm$1.474   & 0.76$\pm$0.008    & 76.05$\pm$44.288  & -0.094     & 0.14       & 2          & 3.33        & 0.009     \\ 
			\hline 
			\multirow{5}{*}{No. 8}  & 0      & 1014    & 36557       & 0.07        & 72.10$\pm$1.633   & 0.46$\pm$0.005    & 49.29$\pm$35.917  & -0.044     & 0.41       & 4          & 1.97       & 0.004     \\ 
			& 1      & 462     & 17316       & 0.16        & 74.96$\pm$1.831   & 0.50$\pm$0.005    & 56.69$\pm$37.010  & -0.042     & 0.33       & 3          & 2.18        & 0.005     \\ 
			& 2      & 233     & 8296        & 0.31        & 71.21$\pm$1.772   & 0.57$\pm$0.006    & 62.83$\pm$37.979  & -0.036     & 0.26       & 3          & 2.49        & 0.005     \\ 
			& 3      & 128     & 3746        & 0.46        & 58.53$\pm$1.570   & 0.66$\pm$0.007    & 66.68$\pm$39.094  & -0.037     & 0.21       & 2          & 2.84        & 0.007     \\ 
			& 4      & 82      & 1851        & 0.56        & 45.15$\pm$1.475   & 0.73$\pm$0.009    & 65.79$\pm$40.188  & -0.062     & 0.18       & 2          & 3.41        & 0.010     \\ 
		\end{tabular}
	\end{ruledtabular}
\end{table*}

\renewcommand{\arraystretch}{0.6} 
\begin{table*}[t] \footnotesize
	\begin{ruledtabular}
		\centering
		\begin{tabular}{*{13}{c}} 
			\thd{Subject}    & \thd{Layer} & \thd{$N$} & \thd{$L$} & \thd{$\rho_l$} & \thd{$\langle k\rangle\pm \text{SEM}$}  &  \thd{$\langle c\rangle\pm \text{SEM}$}  & \thd{$f^{(l)}\pm \text{SEM}$} &\thd{$r_c$} &\thd{$Q$}&\thd{$N_c$} & \thd{$\beta$} & \thd{$\mu$}\\
			\hline
			\multirow{5}{*}{No. 9}  & 0      & 1014    & 38249       & 0.07        & 75.44$\pm$1.613   & 0.43$\pm$0.004    & 55.86$\pm$40.591  & -0.045     & 0.40       & 4          & 1.83       & 0.004     \\ 
			& 1      & 462     & 18352       & 0.17        & 79.45$\pm$1.858   & 0.48$\pm$0.005    & 63.77$\pm$41.589  & -0.043     & 0.33       & 3          & 2.03        & 0.004     \\ 
			& 2      & 233     & 8702        & 0.32        & 74.70$\pm$1.850   & 0.56$\pm$0.006    & 70.08$\pm$42.715  & -0.045     & 0.26       & 3          & 2.19        & 0.005     \\ 
			& 3      & 128     & 3960        & 0.49        & 61.88$\pm$1.712   & 0.66$\pm$0.007    & 73.82$\pm$44.027  & -0.061     & 0.18       & 3          & 2.48        & 0.006     \\ 
			& 4      & 82      & 1922        & 0.58        & 46.88$\pm$1.625   & 0.75$\pm$0.009    & 72.11$\pm$43.995  & -0.105     & 0.13       & 2          & 2.84        & 0.009     \\ 
			\hline 
			\multirow{5}{*}{No. 10}  & 0      & 1014    & 39578       & 0.08        & 78.06$\pm$1.493   & 0.41$\pm$0.004    & 62.41$\pm$43.797  & -0.024     & 0.35       & 3          & 1.90       & 0.004     \\ 
			& 1      & 462     & 18410       & 0.17        & 79.70$\pm$1.721   & 0.48$\pm$0.005    & 71.58$\pm$45.113  & -0.022     & 0.33       & 3          & 2.19        & 0.004     \\ 
			& 2      & 233     & 8589        & 0.32        & 73.73$\pm$1.694   & 0.57$\pm$0.006    & 78.76$\pm$46.197  & -0.022     & 0.27       & 3          & 2.48        & 0.005     \\ 
			& 3      & 128     & 3831        & 0.47        & 59.86$\pm$1.552   & 0.66$\pm$0.007    & 83.60$\pm$47.906  & -0.033     & 0.21       & 3          & 3.23        & 0.007     \\ 
			& 4      & 82      & 1865        & 0.56        & 45.49$\pm$1.430   & 0.73$\pm$0.009    & 82.79$\pm$50.090  & -0.080     & 0.18       & 2          & 3.51        & 0.010     \\ 
			\hline 
			\multirow{5}{*}{No. 11}  & 0      & 1013    & 34436       & 0.07        & 67.99$\pm$1.343   & 0.41$\pm$0.004    & 56.10$\pm$41.192  & -0.025     & 0.38       & 4          & 1.90       & 0.004     \\ 
			& 1      & 462     & 16289       & 0.15        & 70.52$\pm$1.542   & 0.46$\pm$0.004    & 63.75$\pm$42.384  & -0.015     & 0.36       & 4          & 2.12        & 0.005     \\ 
			& 2      & 233     & 7718        & 0.29        & 66.25$\pm$1.546   & 0.54$\pm$0.006    & 70.16$\pm$43.299  & -0.017     & 0.28       & 4          & 2.36        & 0.006     \\ 
			& 3      & 128     & 3527        & 0.43        & 55.11$\pm$1.433   & 0.63$\pm$0.007    & 74.03$\pm$44.312  & -0.036     & 0.22       & 3          & 2.80        & 0.007     \\ 
			& 4      & 82      & 1733        & 0.52        & 42.27$\pm$1.412   & 0.70$\pm$0.009    & 73.19$\pm$45.508  & -0.067     & 0.18       & 2          & 3.03        & 0.010     \\ 
			\hline 
			\multirow{5}{*}{No. 12}  & 0      & 1014    & 39631       & 0.08        & 78.17$\pm$1.541   & 0.42$\pm$0.004    & 55.80$\pm$40.787  & -0.029     & 0.40       & 4          & 1.84       & 0.004     \\ 
			& 1      & 462     & 18863       & 0.18        & 81.66$\pm$1.741   & 0.47$\pm$0.004    & 63.98$\pm$42.180  & -0.013     & 0.34       & 3          & 2.00        & 0.004     \\ 
			& 2      & 233     & 8793        & 0.33        & 75.48$\pm$1.749   & 0.56$\pm$0.006    & 70.85$\pm$43.667  & -0.023     & 0.27       & 3          & 2.46        & 0.005     \\ 
			& 3      & 128     & 3960        & 0.49        & 61.88$\pm$1.583   & 0.66$\pm$0.007    & 76.09$\pm$45.451  & -0.043     & 0.20       & 2          & 2.99        & 0.007     \\ 
			& 4      & 82      & 1944        & 0.59        & 47.41$\pm$1.516   & 0.74$\pm$0.009    & 75.56$\pm$46.732  & -0.090     & 0.15       & 2          & 2.99        & 0.009     \\ 
			\hline 
			\multirow{5}{*}{No. 13}  & 0      & 1014    & 40358       & 0.08        & 79.60$\pm$1.692   & 0.44$\pm$0.004    & 58.05$\pm$40.581  & -0.044     & 0.40       & 4          & 1.97       & 0.004     \\ 
			& 1      & 462     & 19173       & 0.18        & 83.00$\pm$1.886   & 0.49$\pm$0.005    & 66.29$\pm$41.941  & -0.042     & 0.32       & 4          & 2.13        & 0.004     \\ 
			& 2      & 233     & 9043        & 0.33        & 77.62$\pm$1.846   & 0.57$\pm$0.006    & 73.68$\pm$43.391  & -0.042     & 0.23       & 3          & 2.56        & 0.005     \\ 
			& 3      & 128     & 4042        & 0.50        & 63.16$\pm$1.714   & 0.68$\pm$0.007    & 77.88$\pm$44.575  & -0.037     & 0.17       & 3          & 2.64        & 0.006     \\ 
			& 4      & 82      & 1999        & 0.60        & 48.76$\pm$1.592   & 0.76$\pm$0.008    & 76.36$\pm$45.471  & -0.076     & 0.12       & 2          & 3.58        & 0.009     \\ 
			\hline 
			\multirow{5}{*}{No. 14}  & 0      & 1014    & 40821       & 0.08        & 80.51$\pm$1.583   & 0.44$\pm$0.004    & 54.98$\pm$38.809  & -0.040     & 0.43       & 4          & 1.98       & 0.004     \\ 
			& 1      & 462     & 19050       & 0.18        & 82.47$\pm$1.788   & 0.50$\pm$0.005    & 62.97$\pm$40.035  & -0.043     & 0.35       & 4          & 2.19        & 0.004     \\ 
			& 2      & 233     & 8767        & 0.32        & 75.25$\pm$1.749   & 0.57$\pm$0.006    & 69.72$\pm$41.571  & -0.042     & 0.25       & 3          & 2.59        & 0.005     \\ 
			& 3      & 128     & 3934        & 0.48        & 61.47$\pm$1.583   & 0.66$\pm$0.007    & 74.57$\pm$43.268  & -0.060     & 0.18       & 3          & 3.18        & 0.007     \\ 
			& 4      & 82      & 1938        & 0.58        & 47.27$\pm$1.521   & 0.74$\pm$0.009    & 74.58$\pm$45.075  & -0.100     & 0.12       & 3          & 3.24        & 0.009     \\ 
			\hline 
			\multirow{5}{*}{No. 15}  & 0      & 1014    & 37896       & 0.07        & 74.75$\pm$1.563   & 0.41$\pm$0.004    & 61.60$\pm$43.403  & -0.039     & 0.43       & 4          & 1.87       & 0.004     \\ 
			& 1      & 462     & 18119       & 0.17        & 78.44$\pm$1.823   & 0.47$\pm$0.005    & 70.46$\pm$44.242  & -0.040     & 0.35       & 4          & 2.05        & 0.004     \\ 
			& 2      & 233     & 8732        & 0.32        & 74.95$\pm$1.802   & 0.56$\pm$0.006    & 77.63$\pm$45.368  & -0.042     & 0.26       & 3          & 2.47        & 0.005     \\ 
			& 3      & 128     & 4011        & 0.49        & 62.67$\pm$1.602   & 0.66$\pm$0.006    & 82.71$\pm$46.851  & -0.047     & 0.18       & 3          & 2.98        & 0.007     \\ 
			& 4      & 82      & 2035        & 0.61        & 49.63$\pm$1.501   & 0.75$\pm$0.007    & 83.78$\pm$48.511  & -0.074     & 0.13       & 2          & 3.60        & 0.009     \\ 
			\hline 
			\multirow{5}{*}{No. 16}  & 0      & 1014    & 39589       & 0.08        & 78.08$\pm$1.617   & 0.43$\pm$0.004    & 60.62$\pm$43.139  & -0.046     & 0.41       & 4          & 1.95       & 0.004     \\ 
			& 1      & 462     & 18938       & 0.18        & 81.98$\pm$1.837   & 0.49$\pm$0.005    & 69.34$\pm$43.964  & -0.051     & 0.34       & 3          & 2.11        & 0.004     \\ 
			& 2      & 233     & 9018        & 0.33        & 77.41$\pm$1.792   & 0.57$\pm$0.006    & 76.09$\pm$44.730  & -0.055     & 0.26       & 3          & 2.43        & 0.005     \\ 
			& 3      & 128     & 4068        & 0.50        & 63.56$\pm$1.545   & 0.66$\pm$0.006    & 80.33$\pm$46.104  & -0.066     & 0.19       & 3          & 2.80        & 0.006     \\ 
			& 4      & 82      & 2008        & 0.60        & 48.98$\pm$1.383   & 0.74$\pm$0.007    & 79.31$\pm$47.395  & -0.086     & 0.16       & 2          & 3.24        & 0.009     \\ 
			\hline 
			\multirow{5}{*}{No. 17}  & 0      & 1014    & 38009       & 0.07        & 74.97$\pm$1.591   & 0.44$\pm$0.004    & 53.98$\pm$38.401  & -0.026     & 0.43       & 4          & 1.95       & 0.004     \\ 
			& 1      & 462     & 17734       & 0.17        & 76.77$\pm$1.784   & 0.49$\pm$0.005    & 61.72$\pm$39.467  & -0.020     & 0.35       & 4          & 2.21        & 0.005     \\ 
			& 2      & 233     & 8280        & 0.31        & 71.07$\pm$1.758   & 0.57$\pm$0.006    & 67.56$\pm$40.191  & -0.037     & 0.28       & 3          & 2.62        & 0.005     \\ 
			& 3      & 128     & 3815        & 0.47        & 59.61$\pm$1.525   & 0.65$\pm$0.007    & 71.55$\pm$41.296  & -0.052     & 0.20       & 3          & 3.07        & 0.007     \\ 
			& 4      & 82      & 1910        & 0.58        & 46.59$\pm$1.443   & 0.73$\pm$0.009    & 72.47$\pm$43.781  & -0.100     & 0.16       & 2          & 3.64        & 0.009     \\ 
			\hline 
			\multirow{5}{*}{No. 18}  & 0      & 1014    & 46070       & 0.09        & 90.87$\pm$1.707   & 0.41$\pm$0.003    & 62.24$\pm$43.523  & -0.033     & 0.37       & 3          & 1.90       & 0.003     \\ 
			& 1      & 462     & 21730       & 0.20        & 94.07$\pm$1.881   & 0.49$\pm$0.004    & 71.29$\pm$44.918  & -0.021     & 0.30       & 3          & 2.09        & 0.004     \\ 
			& 2      & 233     & 10019       & 0.37        & 86.00$\pm$1.759   & 0.58$\pm$0.005    & 79.50$\pm$46.687  & -0.030     & 0.24       & 3          & 2.56        & 0.004     \\ 
			& 3      & 128     & 4435        & 0.55        & 69.30$\pm$1.475   & 0.68$\pm$0.005    & 85.99$\pm$48.874  & -0.052     & 0.18       & 2          & 3.15        & 0.006     \\ 
			& 4      & 82      & 2155        & 0.65        & 52.56$\pm$1.385   & 0.76$\pm$0.007    & 86.68$\pm$51.335  & -0.103     & 0.14       & 2          & 2.59        & 0.007     \\ 
		\end{tabular}
	\end{ruledtabular}
\end{table*}

\renewcommand{\arraystretch}{0.6} 
\begin{table*}[t] \footnotesize
	\begin{ruledtabular}
		\centering
		\begin{tabular}{*{13}{c}} 
			\thd{Subject}    & \thd{Layer} & \thd{$N$} & \thd{$L$} & \thd{$\rho_l$} & \thd{$\langle k\rangle\pm \text{SEM}$}  &  \thd{$\langle c\rangle\pm \text{SEM}$}  & \thd{$f^{(l)}\pm \text{SEM}$} &\thd{$r_c$} &\thd{$Q$}&\thd{$N_c$} & \thd{$\beta$} & \thd{$\mu$}\\
			\hline
			\multirow{5}{*}{No. 19}  & 0      & 1014    & 41676       & 0.08        & 82.20$\pm$1.639   & 0.42$\pm$0.004    & 57.88$\pm$40.396  & -0.035     & 0.39       & 3          & 1.87       & 0.004     \\ 
			& 1      & 462     & 19773       & 0.19        & 85.60$\pm$1.871   & 0.48$\pm$0.004    & 65.85$\pm$40.966  & -0.031     & 0.32       & 3          & 2.11        & 0.004     \\ 
			& 2      & 233     & 9225        & 0.34        & 79.18$\pm$1.837   & 0.57$\pm$0.005    & 72.99$\pm$42.182  & -0.042     & 0.24       & 3          & 2.29        & 0.005     \\ 
			& 3      & 128     & 4226        & 0.52        & 66.03$\pm$1.615   & 0.68$\pm$0.006    & 78.64$\pm$43.673  & -0.065     & 0.17       & 3          & 3.00        & 0.006     \\ 
			& 4      & 82      & 2066        & 0.62        & 50.39$\pm$1.556   & 0.76$\pm$0.008    & 78.10$\pm$45.409  & -0.112     & 0.12       & 2          & 3.42        & 0.009     \\ 
			\hline 
			\multirow{5}{*}{No. 20}  & 0      & 1013    & 41727       & 0.08        & 82.38$\pm$1.620   & 0.43$\pm$0.004    & 52.87$\pm$37.651  & -0.039     & 0.40       & 4          & 1.95       & 0.004     \\ 
			& 1      & 462     & 19757       & 0.19        & 85.53$\pm$1.800   & 0.48$\pm$0.005    & 60.55$\pm$38.512  & -0.032     & 0.31       & 3          & 2.15        & 0.004     \\ 
			& 2      & 233     & 9241        & 0.34        & 79.32$\pm$1.723   & 0.57$\pm$0.006    & 67.10$\pm$39.748  & -0.034     & 0.26       & 3          & 2.55        & 0.005     \\ 
			& 3      & 128     & 4076        & 0.50        & 63.69$\pm$1.549   & 0.67$\pm$0.006    & 71.76$\pm$41.387  & -0.048     & 0.19       & 2          & 3.09        & 0.007     \\ 
			& 4      & 82      & 1972        & 0.59        & 48.10$\pm$1.455   & 0.74$\pm$0.008    & 70.69$\pm$43.019  & -0.087     & 0.15       & 2          & 3.58        & 0.009     \\ 
			\hline 
			\multirow{5}{*}{No. 21}  & 0      & 1014    & 36659       & 0.07        & 72.31$\pm$1.553   & 0.43$\pm$0.004    & 55.79$\pm$40.647  & -0.041     & 0.43       & 4          & 1.89       & 0.004     \\ 
			& 1      & 462     & 17485       & 0.16        & 75.69$\pm$1.793   & 0.48$\pm$0.005    & 63.86$\pm$41.672  & -0.047     & 0.33       & 3          & 2.07        & 0.004     \\ 
			& 2      & 233     & 8392        & 0.31        & 72.03$\pm$1.760   & 0.56$\pm$0.006    & 70.54$\pm$42.918  & -0.058     & 0.26       & 3          & 2.35        & 0.005     \\ 
			& 3      & 128     & 3884        & 0.48        & 60.69$\pm$1.557   & 0.65$\pm$0.006    & 75.36$\pm$44.246  & -0.083     & 0.17       & 3          & 2.71        & 0.007     \\ 
			& 4      & 82      & 1995        & 0.60        & 48.66$\pm$1.407   & 0.73$\pm$0.007    & 76.37$\pm$46.385  & -0.103     & 0.13       & 2          & 2.72        & 0.008     \\ 
			\hline 
			\multirow{5}{*}{No. 22}  & 0      & 1014    & 38855       & 0.08        & 76.64$\pm$1.514   & 0.40$\pm$0.004    & 65.81$\pm$46.676  & -0.034     & 0.41       & 4          & 1.78       & 0.004     \\ 
			& 1      & 462     & 18783       & 0.18        & 81.31$\pm$1.746   & 0.46$\pm$0.004    & 75.07$\pm$47.732  & -0.031     & 0.29       & 3          & 2.02        & 0.004     \\ 
			& 2      & 233     & 9072        & 0.34        & 77.87$\pm$1.721   & 0.56$\pm$0.005    & 82.94$\pm$48.893  & -0.048     & 0.23       & 3          & 2.31        & 0.005     \\ 
			& 3      & 128     & 4037        & 0.50        & 63.08$\pm$1.590   & 0.67$\pm$0.006    & 87.39$\pm$50.584  & -0.052     & 0.18       & 3          & 2.90        & 0.006     \\ 
			& 4      & 82      & 1966        & 0.59        & 47.95$\pm$1.553   & 0.75$\pm$0.008    & 86.89$\pm$52.990  & -0.091     & 0.12       & 3          & 2.79        & 0.008     \\ 
			\hline 
			\multirow{5}{*}{No. 23}  & 0      & 1014    & 37235       & 0.07        & 73.44$\pm$1.444   & 0.40$\pm$0.004    & 60.43$\pm$43.800  & -0.039     & 0.42       & 4          & 1.87       & 0.004     \\ 
			& 1      & 462     & 17532       & 0.16        & 75.90$\pm$1.645   & 0.46$\pm$0.004    & 68.93$\pm$44.649  & -0.039     & 0.34       & 3          & 2.06        & 0.004     \\ 
			& 2      & 233     & 8467        & 0.31        & 72.68$\pm$1.607   & 0.54$\pm$0.005    & 76.18$\pm$45.622  & -0.049     & 0.27       & 3          & 2.39        & 0.005     \\ 
			& 3      & 128     & 3906        & 0.48        & 61.03$\pm$1.388   & 0.64$\pm$0.005    & 80.67$\pm$46.705  & -0.056     & 0.20       & 3          & 2.95        & 0.007     \\ 
			& 4      & 82      & 1929        & 0.58        & 47.05$\pm$1.286   & 0.71$\pm$0.007    & 80.18$\pm$48.393  & -0.091     & 0.17       & 2          & 3.66        & 0.009     \\ 
			\hline 
			\multirow{5}{*}{No. 24}  & 0      & 1014    & 38407       & 0.07        & 75.75$\pm$1.534   & 0.41$\pm$0.004    & 59.11$\pm$42.707  & -0.040     & 0.39       & 3          & 1.90       & 0.004     \\ 
			& 1      & 462     & 18295       & 0.17        & 79.20$\pm$1.749   & 0.47$\pm$0.004    & 67.13$\pm$43.616  & -0.034     & 0.33       & 3          & 2.08        & 0.004     \\ 
			& 2      & 233     & 8666        & 0.32        & 74.39$\pm$1.687   & 0.56$\pm$0.005    & 73.38$\pm$44.184  & -0.042     & 0.26       & 3          & 2.48        & 0.005     \\ 
			& 3      & 128     & 3926        & 0.48        & 61.34$\pm$1.465   & 0.65$\pm$0.006    & 78.14$\pm$46.105  & -0.049     & 0.19       & 2          & 3.08        & 0.007     \\ 
			& 4      & 82      & 1919        & 0.58        & 46.80$\pm$1.385   & 0.73$\pm$0.008    & 76.37$\pm$46.714  & -0.096     & 0.16       & 2          & 3.57        & 0.009     \\ 
			\hline 
			\multirow{5}{*}{No. 25}  & 0      & 1014    & 43631       & 0.08        & 86.06$\pm$1.659   & 0.42$\pm$0.003    & 57.71$\pm$40.073  & -0.033     & 0.42       & 4          & 1.90       & 0.004     \\ 
			& 1      & 462     & 20366       & 0.19        & 88.16$\pm$1.833   & 0.49$\pm$0.004    & 65.79$\pm$41.023  & -0.021     & 0.30       & 3          & 2.10        & 0.004     \\ 
			& 2      & 233     & 9329        & 0.35        & 80.08$\pm$1.812   & 0.58$\pm$0.006    & 72.29$\pm$41.947  & -0.027     & 0.25       & 3          & 2.56        & 0.005     \\ 
			& 3      & 128     & 4144        & 0.51        & 64.75$\pm$1.651   & 0.69$\pm$0.007    & 77.46$\pm$43.752  & -0.047     & 0.18       & 3          & 3.33        & 0.007     \\ 
			& 4      & 82      & 2033        & 0.61        & 49.59$\pm$1.543   & 0.76$\pm$0.009    & 76.84$\pm$45.405  & -0.100     & 0.14       & 2          & 3.76        & 0.009     \\ 
			\hline 
			\multirow{5}{*}{No. 26}  & 0      & 1014    & 42351       & 0.08        & 83.53$\pm$1.634   & 0.42$\pm$0.004    & 57.06$\pm$40.212  & -0.021     & 0.40       & 4          & 1.90       & 0.004     \\ 
			& 1      & 462     & 20154       & 0.19        & 87.25$\pm$1.843   & 0.48$\pm$0.004    & 65.47$\pm$41.238  & -0.010     & 0.32       & 3          & 2.11        & 0.004     \\ 
			& 2      & 233     & 9465        & 0.35        & 81.24$\pm$1.745   & 0.57$\pm$0.005    & 72.29$\pm$42.137  & -0.011     & 0.23       & 3          & 2.50        & 0.005     \\ 
			& 3      & 128     & 4203        & 0.52        & 65.67$\pm$1.511   & 0.67$\pm$0.005    & 77.28$\pm$43.793  & -0.023     & 0.17       & 3          & 3.01        & 0.006     \\ 
			& 4      & 82      & 2090        & 0.63        & 50.98$\pm$1.412   & 0.75$\pm$0.007    & 77.41$\pm$45.244  & -0.095     & 0.14       & 2          & 2.82        & 0.008     \\ 
			\hline 
			\multirow{5}{*}{No. 27}  & 0      & 1014    & 37875       & 0.07        & 74.70$\pm$1.469   & 0.42$\pm$0.004    & 60.28$\pm$44.013  & -0.025     & 0.43       & 4          & 1.91       & 0.004     \\ 
			& 1      & 462     & 17836       & 0.17        & 77.21$\pm$1.660   & 0.47$\pm$0.004    & 69.29$\pm$45.337  & -0.020     & 0.31       & 3          & 2.14        & 0.004     \\ 
			& 2      & 233     & 8418        & 0.31        & 72.26$\pm$1.619   & 0.56$\pm$0.006    & 75.70$\pm$45.756  & -0.035     & 0.28       & 3          & 2.53        & 0.005     \\ 
			& 3      & 128     & 3834        & 0.47        & 59.91$\pm$1.434   & 0.65$\pm$0.007    & 80.85$\pm$47.362  & -0.047     & 0.22       & 2          & 2.96        & 0.007     \\ 
			& 4      & 82      & 1889        & 0.57        & 46.07$\pm$1.366   & 0.73$\pm$0.009    & 79.79$\pm$49.279  & -0.086     & 0.18       & 2          & 3.25        & 0.009     \\ 
			\hline 
			\multirow{5}{*}{No. 28}  & 0      & 1014    & 35737       & 0.07        & 70.49$\pm$1.478   & 0.42$\pm$0.004    & 56.48$\pm$41.154  & -0.040     & 0.43       & 4          & 1.86       & 0.004     \\ 
			& 1      & 462     & 16984       & 0.16        & 73.52$\pm$1.708   & 0.47$\pm$0.005    & 64.04$\pm$42.062  & -0.039     & 0.33       & 3          & 2.10        & 0.005     \\ 
			& 2      & 233     & 8227        & 0.30        & 70.62$\pm$1.707   & 0.54$\pm$0.006    & 70.74$\pm$43.134  & -0.042     & 0.26       & 3          & 2.36        & 0.005     \\ 
			& 3      & 128     & 3714        & 0.46        & 58.03$\pm$1.579   & 0.65$\pm$0.007    & 74.93$\pm$44.203  & -0.042     & 0.18       & 2          & 2.97        & 0.007     \\ 
			& 4      & 82      & 1844        & 0.56        & 44.98$\pm$1.547   & 0.73$\pm$0.009    & 74.89$\pm$45.887  & -0.077     & 0.14       & 3          & 3.37        & 0.010     \\ 			
		\end{tabular}
	\end{ruledtabular}
\end{table*}

\renewcommand{\arraystretch}{0.6} 
\begin{table*}[t] \footnotesize
	\begin{ruledtabular}
		\centering			
		\begin{tabular}{*{13}{c}} 
			\thd{Subject}    & \thd{Layer} & \thd{$N$} & \thd{$L$} & \thd{$\rho_l$} & \thd{$\langle k\rangle\pm \text{SEM}$}  &  \thd{$\langle c\rangle\pm \text{SEM}$}  & \thd{$f^{(l)}\pm \text{SEM}$} &\thd{$r_c$} &\thd{$Q$} &\thd{$N_c$}& \thd{$\beta$} & \thd{$\mu$}\\
			\hline
			\multirow{5}{*}{No. 29}  & 0      & 1014    & 39125       & 0.08        & 77.17$\pm$1.492   & 0.42$\pm$0.004    & 53.54$\pm$38.013  & -0.033     & 0.45       & 4          & 1.86       & 0.004     \\ 
			& 1      & 462     & 18508       & 0.17        & 80.12$\pm$1.681   & 0.48$\pm$0.004    & 61.19$\pm$39.093  & -0.030     & 0.34       & 3          & 2.19        & 0.004     \\ 
			& 2      & 233     & 8648        & 0.32        & 74.23$\pm$1.640   & 0.57$\pm$0.006    & 67.30$\pm$40.065  & -0.048     & 0.26       & 3          & 2.66        & 0.005     \\ 
			& 3      & 128     & 3898        & 0.48        & 60.91$\pm$1.427   & 0.65$\pm$0.007    & 71.73$\pm$41.755  & -0.054     & 0.21       & 2          & 3.02        & 0.007     \\ 
			& 4      & 82      & 1934        & 0.58        & 47.17$\pm$1.354   & 0.72$\pm$0.008    & 71.67$\pm$43.399  & -0.099     & 0.16       & 2          & 3.40        & 0.009     \\ 
			\hline 
			\multirow{5}{*}{No. 30}  & 0      & 1014    & 41041       & 0.08        & 80.95$\pm$1.596   & 0.43$\pm$0.004    & 55.82$\pm$40.595  & -0.033     & 0.40       & 3          & 1.94       & 0.004     \\ 
			& 1      & 462     & 19264       & 0.18        & 83.39$\pm$1.793   & 0.48$\pm$0.004    & 64.27$\pm$41.937  & -0.030     & 0.34       & 3          & 2.11        & 0.004     \\ 
			& 2      & 233     & 9012        & 0.33        & 77.36$\pm$1.736   & 0.57$\pm$0.006    & 71.01$\pm$43.245  & -0.037     & 0.27       & 3          & 2.45        & 0.005     \\ 
			& 3      & 128     & 4097        & 0.50        & 64.02$\pm$1.548   & 0.66$\pm$0.006    & 76.38$\pm$44.786  & -0.059     & 0.20       & 2          & 3.09        & 0.007     \\ 
			& 4      & 82      & 2034        & 0.61        & 49.61$\pm$1.457   & 0.75$\pm$0.008    & 76.21$\pm$46.418  & -0.096     & 0.15       & 2          & 2.78        & 0.008     \\ 
			\hline 
			\multirow{5}{*}{No. 31}  & 0      & 1014    & 39813       & 0.08        & 78.53$\pm$1.582   & 0.42$\pm$0.004    & 61.38$\pm$43.914  & -0.036     & 0.40       & 4          & 1.84       & 0.004     \\ 
			& 1      & 462     & 18826       & 0.18        & 81.50$\pm$1.805   & 0.47$\pm$0.004    & 70.46$\pm$45.286  & -0.022     & 0.32       & 3          & 2.06        & 0.004     \\ 
			& 2      & 233     & 8861        & 0.33        & 76.06$\pm$1.811   & 0.57$\pm$0.006    & 77.90$\pm$46.474  & -0.030     & 0.24       & 2          & 2.50        & 0.005     \\ 
			& 3      & 128     & 3949        & 0.49        & 61.70$\pm$1.655   & 0.67$\pm$0.007    & 82.78$\pm$47.844  & -0.051     & 0.19       & 2          & 3.16        & 0.007     \\ 
			& 4      & 82      & 1914        & 0.58        & 46.68$\pm$1.539   & 0.75$\pm$0.009    & 81.71$\pm$50.231  & -0.081     & 0.15       & 2          & 3.79        & 0.010     \\ 
			\hline 
			\multirow{5}{*}{No. 32}  & 0      & 1013    & 39161       & 0.08        & 77.32$\pm$1.574   & 0.43$\pm$0.004    & 61.96$\pm$43.540  & -0.033     & 0.43       & 4          & 1.91       & 0.004     \\ 
			& 1      & 462     & 18393       & 0.17        & 79.62$\pm$1.795   & 0.49$\pm$0.005    & 70.68$\pm$44.698  & -0.027     & 0.30       & 3          & 2.20        & 0.004     \\ 
			& 2      & 233     & 8649        & 0.32        & 74.24$\pm$1.768   & 0.57$\pm$0.006    & 78.46$\pm$46.346  & -0.030     & 0.27       & 3          & 2.72        & 0.005     \\ 
			& 3      & 128     & 3827        & 0.47        & 59.80$\pm$1.607   & 0.66$\pm$0.007    & 82.92$\pm$48.538  & -0.058     & 0.20       & 2          & 2.99        & 0.007     \\ 
			& 4      & 82      & 1931        & 0.58        & 47.10$\pm$1.472   & 0.74$\pm$0.008    & 84.74$\pm$52.183  & -0.096     & 0.15       & 2          & 3.22        & 0.009     \\ 
			\hline 
			\multirow{5}{*}{No. 33}  & 0      & 1014    & 40400       & 0.08        & 79.68$\pm$1.602   & 0.41$\pm$0.004    & 60.92$\pm$43.352  & -0.034     & 0.41       & 4          & 1.90       & 0.004     \\ 
			& 1      & 462     & 19127       & 0.18        & 82.80$\pm$1.801   & 0.47$\pm$0.004    & 69.55$\pm$44.780  & -0.026     & 0.29       & 3          & 2.04        & 0.004     \\ 
			& 2      & 233     & 9119        & 0.34        & 78.27$\pm$1.754   & 0.56$\pm$0.005    & 77.32$\pm$46.285  & -0.028     & 0.26       & 3          & 2.29        & 0.005     \\ 
			& 3      & 128     & 4058        & 0.50        & 63.41$\pm$1.579   & 0.66$\pm$0.006    & 82.08$\pm$47.901  & -0.041     & 0.18       & 3          & 2.74        & 0.006     \\ 
			& 4      & 82      & 1988        & 0.60        & 48.49$\pm$1.507   & 0.75$\pm$0.008    & 82.25$\pm$49.821  & -0.072     & 0.15       & 2          & 2.73        & 0.008     \\ 
			\hline 
			\multirow{5}{*}{No. 34}  & 0      & 1014    & 44274       & 0.09        & 87.33$\pm$1.689   & 0.42$\pm$0.003    & 57.56$\pm$39.889  & -0.036     & 0.40       & 3          & 1.87       & 0.003     \\ 
			& 1      & 462     & 20590       & 0.19        & 89.13$\pm$1.871   & 0.49$\pm$0.004    & 65.60$\pm$40.953  & -0.025     & 0.33       & 3          & 2.11        & 0.004     \\ 
			& 2      & 233     & 9377        & 0.35        & 80.49$\pm$1.811   & 0.58$\pm$0.005    & 72.08$\pm$41.786  & -0.032     & 0.26       & 3          & 2.49        & 0.005     \\ 
			& 3      & 128     & 4226        & 0.52        & 66.03$\pm$1.547   & 0.68$\pm$0.006    & 77.22$\pm$43.408  & -0.037     & 0.18       & 3          & 3.22        & 0.006     \\ 
			& 4      & 82      & 2073        & 0.62        & 50.56$\pm$1.417   & 0.76$\pm$0.008    & 76.60$\pm$45.167  & -0.084     & 0.14       & 2          & 3.78        & 0.009     \\ 
			\hline 
			\multirow{5}{*}{No. 35}  & 0      & 1014    & 46025       & 0.09        & 90.78$\pm$1.718   & 0.42$\pm$0.004    & 59.63$\pm$41.389  & -0.039     & 0.42       & 4          & 1.92       & 0.003     \\ 
			& 1      & 462     & 21758       & 0.20        & 94.19$\pm$1.866   & 0.49$\pm$0.005    & 67.87$\pm$42.403  & -0.028     & 0.32       & 4          & 2.19        & 0.004     \\ 
			& 2      & 233     & 10004       & 0.37        & 85.87$\pm$1.779   & 0.58$\pm$0.005    & 75.13$\pm$43.948  & -0.026     & 0.23       & 3          & 2.32        & 0.004     \\ 
			& 3      & 128     & 4407        & 0.54        & 68.86$\pm$1.552   & 0.68$\pm$0.005    & 80.19$\pm$45.905  & -0.030     & 0.15       & 4          & 2.85        & 0.006     \\ 
			& 4      & 82      & 2082        & 0.63        & 50.78$\pm$1.475   & 0.76$\pm$0.007    & 78.59$\pm$47.181  & -0.079     & 0.14       & 2          & 3.11        & 0.008     \\ 
			\hline 
			\multirow{5}{*}{No. 36}  & 0      & 1014    & 37237       & 0.07        & 73.45$\pm$1.452   & 0.40$\pm$0.004    & 66.00$\pm$47.148  & -0.034     & 0.42       & 4          & 1.84       & 0.004     \\ 
			& 1      & 462     & 18008       & 0.17        & 77.96$\pm$1.669   & 0.46$\pm$0.004    & 74.83$\pm$47.827  & -0.032     & 0.33       & 3          & 2.06        & 0.004     \\ 
			& 2      & 233     & 8643        & 0.32        & 74.19$\pm$1.670   & 0.55$\pm$0.006    & 81.89$\pm$48.483  & -0.037     & 0.24       & 3          & 2.34        & 0.005     \\ 
			& 3      & 128     & 3839        & 0.47        & 59.98$\pm$1.594   & 0.66$\pm$0.007    & 85.54$\pm$49.436  & -0.059     & 0.18       & 3          & 3.25        & 0.007     \\ 
			& 4      & 82      & 1882        & 0.57        & 45.90$\pm$1.575   & 0.74$\pm$0.010    & 84.52$\pm$50.913  & -0.105     & 0.14       & 2          & 3.89        & 0.010     \\ 
			\hline 
			\multirow{5}{*}{No. 37}  & 0      & 1014    & 45512       & 0.09        & 89.77$\pm$1.754   & 0.43$\pm$0.003    & 62.17$\pm$43.255  & -0.034     & 0.37       & 3          & 1.90       & 0.003     \\ 
			& 1      & 462     & 21300       & 0.20        & 92.21$\pm$1.907   & 0.49$\pm$0.004    & 71.61$\pm$44.528  & -0.021     & 0.33       & 3          & 2.14        & 0.004     \\ 
			& 2      & 233     & 9778        & 0.36        & 83.93$\pm$1.777   & 0.59$\pm$0.005    & 79.40$\pm$46.037  & -0.030     & 0.26       & 3          & 2.49        & 0.005     \\ 
			& 3      & 128     & 4329        & 0.53        & 67.64$\pm$1.522   & 0.68$\pm$0.005    & 84.30$\pm$47.821  & -0.045     & 0.19       & 2          & 2.97        & 0.006     \\ 
			& 4      & 82      & 2081        & 0.63        & 50.76$\pm$1.403   & 0.76$\pm$0.007    & 82.28$\pm$48.698  & -0.083     & 0.15       & 2          & 4.09        & 0.009     \\ 
			\hline 
			\multirow{5}{*}{No. 38}  & 0      & 1014    & 40209       & 0.08        & 79.31$\pm$1.606   & 0.41$\pm$0.004    & 62.03$\pm$44.458  & -0.034     & 0.39       & 3          & 1.86       & 0.004     \\ 
			& 1      & 462     & 19094       & 0.18        & 82.66$\pm$1.810   & 0.47$\pm$0.004    & 71.12$\pm$45.812  & -0.029     & 0.33       & 3          & 2.05        & 0.004     \\ 
			& 2      & 233     & 9038        & 0.33        & 77.58$\pm$1.724   & 0.55$\pm$0.005    & 79.05$\pm$47.440  & -0.029     & 0.22       & 2          & 2.39        & 0.005     \\ 
			& 3      & 128     & 4024        & 0.50        & 62.88$\pm$1.606   & 0.66$\pm$0.006    & 84.41$\pm$49.498  & -0.039     & 0.16       & 3          & 2.81        & 0.006     \\ 
			& 4      & 82      & 1970        & 0.59        & 48.05$\pm$1.523   & 0.75$\pm$0.008    & 84.05$\pm$51.966  & -0.052     & 0.14       & 2          & 2.87        & 0.008     \\ 
		\end{tabular}
	\end{ruledtabular}
\end{table*}

\renewcommand{\arraystretch}{0.6} 
\begin{table*}[t] \footnotesize
	\begin{ruledtabular}
		\centering
		\begin{tabular}{*{13}{c}} 
			\thd{Subject}    & \thd{Layer} & \thd{$N$} & \thd{$L$} & \thd{$\rho_l$} & \thd{$\langle k\rangle\pm \text{SEM}$}  &  \thd{$\langle c\rangle\pm \text{SEM}$}  & \thd{$f^{(l)}\pm \text{SEM}$} &\thd{$r_c$} &\thd{$Q$} &\thd{$N_c$} & \thd{$\beta$} & \thd{$\mu$}\\
			\hline
			\multirow{5}{*}{No. 39}  & 0      & 1014    & 34947       & 0.07        & 68.93$\pm$1.521   & 0.43$\pm$0.004    & 56.48$\pm$40.483  & -0.047     & 0.44       & 4          & 1.89       & 0.004     \\ 
			& 1      & 462     & 16844       & 0.16        & 72.92$\pm$1.755   & 0.48$\pm$0.005    & 64.16$\pm$41.220  & -0.041     & 0.36       & 4          & 2.09        & 0.005     \\ 
			& 2      & 233     & 8121        & 0.30        & 69.71$\pm$1.767   & 0.56$\pm$0.006    & 70.91$\pm$42.134  & -0.051     & 0.25       & 3          & 2.33        & 0.005     \\ 
			& 3      & 128     & 3703        & 0.46        & 57.86$\pm$1.619   & 0.65$\pm$0.008    & 74.79$\pm$43.341  & -0.062     & 0.18       & 3          & 2.87        & 0.007     \\ 
			& 4      & 82      & 1877        & 0.57        & 45.78$\pm$1.542   & 0.73$\pm$0.009    & 73.59$\pm$44.114  & -0.090     & 0.13       & 3          & 3.23        & 0.009     \\ 
			\hline 
			\multirow{5}{*}{No. 40}  & 0      & 1014    & 38384       & 0.07        & 75.71$\pm$1.547   & 0.41$\pm$0.004    & 55.33$\pm$40.100  & -0.037     & 0.41       & 4          & 1.85       & 0.004     \\ 
			& 1      & 462     & 18201       & 0.17        & 78.79$\pm$1.786   & 0.47$\pm$0.005    & 63.53$\pm$41.379  & -0.032     & 0.29       & 3          & 2.02        & 0.004     \\ 
			& 2      & 233     & 8652        & 0.32        & 74.27$\pm$1.753   & 0.56$\pm$0.005    & 70.13$\pm$42.407  & -0.032     & 0.26       & 3          & 2.42        & 0.005     \\ 
			& 3      & 128     & 3892        & 0.48        & 60.81$\pm$1.611   & 0.66$\pm$0.007    & 73.98$\pm$43.283  & -0.040     & 0.19       & 2          & 2.95        & 0.007     \\ 
			& 4      & 82      & 1916        & 0.58        & 46.73$\pm$1.541   & 0.74$\pm$0.009    & 72.01$\pm$43.377  & -0.079     & 0.15       & 2          & 3.00        & 0.009     \\ 
			\hline 
			\multirow{5}{*}{No. 41}  & 0      & 1014    & 35161       & 0.07        & 69.35$\pm$1.455   & 0.42$\pm$0.004    & 56.84$\pm$41.915  & -0.037     & 0.42       & 4          & 1.90       & 0.004     \\ 
			& 1      & 462     & 16812       & 0.16        & 72.78$\pm$1.695   & 0.46$\pm$0.005    & 64.84$\pm$43.044  & -0.036     & 0.30       & 4          & 1.96        & 0.004     \\ 
			& 2      & 233     & 8086        & 0.30        & 69.41$\pm$1.727   & 0.55$\pm$0.006    & 71.68$\pm$44.498  & -0.050     & 0.26       & 3          & 2.32        & 0.005     \\ 
			& 3      & 128     & 3751        & 0.46        & 58.61$\pm$1.603   & 0.65$\pm$0.007    & 75.60$\pm$45.089  & -0.058     & 0.19       & 3          & 2.77        & 0.007     \\ 
			& 4      & 82      & 1886        & 0.57        & 46.00$\pm$1.542   & 0.74$\pm$0.009    & 75.09$\pm$46.191  & -0.100     & 0.13       & 3          & 2.98        & 0.009     \\ 
			\hline 
			\multirow{5}{*}{No. 42}  & 0      & 1014    & 40731       & 0.08        & 80.34$\pm$1.544   & 0.42$\pm$0.004    & 58.24$\pm$42.013  & -0.037     & 0.42       & 4          & 1.90       & 0.004     \\ 
			& 1      & 462     & 19399       & 0.18        & 83.98$\pm$1.731   & 0.47$\pm$0.005    & 66.42$\pm$43.086  & -0.033     & 0.34       & 3          & 2.13        & 0.004     \\ 
			& 2      & 233     & 9190        & 0.34        & 78.88$\pm$1.677   & 0.56$\pm$0.006    & 73.32$\pm$43.902  & -0.035     & 0.27       & 3          & 2.53        & 0.005     \\ 
			& 3      & 128     & 4147        & 0.51        & 64.80$\pm$1.537   & 0.67$\pm$0.006    & 79.48$\pm$46.699  & -0.052     & 0.19       & 3          & 3.08        & 0.006     \\ 
			& 4      & 82      & 2057        & 0.62        & 50.17$\pm$1.447   & 0.76$\pm$0.008    & 80.02$\pm$49.576  & -0.070     & 0.15       & 2          & 4.14        & 0.009     \\ 
			\hline 
			\multirow{5}{*}{No. 43}  & 0      & 1014    & 37845       & 0.07        & 74.64$\pm$1.535   & 0.43$\pm$0.004    & 56.27$\pm$40.721  & -0.034     & 0.44       & 4          & 1.95       & 0.004     \\ 
			& 1      & 462     & 17927       & 0.17        & 77.61$\pm$1.744   & 0.49$\pm$0.005    & 64.87$\pm$42.105  & -0.034     & 0.32       & 3          & 2.19        & 0.004     \\ 
			& 2      & 233     & 8449        & 0.31        & 72.52$\pm$1.728   & 0.57$\pm$0.006    & 71.61$\pm$43.110  & -0.047     & 0.27       & 4          & 2.64        & 0.005     \\ 
			& 3      & 128     & 3869        & 0.48        & 60.45$\pm$1.536   & 0.66$\pm$0.007    & 77.54$\pm$44.988  & -0.057     & 0.16       & 3          & 2.81        & 0.007     \\ 
			& 4      & 82      & 1898        & 0.57        & 46.29$\pm$1.416   & 0.73$\pm$0.008    & 76.80$\pm$46.405  & -0.088     & 0.12       & 2          & 3.39        & 0.009     \\ 			
		\end{tabular}
	\end{ruledtabular}
\end{table*}

\clearpage
\newpage
\renewcommand{\arraystretch}{0.55} 

\begin{table*}
	\begin{ruledtabular}
		\centering	
		\caption{The dispersion between empirical subjects for degrees, number of triangles of nodes, and sum of the degrees of neighbors in layer $0$ for UL dataset. For each brain region, we calculate the mean and standard deviation $\sigma$ of the three quantites over all subjects. Then we obtain the pearson correlation coefficient $\rho$ and the $\chi^2$ test ($\chi^2=\sum_{i}^{N}(\frac{value_{real}-value_{group}}{\sigma_{group}})^2$) between specific subject and the average in the cohort. The quantity $\zeta$ corresponds to the fraction of nodes for which the value measured on the specific network lies outside the $2\sigma$ confidence interval around the average.}			
		\begin{tabular}{*{10}{c}}
			\multirow{2}{*}{Subject} &		
			\multicolumn{3}{c}{degree} &
			\multicolumn{3}{c}{number of triangles} &
			\multicolumn{3}{c}{sum degree of neighbors}\\
			\cline{2-4}
			\cline{5-7}
			\cline{8-10}	
			& {$\rho$} & {$\chi^2/N$} & {$\zeta$} 
			& {$\rho$} & {$\chi^2/N$} & {$\zeta$} 
			& {$\rho$} & {$\chi^2/N$} & {$\zeta$}\\
			\midrule
			0     &0.812 &1.034 &0.056 &0.817 &1.117 &0.067 &0.789 &0.865 &0.037  \\ 
			1     &0.826 &1.018 &0.050 &0.847 &1.075 &0.062 &0.787 &1.170 &0.066  \\ 
			2     &0.795 &0.959 &0.037 &0.825 &0.916 &0.044 &0.773 &0.835 &0.026  \\ 
			3     &0.823 &0.862 &0.028 &0.853 &0.692 &0.015 &0.820 &0.716 &0.007  \\ 
			4     &0.852 &1.031 &0.060 &0.891 &1.084 &0.064 &0.841 &1.352 &0.088  \\ 
			5     &0.859 &0.850 &0.032 &0.890 &0.719 &0.025 &0.831 &0.823 &0.037  \\ 
			6     &0.818 &0.916 &0.032 &0.835 &0.841 &0.034 &0.763 &0.864 &0.024  \\ 
			7     &0.853 &0.909 &0.033 &0.874 &0.849 &0.033 &0.815 &0.870 &0.035  \\ 
			8     &0.770 &1.121 &0.061 &0.799 &1.080 &0.055 &0.756 &1.033 &0.051  \\ 
			9     &0.845 &0.737 &0.024 &0.856 &0.676 &0.019 &0.822 &0.659 &0.011  \\ 
			10    &0.813 &0.983 &0.055 &0.817 &1.106 &0.066 &0.789 &0.926 &0.043  \\ 
			11    &0.784 &1.154 &0.056 &0.810 &1.181 &0.057 &0.735 &1.175 &0.060  \\ 
			12    &0.780 &1.044 &0.049 &0.804 &1.079 &0.057 &0.756 &0.977 &0.045  \\ 
			13    &0.812 &1.118 &0.055 &0.840 &1.228 &0.067 &0.783 &1.240 &0.070  \\ 
			14    &0.830 &1.057 &0.052 &0.846 &1.194 &0.076 &0.815 &1.080 &0.062  \\ 
			15    &0.818 &1.271 &0.077 &0.844 &1.726 &0.121 &0.787 &1.499 &0.106  \\ 
			16    &0.849 &0.895 &0.036 &0.878 &0.891 &0.040 &0.818 &0.994 &0.049  \\ 
			17    &0.831 &0.907 &0.036 &0.844 &0.850 &0.038 &0.795 &0.844 &0.032  \\ 
			18    &0.834 &0.906 &0.026 &0.850 &0.762 &0.026 &0.792 &0.806 &0.021  \\ 
			19    &0.805 &1.153 &0.067 &0.837 &1.119 &0.072 &0.779 &1.202 &0.071  \\ 
			20    &0.759 &1.181 &0.058 &0.784 &1.043 &0.047 &0.724 &1.002 &0.036  \\ 
			21    &0.757 &1.229 &0.051 &0.761 &1.060 &0.043 &0.695 &1.025 &0.026  \\ 
			22    &0.834 &0.798 &0.028 &0.849 &0.698 &0.023 &0.814 &0.699 &0.019  \\ 
			23    &0.793 &1.116 &0.058 &0.784 &1.362 &0.081 &0.744 &1.125 &0.062  \\ 
			24    &0.825 &0.910 &0.041 &0.841 &0.823 &0.038 &0.797 &0.850 &0.030  \\ 
			25    &0.838 &0.982 &0.045 &0.847 &1.152 &0.068 &0.801 &1.293 &0.077  \\ 
			26    &0.796 &1.163 &0.058 &0.791 &1.137 &0.045 &0.717 &1.118 &0.043  \\ 
			27    &0.808 &0.848 &0.026 &0.830 &0.718 &0.016 &0.777 &0.768 &0.014  \\ 
			28    &0.822 &1.079 &0.055 &0.831 &1.268 &0.073 &0.775 &1.391 &0.084  \\ 
			29    &0.814 &0.912 &0.028 &0.839 &0.837 &0.036 &0.785 &0.790 &0.022  \\ 
			30    &0.848 &0.875 &0.035 &0.879 &0.891 &0.051 &0.816 &0.912 &0.043  \\ 
			31    &0.815 &1.231 &0.074 &0.842 &1.474 &0.090 &0.778 &1.580 &0.114  \\ 
			32    &0.807 &0.958 &0.033 &0.811 &0.874 &0.027 &0.751 &0.942 &0.032  \\ 
			33    &0.821 &1.075 &0.055 &0.847 &1.169 &0.062 &0.813 &1.037 &0.049  \\ 
			34    &0.815 &1.088 &0.057 &0.831 &1.055 &0.054 &0.789 &1.104 &0.058  \\ 
			35    &0.851 &0.809 &0.025 &0.863 &0.706 &0.018 &0.816 &0.785 &0.013  \\ 
			36    &0.832 &0.958 &0.043 &0.837 &0.988 &0.058 &0.814 &0.924 &0.044  \\ 
			37    &0.834 &0.908 &0.035 &0.852 &0.812 &0.037 &0.808 &0.899 &0.038  \\ 
			38    &0.804 &1.021 &0.048 &0.838 &0.983 &0.049 &0.792 &0.968 &0.033  \\ 
			39    &0.848 &0.937 &0.029 &0.881 &0.773 &0.022 &0.806 &0.860 &0.026  \\ 
			
		\end{tabular}
	\end{ruledtabular}
\end{table*}

\clearpage
\newpage
\renewcommand{\arraystretch}{0.5} 

\begin{table*}
	\begin{ruledtabular}
		\centering	
		\caption{The dispersion between empirical subjects for degrees, number of triangles of nodes, and sum of the degrees of neighbors in layer $0$ for HCP dataset. For each brain region, we calculate the mean and standard deviation $\sigma$ of the three quantites over all subjects. Then we obtain the pearson correlation coefficient $\rho$ and the $\chi^2$ test ($\chi^2=\sum_{i}^{N}(\frac{value_{real}-value_{group}}{\sigma_{group}})^2$) between specific subject and the average in the cohort. The quantity $\zeta$ corresponds to the fraction of nodes for which the value measured on the specific network lies outside the $2\sigma$ confidence interval around the average.}			
		\begin{tabular}{*{10}{c}}
			\multirow{2}{*}{Subject} &		
			\multicolumn{3}{c}{degree} &
			\multicolumn{3}{c}{number of triangles} &
			\multicolumn{3}{c}{sum degree of neighbors}\\
			\cline{2-4}
			\cline{5-7}
			\cline{8-10}	
			& {$\rho$} & {$\chi^2/N$} & {$\zeta$} 
			& {$\rho$} & {$\chi^2/N$} & {$\zeta$} 
			& {$\rho$} & {$\chi^2/N$} & {$\zeta$}\\
			\midrule
			0     &0.875 &0.937 &0.035 &0.913 &0.805 &0.017 &0.873 &0.858 &0.022  \\ 
			1     &0.909 &0.859 &0.027 &0.938 &0.779 &0.025 &0.899 &0.796 &0.029  \\ 
			2     &0.904 &0.802 &0.023 &0.942 &0.713 &0.019 &0.897 &0.712 &0.016  \\ 
			3     &0.894 &1.037 &0.044 &0.929 &1.017 &0.050 &0.888 &0.988 &0.044  \\ 
			4     &0.888 &1.180 &0.068 &0.932 &1.240 &0.079 &0.875 &1.353 &0.084  \\ 
			5     &0.869 &1.093 &0.059 &0.922 &1.047 &0.053 &0.853 &0.981 &0.051  \\ 
			6     &0.897 &0.864 &0.028 &0.936 &0.803 &0.022 &0.881 &0.763 &0.026  \\ 
			7     &0.899 &1.131 &0.065 &0.942 &1.421 &0.105 &0.892 &1.388 &0.088  \\ 
			8     &0.888 &1.064 &0.034 &0.918 &0.952 &0.030 &0.870 &0.934 &0.026  \\ 
			9     &0.866 &1.208 &0.060 &0.905 &1.097 &0.047 &0.840 &1.075 &0.049  \\ 
			10    &0.889 &0.880 &0.026 &0.935 &0.771 &0.022 &0.882 &0.737 &0.017  \\ 
			11    &0.886 &0.933 &0.027 &0.927 &0.841 &0.010 &0.881 &1.159 &0.034  \\ 
			12    &0.888 &0.969 &0.039 &0.932 &0.886 &0.037 &0.883 &0.829 &0.024  \\ 
			13    &0.895 &1.005 &0.043 &0.928 &1.033 &0.049 &0.882 &0.967 &0.046  \\ 
			14    &0.873 &1.052 &0.058 &0.910 &1.086 &0.062 &0.855 &0.948 &0.045  \\ 
			15    &0.888 &0.975 &0.044 &0.932 &0.846 &0.025 &0.870 &0.842 &0.022  \\ 
			16    &0.899 &0.882 &0.029 &0.934 &0.820 &0.029 &0.876 &0.799 &0.019  \\ 
			17    &0.889 &0.976 &0.039 &0.913 &0.965 &0.037 &0.877 &0.916 &0.030  \\ 
			18    &0.860 &1.616 &0.116 &0.906 &2.177 &0.170 &0.858 &2.041 &0.175  \\ 
			19    &0.884 &1.060 &0.056 &0.929 &1.083 &0.064 &0.872 &1.069 &0.061  \\ 
			20    &0.881 &1.032 &0.052 &0.924 &1.094 &0.054 &0.870 &0.995 &0.050  \\ 
			21    &0.884 &1.024 &0.039 &0.922 &0.888 &0.025 &0.865 &0.923 &0.025  \\ 
			22    &0.893 &0.851 &0.023 &0.934 &0.747 &0.020 &0.880 &0.724 &0.011  \\ 
			23    &0.891 &0.829 &0.018 &0.931 &0.719 &0.009 &0.874 &0.797 &0.019  \\ 
			24    &0.897 &0.837 &0.024 &0.932 &0.744 &0.020 &0.884 &0.732 &0.020  \\ 
			25    &0.896 &1.082 &0.062 &0.940 &1.242 &0.085 &0.886 &1.224 &0.076  \\ 
			26    &0.884 &1.077 &0.052 &0.926 &1.151 &0.069 &0.885 &1.124 &0.061  \\ 
			27    &0.894 &0.832 &0.023 &0.928 &0.731 &0.014 &0.884 &0.773 &0.012  \\ 
			28    &0.892 &0.913 &0.019 &0.929 &0.804 &0.009 &0.878 &0.914 &0.020  \\ 
			29    &0.883 &0.885 &0.034 &0.928 &0.773 &0.019 &0.875 &0.725 &0.017  \\ 
			30    &0.891 &0.956 &0.038 &0.928 &0.974 &0.055 &0.872 &0.927 &0.031  \\ 
			31    &0.895 &0.852 &0.026 &0.934 &0.772 &0.023 &0.886 &0.734 &0.019  \\ 
			32    &0.894 &0.903 &0.028 &0.934 &0.815 &0.029 &0.884 &0.788 &0.021  \\ 
			33    &0.894 &0.943 &0.036 &0.936 &0.860 &0.037 &0.880 &0.858 &0.027  \\ 
			34    &0.901 &1.144 &0.061 &0.944 &1.353 &0.088 &0.893 &1.382 &0.084  \\ 
			35    &0.884 &1.494 &0.095 &0.925 &2.074 &0.167 &0.870 &2.204 &0.185  \\ 
			36    &0.872 &0.942 &0.029 &0.915 &0.851 &0.016 &0.852 &0.889 &0.022  \\ 
			37    &0.885 &1.383 &0.095 &0.926 &2.008 &0.141 &0.878 &1.977 &0.159  \\ 
			38    &0.893 &0.951 &0.042 &0.932 &0.948 &0.048 &0.880 &0.887 &0.032  \\ 
			39    &0.915 &0.885 &0.020 &0.946 &0.782 &0.013 &0.899 &0.874 &0.013  \\ 
			40    &0.890 &0.866 &0.029 &0.932 &0.744 &0.019 &0.877 &0.749 &0.016  \\ 
			41    &0.894 &0.938 &0.025 &0.929 &0.839 &0.010 &0.879 &0.981 &0.015  \\ 
			42    &0.880 &0.963 &0.042 &0.918 &0.940 &0.042 &0.860 &0.875 &0.036  \\ 
			43    &0.888 &0.897 &0.028 &0.924 &0.764 &0.019 &0.882 &0.789 &0.017  \\ 	
			
		\end{tabular}
	\end{ruledtabular}
\end{table*}

\clearpage
\newpage
\renewcommand{\arraystretch}{0.55} 

\begin{table*}
	\begin{ruledtabular}
		\centering	
		\caption{The goodness of fit between
			the predictions of our model and real values for degrees, number
			of triangles of nodes, and sum of the degrees of neighbors in layer $0$ for UL dataset. The Pearson correlation coefficient $\rho$, the $\chi^2$ test ($\chi^2=\sum_{i}^{N}(\frac{x_{original}-x_{inferred}}{\sigma})^2$), the quantity $\zeta$ corresponds to the fraction of nodes for which the value measured on the original network lies outside the $2\sigma$ confidence interval.}			
		\begin{tabular}{*{10}{c}}
			\multirow{2}{*}{Subject} &		
			\multicolumn{3}{c}{degree} &
			\multicolumn{3}{c}{number of triangles} &
			\multicolumn{3}{c}{sum degree of neighbors}\\
			\cline{2-4}
			\cline{5-7}
			\cline{8-10}	
			& {$\rho$} & {$\chi^2/N$} & {$\zeta$} 
			& {$\rho$} & {$\chi^2/N$} & {$\zeta$} 
			& {$\rho$} & {$\chi^2/N$} & {$\zeta$}\\
			\midrule
			0     &1.000 &0.011 &0.000 &0.977 &0.944 &0.049 &0.963 &1.419 &0.082  \\ 
			1     &1.000 &0.010 &0.000 &0.979 &0.932 &0.053 &0.963 &1.728 &0.127  \\ 
			2     &1.000 &0.010 &0.000 &0.981 &0.824 &0.034 &0.964 &1.465 &0.096  \\ 
			3     &1.000 &0.010 &0.000 &0.985 &0.769 &0.033 &0.955 &1.721 &0.116  \\ 
			4     &1.000 &0.010 &0.000 &0.980 &0.888 &0.034 &0.968 &1.736 &0.123  \\ 
			5     &1.000 &0.011 &0.000 &0.983 &0.832 &0.038 &0.959 &1.545 &0.088  \\ 
			6     &1.000 &0.011 &0.000 &0.983 &0.751 &0.029 &0.956 &1.647 &0.110  \\ 
			7     &1.000 &0.010 &0.000 &0.983 &0.843 &0.041 &0.956 &1.662 &0.116  \\ 
			8     &1.000 &0.010 &0.000 &0.980 &0.877 &0.040 &0.964 &1.547 &0.089  \\ 
			9     &1.000 &0.011 &0.000 &0.976 &0.713 &0.022 &0.960 &1.545 &0.076  \\ 
			10    &1.000 &0.010 &0.000 &0.973 &1.092 &0.064 &0.966 &1.328 &0.073  \\ 
			11    &1.000 &0.010 &0.000 &0.979 &1.094 &0.067 &0.965 &1.352 &0.087  \\ 
			12    &1.000 &0.010 &0.000 &0.974 &1.016 &0.052 &0.968 &1.496 &0.104  \\ 
			13    &1.000 &0.010 &0.000 &0.989 &0.833 &0.045 &0.969 &1.562 &0.105  \\ 
			14    &1.000 &0.010 &0.000 &0.982 &0.787 &0.035 &0.967 &1.461 &0.089  \\ 
			15    &1.000 &0.009 &0.000 &0.983 &0.823 &0.034 &0.968 &1.416 &0.080  \\ 
			16    &1.000 &0.009 &0.000 &0.985 &0.894 &0.041 &0.966 &1.482 &0.104  \\ 
			17    &1.000 &0.010 &0.000 &0.983 &0.749 &0.032 &0.970 &1.265 &0.074  \\ 
			18    &1.000 &0.010 &0.000 &0.986 &0.653 &0.024 &0.960 &1.387 &0.086  \\ 
			19    &1.000 &0.010 &0.000 &0.983 &0.811 &0.036 &0.962 &1.723 &0.114  \\ 
			20    &1.000 &0.011 &0.000 &0.987 &0.584 &0.013 &0.965 &1.299 &0.072  \\ 
			21    &1.000 &0.010 &0.000 &0.985 &0.665 &0.017 &0.950 &1.865 &0.120  \\ 
			22    &1.000 &0.010 &0.000 &0.978 &0.898 &0.040 &0.958 &1.508 &0.091  \\ 
			23    &1.000 &0.010 &0.000 &0.987 &0.731 &0.025 &0.967 &1.526 &0.091  \\ 
			24    &1.000 &0.011 &0.000 &0.981 &0.698 &0.030 &0.964 &1.351 &0.081  \\ 
			25    &1.000 &0.010 &0.000 &0.989 &0.656 &0.028 &0.969 &1.649 &0.103  \\ 
			26    &1.000 &0.010 &0.000 &0.986 &0.712 &0.022 &0.963 &1.360 &0.089  \\ 
			27    &1.000 &0.010 &0.000 &0.983 &0.669 &0.022 &0.958 &1.439 &0.091  \\ 
			28    &1.000 &0.010 &0.000 &0.982 &1.038 &0.059 &0.968 &1.603 &0.113  \\ 
			29    &1.000 &0.011 &0.000 &0.979 &0.762 &0.033 &0.960 &1.621 &0.102  \\ 
			30    &1.000 &0.011 &0.000 &0.982 &0.732 &0.028 &0.963 &1.540 &0.103  \\ 
			31    &1.000 &0.010 &0.000 &0.986 &0.948 &0.057 &0.970 &1.625 &0.103  \\ 
			32    &1.000 &0.011 &0.000 &0.984 &0.980 &0.053 &0.963 &1.582 &0.104  \\ 
			33    &1.000 &0.011 &0.000 &0.984 &0.912 &0.047 &0.961 &1.629 &0.109  \\ 
			34    &1.000 &0.010 &0.000 &0.982 &0.892 &0.044 &0.964 &1.581 &0.102  \\ 
			35    &1.000 &0.010 &0.000 &0.987 &0.710 &0.032 &0.956 &1.704 &0.135  \\ 
			36    &1.000 &0.010 &0.000 &0.977 &0.939 &0.051 &0.961 &1.623 &0.106  \\ 
			37    &1.000 &0.010 &0.000 &0.980 &0.946 &0.048 &0.962 &1.430 &0.084  \\ 
			38    &1.000 &0.010 &0.000 &0.986 &0.738 &0.033 &0.966 &1.791 &0.110  \\ 
			39    &1.000 &0.010 &0.000 &0.985 &0.862 &0.044 &0.941 &1.916 &0.145  \\ 
			
		\end{tabular}
	\end{ruledtabular}
\end{table*}

\clearpage
\newpage
\renewcommand{\arraystretch}{0.55} 
\begin{table*}
	\begin{ruledtabular}
		\centering	
		\caption{The goodness of fit between
			the predictions of our model and real values for degrees, number
			of triangles of nodes, and sum of the degrees of neighbors in layer $0$ for HCP dataset. The Pearson correlation coefficient $\rho$, the $\chi^2$ test ($\chi^2=\sum_{i}^{N}(\frac{x_{original}-x_{inferred}}{\sigma})^2$), the quantity $\zeta$ corresponds to the fraction of nodes for which the value measured on the original network lies outside the $2\sigma$ confidence interval.}			
		\begin{tabular}{*{10}{c}}
			\multirow{2}{*}{Subject} &		
			\multicolumn{3}{c}{degree} &
			\multicolumn{3}{c}{number of triangles} &
			\multicolumn{3}{c}{sum degree of neighbors}\\
			\cline{2-4}
			\cline{5-7}
			\cline{8-10}	
			& {$\rho$} & {$\chi^2/N$} & {$\zeta$} 
			& {$\rho$} & {$\chi^2/N$} & {$\zeta$} 
			& {$\rho$} & {$\chi^2/N$} & {$\zeta$}\\
			\midrule
			0     &1.000 &0.010 &0.000 &0.987 &1.396 &0.079 &0.983 &1.476 &0.100  \\ 
			1     &1.000 &0.010 &0.000 &0.991 &1.205 &0.055 &0.988 &1.200 &0.069  \\ 
			2     &1.000 &0.011 &0.000 &0.989 &1.252 &0.066 &0.986 &1.232 &0.067  \\ 
			3     &1.000 &0.011 &0.000 &0.991 &1.232 &0.071 &0.988 &1.145 &0.061  \\ 
			4     &1.000 &0.011 &0.000 &0.988 &1.498 &0.104 &0.989 &1.047 &0.047  \\ 
			5     &1.000 &0.010 &0.000 &0.990 &1.212 &0.064 &0.986 &1.158 &0.066  \\ 
			6     &1.000 &0.010 &0.000 &0.991 &1.105 &0.049 &0.987 &1.073 &0.055  \\ 
			7     &1.000 &0.010 &0.000 &0.991 &1.240 &0.073 &0.989 &1.077 &0.053  \\ 
			8     &1.000 &0.011 &0.000 &0.991 &1.280 &0.078 &0.989 &1.152 &0.052  \\ 
			9     &1.000 &0.010 &0.000 &0.989 &1.217 &0.071 &0.987 &1.136 &0.058  \\ 
			10    &1.000 &0.011 &0.000 &0.987 &1.583 &0.093 &0.985 &1.260 &0.066  \\ 
			11    &1.000 &0.010 &0.000 &0.986 &1.117 &0.052 &0.981 &1.341 &0.084  \\ 
			12    &1.000 &0.010 &0.000 &0.988 &1.368 &0.085 &0.984 &1.330 &0.076  \\ 
			13    &1.000 &0.011 &0.000 &0.992 &0.955 &0.035 &0.991 &0.931 &0.037  \\ 
			14    &1.000 &0.010 &0.000 &0.990 &1.417 &0.084 &0.990 &0.934 &0.038  \\ 
			15    &1.000 &0.010 &0.000 &0.990 &1.103 &0.060 &0.983 &1.480 &0.099  \\ 
			16    &1.000 &0.010 &0.000 &0.993 &0.985 &0.045 &0.987 &1.129 &0.061  \\ 
			17    &1.000 &0.010 &0.000 &0.981 &1.949 &0.094 &0.987 &1.265 &0.074  \\ 
			18    &1.000 &0.010 &0.000 &0.988 &1.479 &0.107 &0.988 &1.097 &0.053  \\ 
			19    &1.000 &0.010 &0.000 &0.990 &1.229 &0.069 &0.987 &1.107 &0.060  \\ 
			20    &1.000 &0.010 &0.000 &0.992 &1.037 &0.047 &0.989 &0.991 &0.040  \\ 
			21    &1.000 &0.010 &0.000 &0.992 &1.156 &0.061 &0.988 &1.026 &0.050  \\ 
			22    &1.000 &0.011 &0.000 &0.992 &1.020 &0.036 &0.987 &1.051 &0.058  \\ 
			23    &1.000 &0.010 &0.000 &0.989 &1.238 &0.074 &0.981 &1.425 &0.084  \\ 
			24    &1.000 &0.011 &0.000 &0.989 &1.189 &0.072 &0.981 &1.470 &0.100  \\ 
			25    &1.000 &0.011 &0.000 &0.992 &1.114 &0.066 &0.989 &0.971 &0.047  \\ 
			26    &1.000 &0.011 &0.000 &0.991 &1.061 &0.054 &0.990 &1.016 &0.046  \\ 
			27    &1.000 &0.010 &0.000 &0.987 &1.294 &0.076 &0.984 &1.327 &0.087  \\ 
			28    &1.000 &0.010 &0.000 &0.992 &1.088 &0.060 &0.986 &1.058 &0.049  \\ 
			29    &1.000 &0.010 &0.000 &0.991 &1.231 &0.077 &0.990 &0.834 &0.030  \\ 
			30    &1.000 &0.010 &0.000 &0.990 &1.063 &0.058 &0.988 &1.045 &0.055  \\ 
			31    &1.000 &0.011 &0.000 &0.992 &1.224 &0.068 &0.987 &1.067 &0.053  \\ 
			32    &1.000 &0.011 &0.000 &0.988 &1.788 &0.142 &0.984 &1.437 &0.095  \\ 
			33    &1.000 &0.010 &0.000 &0.991 &1.176 &0.057 &0.983 &1.536 &0.110  \\ 
			34    &1.000 &0.010 &0.000 &0.991 &1.178 &0.065 &0.989 &0.986 &0.035  \\ 
			35    &1.000 &0.010 &0.000 &0.986 &1.548 &0.100 &0.988 &1.122 &0.053  \\ 
			36    &1.000 &0.010 &0.000 &0.990 &0.998 &0.047 &0.988 &0.970 &0.042  \\ 
			37    &1.000 &0.010 &0.000 &0.988 &1.647 &0.119 &0.988 &1.182 &0.062  \\ 
			38    &1.000 &0.011 &0.000 &0.994 &0.930 &0.036 &0.986 &1.204 &0.060  \\ 
			39    &1.000 &0.010 &0.000 &0.991 &1.144 &0.060 &0.986 &1.064 &0.051  \\ 
			40    &1.000 &0.011 &0.000 &0.987 &1.494 &0.077 &0.984 &1.282 &0.076  \\ 
			41    &1.000 &0.010 &0.000 &0.990 &1.183 &0.066 &0.985 &1.172 &0.065  \\ 
			42    &1.000 &0.010 &0.000 &0.991 &1.008 &0.046 &0.988 &0.998 &0.041  \\ 
			43    &1.000 &0.010 &0.000 &0.990 &1.385 &0.088 &0.985 &1.409 &0.097  \\ 
			
		\end{tabular}
	\end{ruledtabular}
\end{table*}
\clearpage
\newpage
\bibliography{reference}

\end{document}